\documentclass[12pt]{article}

\RequirePackage[english]{babel}
\usepackage{charter}

\usepackage[a4paper, total={6.5in, 9in}]{geometry}

\usepackage{amsmath} 
\usepackage{amssymb} 
\usepackage{mathrsfs} 
\usepackage{bbm}

\usepackage{enumitem}

\usepackage{nicefrac}
\usepackage{centernot}

\usepackage{multicol}

\usepackage{amsthm}
\newtheorem{theorem}{Theorem}[section]
\newtheorem{result}{Result}
\newtheorem{proposition}[theorem]{Proposition}
\newtheorem{corollary}[theorem]{Corollary}
\newtheorem{lemma}[theorem]{Lemma}

\theoremstyle{definition}
\newtheorem{definition}[theorem]{Definition}
\theoremstyle{remark}
\newtheorem*{example*}{Example}
\newtheorem{example}[theorem]{Example}

\theoremstyle{remark}
\newtheorem{remark}[theorem]{Remark}

\RequirePackage{graphicx} 

\RequirePackage{hyperref} 
\hypersetup{   
    linkbordercolor = [rgb]{1, 1, 1}, 
    filebordercolor = [rgb]{1, 1, 1},
    citebordercolor = [rgb]{1, 1, 1},
    menubordercolor = [rgb]{1, 1, 1},
    urlbordercolor = [rgb]{1, 1, 1}, 
    colorlinks = true,
    linkcolor = [rgb]{0, 0.2, 0.5},
    citecolor = [rgb]{0, 0.35, 0},
}

\usepackage{color, colortbl}
\definecolor{gray1}{gray}{0.9}

\usepackage{authblk} 
\usepackage[backend=bibtex, style=numeric, url=false, maxnames=10, sortcites=true]{biblatex} 
\AtBeginBibliography{\footnotesize}
\renewbibmacro*{doi+eprint+url}{%
    \printfield{doi}%
    \newunit\newblock%
    \iftoggle{bbx:eprint}{%
        \usebibmacro{eprint}%
    }{}%
    \newunit\newblock%
    \iffieldundef{doi}{%
        \usebibmacro{url+urldate}}%
        {}%
    }
\bibliography{Bibliography}
\newboolean{printBibInSubfiles}
\setboolean{printBibInSubfiles}{true}
\def\bib{\ifthenelse{\boolean{printBibInSubfiles}}
        {\newpage
	\printbibliography
        }
        {}
    }

\usepackage{graphicx}
\graphicspath{ {./Images/} }
\usepackage{tikz-cd}  
\usepackage{pgf,tikz}
\usetikzlibrary{arrows}
\usetikzlibrary{positioning, decorations.pathreplacing}
    
\usepackage{xcolor}
\definecolor{myblue}{RGB}{30, 70, 120}

\newcommand{\refprop}[2][]{{\normalfont\hyperref[#2]{\if #1\empty \else#1~\fi\ref{#2}\!\!}}
} 
\newcommand{\eqrefprop}[2][]{{\normalfont\hyperref[#2]{\if #1\empty \else#1~\fi(\ref{#2})\!\!}}
}

\newcommand{\0}{\mathbf 0}
\newcommand{\1}{\mathbf 1}
\newcommand{\A}{\mathcal{A}}
\newcommand{\aka}{\textit{a.k.a.}~}
\newcommand{\C}{\mathcal{C}}
\newcommand{\CC}{\mathbb{C}}
\newcommand{\CHSH}{\mathtt{CHSH}}
\newcommand{\Cpartition}{\mathscr{C}}
\newcommand{\Dpartition}{\mathscr{D}}
\newcommand{\EE}{\scalebox{1.4}{$\mathbb{E}$}}
\newcommand{\G}{\mathcal{G}}
\renewcommand{\geq}{\geqslant}
\renewcommand{\H}{\mathcal{H}}
\DeclareMathOperator{\Hom}{Hom}

\newcommand{\IdentityMatrix}{\mathbf{I}}
\newcommand{\ie}{\textit{i.e.}~}
\renewcommand{\iff}{if and only if~}
\newcommand{\K}{\mathcal{K}}
\renewcommand{\L}{\mathcal{L}}
\renewcommand{\leq}{\leqslant}
\newcommand{\N}{\mathbb{N}}
\newcommand{\NN}{\mathbb{N}}
\newcommand{\NS}{\mathcal{N\!S}}
\renewcommand{\P}{\mathtt{P}}
\newcommand{\Path}{\mathcal{P}}
\newcommand{\pp}{\mathfrak{p}}
\newcommand{\PP}{\mathbb{P}}
\newcommand{\PR}{\mathtt{PR}}

\newcommand{\Ptilde}{\tilde{\mathtt{P}}}
\newcommand{\Q}{\mathcal{Q}}
\newcommand{\qq}{\mathfrak{q}}
\newcommand{\R}{\mathbb{R}}
\renewcommand{\S}{\mathcal S}
\newcommand{\SR}{\mathtt{SR}}
\newcommand{\Sym}{\mathfrak S}

\DeclareMathOperator{\Aut}{Aut}
\DeclareMathOperator{\diam}{diam}
\DeclareMathOperator{\rel}{rel}

\definecolor{lime}{HTML}{A6CE39}
\DeclareRobustCommand{\orcidicon}{
    \begin{tikzpicture}[every node/.style={scale=0.8}, scale=1]
    \draw[lime, fill=lime] (0,0)
       circle [radius=0.16]
       node[white] {{\fontfamily{qag}\selectfont\tiny\,ID}};
    \draw[white, fill=white] (-0.048,0.095)
       circle [radius=0.007];
    \end{tikzpicture}
}
\newcommand{\myorcid}[1]{
	\hspace{-0.5cm}
	\href{https://orcid.org/#1}{\orcidicon}
	\hspace{-0.4cm}
    }

\usepackage{subfiles} 

\title{\bf Communication Complexity of Graph Isomorphism, Coloring, and Distance Games}
\date{}

\author[1]{Pierre Botteron~\myorcid{0000-0002-3861-0934}\thanks{Pierre Botteron \hspace{0.1cm}\myorcid{0000-0002-3861-0934}\hspace{0.1cm}: {pierre.botteron@math.univ-toulouse.fr}}}

\author[2]{Moritz Weber\thanks{Moritz Weber: weber@math.uni-sb.de}}

\affil[1]{\small Institut de Mathématiques de Toulouse, Université de Toulouse, France.}
\affil[2]{\small Fachbereich Mathematik, Saarland University, Postfach 151150, 66041 Saarbrücken, Germany.}

\begin{document}
\setboolean{printBibInSubfiles}{false}

\maketitle

\vspace{-0.7cm}
\begin{abstract}
In quantum information,
nonlocal games are particularly useful for differentiating classical, quantum, and non-signalling correlations.
An example of differentiation is given by the principle of no-collapse of communication complexity, which is often interpreted as necessary for a feasible physical theory. It
is satisfied by quantum correlations but violated by some non-signalling ones.

In this work, we investigate this principle in the context of three nonlocal games related to graph theory, starting from the well-known graph isomorphism and graph coloring games, and introducing a new game, the vertex distance game, with a parameter $D\in\mathbb N$, that generalizes the former two to some extent. 
For these three games, we prove that perfect non-signalling strategies collapse communication complexity under favorable conditions. 
We also define a refinement of fractional isomorphism of graphs, namely $D$-fractional isomorphisms, and we show that this characterizes perfect non-signalling strategies for the vertex distance game. 
Surprisingly, we observe that non-signalling strategies provide a finer distinction for the new game compared to classical and quantum strategies since the parameter $D$ is visible only in the non-signalling setting.
\end{abstract}

\setcounter{tocdepth}{2}
\tableofcontents

	\section*{Introduction}
	\addcontentsline{toc}{section}{Introduction}	

		\subsection*{Motivation}
		\addcontentsline{toc}{subsection}{Motivation}

The non-signalling theory~($\NS$) is a generalization of the quantum theory~($\Q$), which allows any correlations satisfying the no-faster-than-light communication principle.
Although many physical experiments proved the relevance of the quantum theory~\cite{FC72,AGR82,HBD+15}, the non-signalling theory might be too general and include unrealistic states.

This is where the principle of communication complexity~(CC) comes into play: a collapse of CC is strongly believed to be unachievable in nature~\cite{vD99, BBLMTU06, BS09, BG15}, so one may view this no-collapse principle as an axiom for a feasible physical theory.
In contrast with the quantum theory which does satisfy the no-collapse principle, some of the correlations predicted by the non-signalling theory violate it and thus seem unlikely to exist in nature. In this sense, showing that non-signalling correlations collapse CC contributes to showing that the quantum theory is already ``the best theory'' to some extent, see also the discussion in the \hyperref[{sec:conclusion}]{Conclusion}.

For most of the non-signalling correlations, we do not know whether they collapse CC, the question is open.
In this manuscript, we present new ways to determine collapsing non-signalling correlations, with techniques based on nonlocal games and graphs.

		\subsection*{Preliminary Definitions}
		\addcontentsline{toc}{subsection}{Preliminary Definitions}

Our results connect three notions that we recall below: nonlocal games, nonlocal boxes and communication complexity.
	
\paragraph{Nonlocal Games.} A two-player nonlocal game is a cooperative game played by two characters, commonly named Alice and Bob, who agree on a common strategy $\S$ beforehand, but who are space-like separated during the game, meaning that communication is forbidden. Each of the players is provided with a ``question" $h_A$ (resp. $h_B$) by a referee, and the players prepare their ``answer" $g_A$ (resp. $g_B$) based on the chosen strategy $\S$, possibly using ``shared resources" (like shared randomness or a pair of entangled particles).
Finally, the referee verifies whether the players won the game: he computes a deterministic Boolean function depending on the questions and answers, called the ``rule" of the game.
The goal of Alice and Bob is to maximize the winning probability. If they win almost surely, \ie with probability $1$, then we say that their strategy $\S$ is a \emph{perfect strategy}.

\begin{example*}
	The best-known example of a nonlocal game is the $\CHSH$ game, named after Clauser, Horne, Shimony, and Holt~\cite{CHSH69}. In this game, the questions $h_A,h_B$ and the answers $g_A,g_B$ are classical bits in $\{0,1\}$, and Alice and Bob win if and only if $g_A\oplus g_B = h_A\cdot h_B$, where ``$\oplus$'' is the sum modulo $2$ and ``$\cdot$'' is the product.
	We will comment on the perfect strategies for this game below.
\end{example*}

\paragraph{Strategies and Nonlocal Boxes.} As mentioned above, Alice and Bob are allowed to use shared resources in their strategy $\S$. Depending on the type of allowed resource, the strategy will belong to a certain set of correlations. 
Independently of what type of resource is allowed, the resource can always be modeled by a black box, that is called a \emph{nonlocal box} $\P$, which is shared between Alice and Bob, and which produces some outputs $a,b$ given some inputs $x,y$ based on the joint conditional probability distribution $\P(a,b\,|\,x,y)$, also abbreviated as $\P(ab|xy)$; see a representation in \refprop[Figure]{fig:NLB}. In their strategy $\S$, Alice and Bob can pre-process their data $h_A, h_B$ before inputting $x,y$ in the box, and post-process the outputs $a,b$ of the box in order to produce $g_A, g_B$, but it is equivalent to think of $\P$ as the strategy $\S$ without pre-processing and post-processing.
Find in \refprop[Figure]{fig:famous-types-of-strategies} some famous examples of strategy types. Note that the strict inclusions $\L\subsetneq\Q\subsetneq\NS$ hold due to~\cite{Bell64,PR94}.

\begin{figure}[h]
	\centering
	\begin{tikzpicture}[line cap=round,line join=round,>=triangle 45,x=1.0cm,y=1.0cm,every node/.style={scale=0.9}, scale=0.8]

\newcommand{\mynumber}{0.15}

\draw [line width=2.pt] (-0.2,-1.2 + \mynumber)-- (-0.1,-1.04 + \mynumber);
\draw [line width=2.pt] (-0.32,-1.06 + \mynumber)-- (-0.2,-1.2 + \mynumber);
\draw [shift={(-0.8,-1.2)},line width=2.pt]  plot[domain=0.:1.5707963267948966,variable=\t]({1.*0.6*cos(\t r)+0.*0.6*sin(\t r)},{0.*0.6*cos(\t r)+1.*0.6*sin(\t r) + \mynumber});
\draw [line width=2.pt] (3.2,-1.2 + \mynumber)-- (3.1,-1.04 + \mynumber);
\draw [line width=2.pt] (3.2,-1.2 + \mynumber)-- (3.32,-1.06 + \mynumber);
\draw [shift={(3.8,-1.2)},line width=2.pt]  plot[domain=1.5707963267948966:3.141592653589793,variable=\t]({1.*0.6*cos(\t r)+0.*0.6*sin(\t r)},{0.*0.6*cos(\t r)+1.*0.6*sin(\t r) +  \mynumber});
\draw [line width=2.pt] (-0.8,-2.6)-- (-0.68,-2.48);
\draw [line width=2.pt] (-0.8,-2.6)-- (-0.66,-2.7);
\draw [shift={(3.8,-2.)},line width=2.pt]  plot[domain=3.141592653589793:4.71238898038469,variable=\t]({1.*0.6*cos(\t r)+0.*0.6*sin(\t r)},{0.*0.6*cos(\t r)+1.*0.6*sin(\t r)});
\draw [line width=2.pt] (3.8,-2.6)-- (3.68,-2.48);
\draw [line width=2.pt] (3.8,-2.6)-- (3.66,-2.7);
\draw [shift={(-0.8,-2.)},line width=2.pt]  plot[domain=-1.5707963267948966:0.,variable=\t]({1.*0.6*cos(\t r)+0.*0.6*sin(\t r)},{0.*0.6*cos(\t r)+1.*0.6*sin(\t r)});

\fill[line width=2.pt, color=myblue, fill=myblue, fill opacity=0.15000000596046448] (-1.,-2.) -- (4.,-2.) -- (4.,-1.2) -- (-1.,-1.2) -- cycle;
\draw [line width=2.pt,color=myblue] (-1.,-2.)-- (4.,-2.);
\draw [line width=2.pt,color=myblue] (4.,-2.)-- (4.,-1.2);
\draw [line width=2.pt,color=myblue] (4.,-1.2)-- (-1.,-1.2);
\draw [line width=2.pt,color=myblue] (-1.,-1.2)-- (-1.,-2.);

\draw [line width=1.pt, dashed, color=black] (1.45, 0.)-- (1.45, -1.);
\renewcommand{\mynumber}{-0.18}
\draw [line width=1.pt, dashed, color=black] (1.45, -3.+ \mynumber)-- (1.45, -2.+ \mynumber);

\draw (-1.0, -0.4) node[anchor=east] {$x$};
\draw (3.9, -0.43) node[anchor=west] {$y$};
\draw (-1.0, -2.6) node[anchor=east] {$a$};
\draw (3.9, -2.6) node[anchor=west] {$b$};

\draw [color=myblue](1.45, -1.6) node {\textbf{Nonlocal box}};
\draw[color=black] (-2.2, -1.6) node[anchor=east] {{$\mathscr Alice$}};
\draw[color=black] (5., -1.6) node[anchor=west] {{$\mathscr Bob$}};

\end{tikzpicture}
	\caption{Representation of a nonlocal box.}
	\label{fig:NLB}
\end{figure}
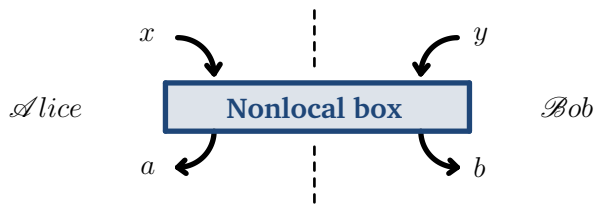

\begin{example*}
	For the $\CHSH$ game mentioned above, optimal classical strategies yield a winning probability of $75\%$; the best quantum strategies yield $\cos^2(\pi/8)\approx85\%$; and the $\PR$ box provides the best winning strategy among non-signalling ones, winning with $100\%$, \ie it is a perfect strategy. In each case, examples of optimal strategies are exhibited in \refprop[Figure]{fig:famous-types-of-strategies}.
\end{example*}

\newcommand{\rowlength}{8cm}
\newcommand{\rrowlength}{4.3cm}
\newcommand{\vadjustement}{0.1cm}
\newcommand{\vvadjustement}{-0.2cm}
\begin{figure}[h]
	\footnotesize
	\centering
	\begin{tabular}{|c|c|c|}
		\hline
		\rowcolor{gray1}
		\begin{tabular}{c}
			\small\bf Strategy \\
			\small\bf Type 
		\end{tabular}
		& 
		\small\bf Definition 
		& 
		\small\bf Examples\\
		\hline
		\hline
		Deterministic 
		& 
		\parbox{\rowlength}{
			\vspace{\vadjustement}
			The outputs $a,b$ are the images by deterministic functions $f, g$ of the inputs $x,y$: 
			\vspace{\vvadjustement}
			\[
				\P(ab|xy)
				\,=\,
				\delta_{a=f(x)} \delta_{b=g(y)}\,,
				\vspace{\vvadjustement}
			\]
			where $\delta$ is the Kronecker delta.
			\vspace{\vadjustement}
		} 
		& 
		\parbox{\rrowlength}{
			\vspace{\vadjustement}
			$\P_0,\P_1$ always output $(0,0),(1,1)$ independently of the inputs:
			\vspace{-0.3cm}
			$$\P_0(ab|xy):=\delta_{a=b=0}\,.$$
			\vspace{-0.8cm}
			$$\P_1(ab|xy):=\delta_{a=b=1}\,.$$
		} 
		\\
		\hline
		\begin{tabular}{c}
			Classical\\
			$\L$
		\end{tabular}
		&
		\parbox{\rowlength}{
			\vspace{\vadjustement}
			Alice and Bob share randomness:
			\vspace{\vvadjustement}
			\[
				\P(ab|xy)=\int_\lambda \mathbb P_A(a\,|\, x, \lambda)\, \mathbb P_B(b\,|\,y, \lambda)\, \mu(\lambda)\,,
				\vspace{\vvadjustement}
			\]
			where $\mathbb P_A, \mathbb P_B, \mu$ are probability distributions, and where the parameter $\lambda$ represents the ``shared randomness".
			\vspace{\vadjustement}
		}
		&
		\parbox{\rrowlength}{
			\vspace{\vadjustement}
			$\SR$ is the \emph{shared randomness box}, outputting $(0,0)$ or $(1,1)$ uniformly:
			\vspace{-0.3cm}
			$$\SR:=\textstyle\frac12\big(\P_0+\P_1\big)\,.$$
		} 
		\\
		\hline
		\begin{tabular}{c}
			Quantum\\
			(commuting)\\
			$\Q$
		\end{tabular}
		&
		\parbox{\rowlength}{
			\vspace{\vadjustement}
			Alice and Bob share a possibly infinite dimensional quantum state $|\Psi\rangle$, \ie $\P$ is of the form
			\[
				\P(ab|xy)= \langle \Psi| E^x_a \circ F^y_b  |\Psi\rangle \,,
			\]
			where $\{E^x_a\}_a,\{F^y_b\}_b$ are commuting POVMs over a Hilbert space $\H_{AB}$,
			and $|\Psi\rangle$ is a norm-$1$ vector of $\H_{AB}$.
			\vspace{\vadjustement}
		}
		&
		\parbox{\rrowlength}{
			\vspace{\vadjustement}
			Take the entangled state $|\Psi\rangle=\frac{1}{\sqrt{2}}\big(|00\rangle + |11\rangle\big)$ and take $E^x_a=A^x_a\otimes I$, $F^y_b=I\otimes B^y_b$, with
			$A^0_a, A^1_a,B^0_b,B^1_b$ the measurements in the rotated computational basis $B_\theta$ 
			with angles $\theta=0,\pi/4,\pi/8,-\pi/8$ respectively.
			\vspace{\vadjustement}
		}
		\\
		\hline
		\begin{tabular}{c}
			Non-\\
			signalling\\
			$\NS$
		\end{tabular}
		&
		\parbox{\rowlength}{
			\vspace{\vadjustement}
			Alice and Bob share any resource without faster-than-light communication, \ie $\P$ satisfies the condition that the marginal of Alice's output is independent of Bob's input:
			\vspace{\vvadjustement}
			$$
				\sum_{b}\P(ab|xy) = \!\sum_{b}\P(ab|xy') =: \P(a|x)\,,
				\vspace{\vvadjustement}
			$$
			for all $a,x,y,y'$,
			and similarly for $\P(b|y)$,
			in addition to $\P$ being a well-defined probability distribution: $\P(ab|xy)\geq0$ and $\sum_{ab} \P(ab|xy) = 1$.
			\vspace{\vadjustement}
		}
		&
		\parbox{\rrowlength}{
			\vspace{\vadjustement}
			The $\PR$ box, named after Popescu and Rohrlich~\cite{PR94}, is designed to perfectly win the $\CHSH$ game:
			$$
				\PR(ab|xy)
				=
				\textstyle \frac12 \delta_{a\oplus b=x\cdot y}\,,
			$$
			where ``$\oplus$'' is the sum modulo $2$ and ``$\cdot$'' is the product.
			\vspace{\vadjustement}
		} 
		\\
		\hline
	\end{tabular}
	\caption{Typical types of strategies.}
	\label{fig:famous-types-of-strategies}
\end{figure}

\paragraph{Communication Complexity.}
\emph{Communication complexity} (CC) is a notion that was introduced by Yao in~\cite{Yao79}
and later reviewed in~\cite{KN96, RY20}. 
It quantifies the difficulty of performing a distributed computation.
Say we have access to two distant computers and we want to compute the value of a Boolean function $f:\{0,1\}^n\times\{0,1\}^m\to\{0,1\}$ evaluated at some $(X,Y)$, where the $X\in\{0,1\}^n$ is given to the first computer and $Y\in\{0,1\}^m$ to the other, with the constraint of minimizing the cost of communication bits between the computers.
The CC of the function $f$ is then defined as the minimal number of bits that the computers need to communicate so that the first computer is able to compute the value $f(X,Y)\in\{0,1\}$.
For instance, when $n=m=2$, $X=(x_1, x_2)$, $Y=(y_1, y_2)$, the CC of $f_1:= x_1\cdot(y_1 \oplus y_2)$ equals~$1$, using the communication bit $y_1 \oplus y_2$, which can be computed by the second computer and then communicated to the first computer as it has access to $Y$. However, it is possible to show that the CC of $f_2:= (x_1\cdot y_1) \oplus (x_2\cdot y_2)$ equals~$2$, using communication bits $y_1$ and $y_2$, which then means that $f_2$ is more complex than $f_1$ in the sense of CC.
Yao also introduced in~\cite{Yao79} a probabilistic version of CC, in which the computers can access shared randomness, and where the condition is relaxed: there exists $p>1/2$ such that for all $X,Y$ the first computer has to output the correct value of $f(X,Y)$ with probability at least $p$.
This notion is linked with nonlocal boxes: if a nonlocal box $\P$ is used in the protocol to compute the value $f(X,Y)$ and if the probabilistic CC of \emph{any} function $f$ is at most $1$ independently of its entry size and with fixed $p$, then we say that the box $\P$ \emph{collapses} communication complexity.
In this definition, an arbitrary number of copies of the box $\P$ can be used in the protocol.
Such a collapse is strongly believed to be unachievable in nature since it would imply the absurdity that a single bit of communication is sufficient to distantly estimate any value of any Boolean function $f$~\cite{vD99, BBLMTU06, BS09, BG15}.

\begin{example*}
	The $\PR$ box is known to collapse CC~\cite{vD99}, so this correlation is physically unfeasible according to the principle of communication complexity.
	More generally, we know that some noisy versions of the $\PR$ box also collapse CC for different types of noise~\cite{BBLMTU06, BS09, BMRC19, Botteron22, BBP23, EWC22PRL,BBCNP23}.
	On the other hand, we know that quantum correlations do \emph{not} collapse communication complexity~\cite{CvDNT99}, and neither does a slightly wider set named ``almost quantum correlations"~\cite{NGHA15}.
	To this day, the question is still open whether the remaining non-signalling boxes are collapsing, meaning that there is still a gap to be filled.
	We refer to \cite[Fig. 2]{BBP23} for a figure that summarizes the contributions.
\end{example*}

		\subsection*{Main Results}
		\addcontentsline{toc}{subsection}{Main Results}
		
In this work, we study three nonlocal games related to graphs: the graph isomorphism game (\refprop[Section]{section:graph-isom-game}), the graph coloring game (\refprop[Section]{section:graph-homomorphism-game}), and we also define a new game, the vertex distance game (\refprop[Section]{section:vertex-distance-game}), depending on a parameter $D\in\NN$. All graphs are always assumed to be non-empty, finite, undirected, and loopless. In the sequel, let $\G$ be a given graph, with vertex set $V(\G)$, and write $g\sim g'$ if two vertices $g,g'\in V(\G)$ are linked by an edge. 
Let $\H$ be graph with disjoint vertex set $V(\G)\cap V(\H)=\emptyset$.
We refer to~\cite{Godsil-Royle-01} for more background on graph theory.

\paragraph{Graph Isomorphism Game~\cite{AMRSSV19}.} For the well-known \emph{graph isomorphism game} $(\G,\H)$, Alice and Bob receive vertices $x_A,x_B\in V=V(\G)\cup V(\H)$ and they respond with vertices $y_A,y_B\in V$. The first winning condition is that the set $\{x_A, y_A\}$ consists in exactly one vertex from $V(\G)$, that we call $g_A\in V(\G)$, and the other from $V(\H)$, called $h_A\in V(\H)$; 
and similarly for $\{x_B,y_B\}$ giving rise to $g_B\in V(\G)$ and $h_B\in V(\H)$.
The second winning condition condition is that $g_A$ and $g_B$ are related in the same way as $h_A$ and $h_B$ are related, in the sense that\vspace{-5pt}
\begin{enumerate}[label=(\roman*),itemsep=-3pt]
	\item \label{item:isom-game-rule-1} if $g_A=g_B$, then $h_A=h_B$, 
	\item \label{item:isom-game-rule-2} if $g_A\sim g_B$, then $h_A\sim h_B$,
	\item \label{item:isom-game-rule-3} if $g_A\not\simeq g_B$, then $h_A\not\simeq h_B$,
\end{enumerate}
where the symbol $\simeq$ means equal or linked by an edge. Note that the three implications in items~\ref{item:isom-game-rule-1},~\ref{item:isom-game-rule-2}, and~\ref{item:isom-game-rule-3}, are actually equivalences. See \refprop[Section]{section:graph-isom-game} for details. 
For the graph isomorphism game, we prove the following result:

\begin{result}[Thm.~\ref{theo:toy-example-thm}, Cor.~\ref{coro:toy-example-thm}, Thm.~\ref{theo: collapse of CC}, Thm.~\ref{theo:collapse-for-ALL-strategies}, Cor.~\ref{coro:cannot-be-quantum}]
Given $\G$ and $\H$ for the graph isomorphism game, we have:\vspace{-3pt}
\begin{enumerate}[label=(\alph*),itemsep=-3pt]
	\item If the diameter of $\G$ satisfies $\diam(\G)\geq2$ and if $\H$ has exactly two connected components which are both complete, then any perfect non-signalling strategy collapses CC. This may be weakened to strategies winning with probability $\pp>\frac{3+\sqrt 6}{6}\approx 0.91$.
	\item If $\diam(\G)\geq2$, if $\H$ is not connected, and if there is a common equitable partition with an additional technical assumption~\ref{eq: assumption sufficient condition proposition}, then there is a non-signalling strategy which collapses CC. 
	\item With the same conditions as in the previous item, and if $\H$ is additionally strongly transitive (a generalization of the notion of \emph{transitivity} from graph automorphism theory) and $d$-regular, and if Alice and Bob share randomness, then any perfect non-signalling strategy collapses CC. As a consequence, these strategies cannot be quantum.
\end{enumerate}
\end{result}

\paragraph{Graph Coloring Game~\cite{Cameron-Newman-Montanaro-etal-06}.} The graph isomorphism game can be relaxed to a \emph{graph homomorphism game} $\G\to\H$ omitting item~\ref{item:isom-game-rule-3} in the above game and with questions $x_A,x_B$ always lying in $V(\G)$. If $\H=\K_N$, the complete graph on $N$ points, then the graph homomorphism game $\G\to\K_N$  is called the \emph{graph coloring game}\footnote{Recent works by Assadi, Chakrabarti, Ghosh, and Stoeckl~\cite{Assadi-Chakrabarti-Ghosh-Stoeckl-23} and then by Flin and Mittal~\cite{Flin-Mittal-24} also study the link between the graph coloring game and CC: the authors upper-bound the minimal number of communication bits required to compute a coloration of a graph with two distant parties.}. For this game, we show the following result:

\begin{result}[Thm.~\ref{thm:collapse-of-CC-for-the-homomorphism-game}, Thm.~\ref{thm:combining-isomorphism-game-and-coloring-game}]
Given $\G$ and $\H$ for the graph homomorphism game resp. the graph coloring game, we have:\vspace{-3pt}
\begin{enumerate}[label=(\alph*),itemsep=-3pt]
	\item For any non-signalling strategy winning the homomorphism game  $\K_3\to\G$ with probability $\pp$, together with another non-signalling strategy winning the graph coloring game $\G\to\K_2$ with probability $\qq$, such that $\pp\qq>\frac{3+\sqrt 6}{6}\approx 0.91$, there is a collapse of CC.
	\item If $\diam(\G)\geq2$ and if $\H$ has exactly $N$ connected components all of which are complete, then for any non-signalling strategy winning the graph isomorphism game $\G\to\H$ with probability $\pp$, combined with any non-signalling strategy winning the coloring game $\K_N\to\K_2$ with probability $\qq$, such that $\pp\qq>\frac{3+\sqrt 6}{6}\approx 0.91$, there is a collapse of CC.
\end{enumerate}
\end{result}

\paragraph{Vertex Distance Game [new].} In \refprop[Section]{section:vertex-distance-game}, we describe a new game that we call \emph{vertex $D$-distance game}, with a parameter $D\in\NN$. This is a generalization of the graph isomorphism game, changing the winning conditions~\ref{item:isom-game-rule-1},~\ref{item:isom-game-rule-2},~\ref{item:isom-game-rule-3} into: \vspace{-5pt}
\begin{enumerate}[label=(\roman*),itemsep=-3pt]
	\item if $d(g_A,g_B)=t\leq D$, then $d(h_A,h_B)=t$,
	\item if $d(g_A,g_B)>D$, then $d(h_A,h_B)>D$.
\end{enumerate}
For $D=0$, if we consider the graphs $\G=\K_M$ and $\H=\K_N$, this exactly corresponds to the $N$-coloring game of $\K_M$.
For $D=1$, this is the graph isomorphism game. We show that neither classical nor quantum strategies may distinguish the vertex $D$-distance games for different parameters $D$:

\begin{result}[Prop.~\ref{prop:same-perfect-classical-and-quantum-strategies}, Thm.~\ref{theo:strict-ns-D-isomorphisms}]
	For any $D\geq 1$, perfect classical and quantum strategies are precisely the same for the $D$-distance game as for the isomorphism game. However, in the non-signalling setting, the perfect strategies for the two games differ.
\end{result}

More precisely, for any $D\in \NN$, there is a pair of graphs which admits a perfect non-signalling strategy for the vertex $D$-distance game but not for the vertex $(D+1)$-distance game, see \refprop[Proposition]{prop:ExampleDButNotDPlusOne}. So, non-signalling strategies provide a finer tool for distinguishing nonlocal games. 
Moreover, our definition of a vertex $D$-distance game produces a notion of \emph{$D$-fractional isomorphism}, see details in \refprop[Subsection]{subsec:perfect-NS-strategies}, and we obtain the chain of strict implications drawn in \refprop[Figure]{fig:chain-of-strict-implications-INTRO}.

\begin{figure}[h]
	\centering
	\begin{tikzpicture}[line cap=round,line join=round,>=triangle 45,x=1.0cm,y=1.0cm,every node/.style={scale=0.8}, scale=0.7]

\newcommand{\myheight}{1.8}
\newcommand{\mywidthA}{2.1}
\newcommand{\mywidthB}{2.1}
\newcommand{\mywidthC}{2.6}
\newcommand{\mywidthD}{2.7}
\newcommand{\mywidthE}{2.6}
\newcommand{\mywidthF}{2.6}
\newcommand{\mysep}{1.7}
\newcommand{\myyshift}{3.4em}
\newcommand{\myxshift}{3.4em}

\definecolor{mycolor}{RGB}{30, 70, 120} 
\definecolor{mycolor2}{HTML}{53257F} 
\definecolor{mycolor3}{RGB}{191, 95, 0}
\definecolor{mycolor4}{RGB}{0, 96, 81}  
\definecolor{mycolor5}{RGB}{96, 0, 81}
\definecolor{mycolor6}{RGB}{150, 30, 30} 

\draw[line width=2.pt, color=mycolor4, fill=mycolor4, fill opacity=0.15] (0,0) rectangle ++(\mywidthA,\myheight) ;
\draw[line width=2.pt, color=mycolor4] (0,0) rectangle node {\color{mycolor4}$\G\cong\!\H$} ++(\mywidthA,\myheight) ;
\node (A) at (\mywidthA+0.5*\mysep, 0.7*\myheight) {$\Longrightarrow$};
\node[above=of A, yshift=-\myyshift] {\scriptsize$\L\subseteq\Q$};
\node (A') at (\mywidthA+0.5*\mysep, 0.3*\myheight) {$\centernot\Longleftarrow$};
\node[below=of A', yshift=\myyshift] {\scriptsize\cite{AMRSSV19}};

\draw[line width=2.pt, color=mycolor4, fill=mycolor4, fill opacity=0.15] (0,-\myheight-\mysep) rectangle ++(\mywidthA,\myheight) ;
\draw[line width=2.pt, color=mycolor4] (0,-\myheight-\mysep) rectangle node {\color{mycolor4}$\begin{matrix}~\\\G\cong^{D}\!\H\\\text{\scriptsize$\forall D\in\NN$}\end{matrix}$} ++(\mywidthA,\myheight) ;
\node (A'') at (0.5*\mywidthA, -0.5*\mysep) {$\Big\Updownarrow$};
\node[right=of A'', xshift=-\myxshift] {\scriptsize Prop~\ref{prop:same-perfect-classical-and-quantum-strategies}};

\draw[line width=2.pt, color=mycolor6, fill=mycolor6, fill opacity=0.15] (\mywidthA+\mysep,0) rectangle ++(\mywidthB,\myheight) ;
\draw[line width=2.pt, color=mycolor6] (\mywidthA+\mysep,0) rectangle node {\color{mycolor6}$\G\cong_q\!\H$} ++(\mywidthB,\myheight) ;
\node (A') at (\mywidthA+\mywidthB+1.5*\mysep, 0.5*\myheight) {$\centernot\Longleftarrow$};
\node[above=of A', yshift=-\myyshift] {\footnotesize??};

\draw[line width=2.pt, color=mycolor6, fill=mycolor6, fill opacity=0.15] (\mywidthA+\mysep,-\myheight-\mysep) rectangle ++(\mywidthA,\myheight) ;
\draw[line width=2.pt, color=mycolor6] (\mywidthA+\mysep,-\myheight-\mysep) rectangle node {\color{mycolor6}$\begin{matrix}~\\\G\cong_q^{D}\!\H\\\text{\scriptsize$\forall D\in\NN$}\end{matrix}$} ++(\mywidthA,\myheight) ;
\node (A'') at (\mywidthA+\mysep+0.5*\mywidthB, -0.5*\mysep) {$\Big\Updownarrow$};
\node[right=of A'', xshift=-\myxshift] {\scriptsize Prop~\ref{prop:same-perfect-classical-and-quantum-strategies}};

\draw[>=triangle 45, -{to[scale=2]}, double] (\mywidthA+\mywidthB+\mysep+0.3, -0.5*\myheight-\mysep+0.1) to [out=0,in=270] node[below right] {\scriptsize $\Q\subseteq\NS$} (\mywidthA+\mywidthB+0.5*\mywidthC+2*\mysep-0.1, -0.3);

\draw[line width=2.pt, color=mycolor2, fill=mycolor2, fill opacity=0.15] (\mywidthA+\mywidthB+2*\mysep,0) rectangle ++(\mywidthC,\myheight) ;
\draw[line width=2.pt, color=mycolor2] (\mywidthA+\mywidthB+2*\mysep,0) rectangle node {\color{mycolor2}$\begin{matrix}~\\\G\cong_{frac}^{D}\!\H\\\text{\scriptsize$\forall D\in\NN$}\end{matrix}$} ++(\mywidthC,\myheight) ;
\node (A) at (\mywidthA+\mywidthB+\mywidthC+2.5*\mysep, 0.7*\myheight) {$\Longrightarrow$};
\node[above=of A, yshift=-\myyshift] {\scriptsize By def.};
\node (A') at (\mywidthA+\mywidthB+\mywidthC+2.5*\mysep, 0.3*\myheight) {$\centernot\Longleftarrow$};
\node[below=of A', yshift=\myyshift] {\scriptsize Prop~\ref{prop:ExampleDButNotDPlusOne}};

\draw[line width=2.pt, color=mycolor2, fill=mycolor2, fill opacity=0.15] (\mywidthA+\mywidthB+\mywidthC+3*\mysep,0) rectangle ++(\mywidthD,\myheight) ;
\draw[line width=2.pt, color=mycolor2] (\mywidthA+\mywidthB+\mywidthC+3*\mysep,0) rectangle node {\color{mycolor2}$\G\cong_{frac}^{D_0+1}\!\H$} ++(\mywidthD,\myheight) ;
\node (A) at (\mywidthA+\mywidthB+\mywidthC+\mywidthD+3.5*\mysep, 0.7*\myheight) {$\Longrightarrow$};
\node[above=of A, yshift=-\myyshift] {\scriptsize By def.};
\node (A') at (\mywidthA+\mywidthB+\mywidthC+\mywidthD+3.5*\mysep, 0.3*\myheight) {$\centernot\Longleftarrow$};
\node[below=of A', yshift=\myyshift] {\scriptsize Prop~\ref{prop:ExampleDButNotDPlusOne}};

\draw[line width=2.pt, color=mycolor2, fill=mycolor2, fill opacity=0.15] (\mywidthA+\mywidthB+\mywidthC+\mywidthD+4*\mysep,0) rectangle ++(\mywidthE,\myheight) ;
\draw[line width=2.pt, color=mycolor2] (\mywidthA+\mywidthB+\mywidthC+\mywidthD+4*\mysep,0) rectangle node {\color{mycolor2}$\G\cong_{frac}^{D_0}\!\H$} ++(\mywidthE,\myheight) ;
\node (A) at (\mywidthA+\mywidthB+\mywidthC+\mywidthD+\mywidthE+4.5*\mysep, 0.7*\myheight) {$\Longrightarrow$};
\node[above=of A, yshift=-\myyshift] {\scriptsize By def.};
\node (A') at (\mywidthA+\mywidthB+\mywidthC+\mywidthD+\mywidthE+4.5*\mysep, 0.3*\myheight) {$\centernot\Longleftarrow$};
\node[below=of A', yshift=\myyshift] {\scriptsize Prop~\ref{prop:ExampleDButNotDPlusOne}};

\draw[line width=2.pt, color=mycolor, fill=mycolor, fill opacity=0.15] (\mywidthA+\mywidthB+\mywidthC+\mywidthD+\mywidthE+5*\mysep,0) rectangle ++(\mywidthF,\myheight) ;
\draw[line width=2.pt, color=mycolor] (\mywidthA+\mywidthB+\mywidthC+\mywidthD+\mywidthE+5*\mysep,0) rectangle node {\color{mycolor}$\begin{matrix}~\\\G\cong_{frac}\!\H\\\text{\scriptsize($D_0=1$)}\end{matrix}$} ++(\mywidthF,\myheight) ;

\end{tikzpicture}
	\caption{Chain of strict implications, with $D_0\geq2$ fixed.
	}
	\label{fig:chain-of-strict-implications-INTRO}
\end{figure}
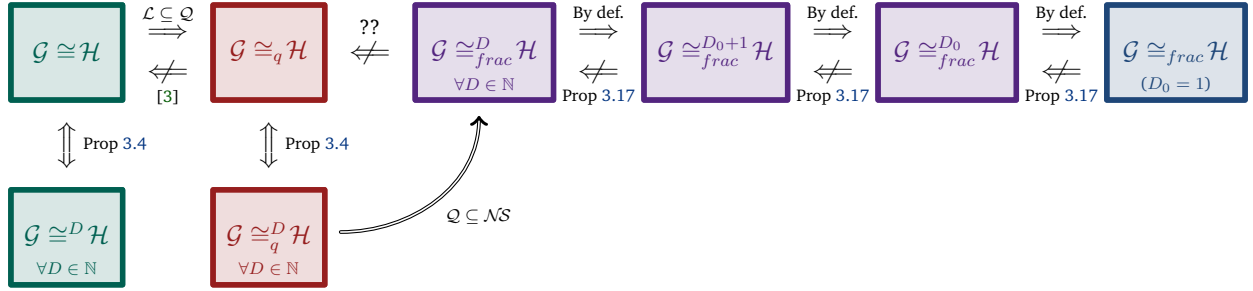

We also characterize perfect non-signalling strategies for the vertex $D$-distance game by adapting results from~\cite{AMRSSV19, RSU94}:

\begin{result}[Thm.~\ref{thm:Characterization-of-perfect-NS-strategies}]
The following are equivalent:\vspace{-5pt}
\begin{enumerate}[label=(\roman*),itemsep=-3pt]
	\item The graphs $\G$ and $\H$ are $\NS$-isomorphic in the sense of the $D$-distance game.
	\item The graphs $\G$ and $\H$ are $D$-fractionally isomorphic.
	\item There exists a $D$-common equitable partition of $\G$ and $\H$.
\end{enumerate}
\end{result}

This characterization allows us to finally study the collapse of CC for this game:

\begin{result}[Thm.~\ref{theo:existence-of-collapsing-strategy-for-the-D-distance-game}, Prop.~\ref{prop:easy-case-D-distance-game}, Prop.~\ref{prop:easy-case-D-distance-game-2}, Thm.~\ref{theo:collapse-for-ALL-strategies-D-distance-game}]
Given $\G$ and $\H$ for the vertex $D$-distance game:\vspace{-3pt}
\begin{enumerate}[label=(\alph*),itemsep=-3pt]
	\item If $1\leq D<\diam(\G)$, if $\H$ is not connected, and if the graphs $(\G,\H)$ admit a $D$-common equitable partition with technical assumption~\ref{eq:H-prime}, then there exists a perfect strategy collapsing CC.
	\item If $1\leq D \leq \diam(\H) < \diam(\G)$ and if $\H$ admits exactly two connected components, then any perfect non-signalling strategy collapses CC. This may be weakened to strategies winning with probability $\pp>\frac{3+\sqrt{6}}{6} \approx 0.91$.
	\item If $1\leq D \leq \diam(\H) < \diam(\G)$ and if $\H$ admits exactly $N$ connected components, then for any perfect non-signalling strategy for the $D$-distance game, combined with a perfect non-signalling strategy for the coloring game $\K_N\to\K_2$, there is a collapse of CC. This may be weakened to strategies winning with respective probabilities $\pp$ and $\qq$ such that $\pp\qq>\frac{3+\sqrt{6}}{6} \approx 0.91$.
	\item If $2\leq \diam(\G)$, if $\H$ is not connected, if the graphs $(\G,\H)$ admit a $D$-common equitable partition with technical assumption~\ref{eq:H-prime}, and if $\H$ is strongly transitive and regular, then any perfect strategy collapses CC.
\end{enumerate}
\end{result}

\bib

\section{Graph Isomorphism Game}
\label{section:graph-isom-game}

In this section, we show the collapse of communication complexity for some perfect strategies for the graph isomorphism game. 
In the coming subsections, after presenting the definition of the game in \refprop[Subsection]{subsec:definition-of-graph-isom-game}, 
we prove a key preliminary result in \refprop[Subsection]{subsec:preliminary-result-for-graph-isom-game}, 
then we introduce the new family of symmetric strategies in \refprop[Subsection]{subsec:symmetric-strategies},
and finally we state and prove our main theorems of this section in \refprop[Subsections]{subsec:link-with-communication-complexity} and~\ref{para:All-Perfect-Strategies-Collapse-CC}.

	\subsection{Definition of the Game}
	\label{subsec:definition-of-graph-isom-game}

The graph isomorphism game was introduced by Atserias, Man\v{c}inska, Roberson, \v{S}ámal, Severini, and Varvitsiotis in~\cite{AMRSSV19}.
This is a nonlocal game based on two graphs, namely $\G$ and $\H$, in which the players Alice and Bob try to pretend to a referee that the two graphs are isomorphic in the classical sense. Recall that $\G$ is said \emph{isomorphic} to $\H$, denoted $\G\cong\H$, if there exists a bijection $\varphi$ from the vertex set $V(\G)$ to the vertex set $V(\H)$ such that adjacency is preserved in both ways:
\begin{equation}  \label{eq:def-of-a-graph-isomorphism}
	\forall g,g'\in\G, \quad\quad
	g \sim g'
	\quad\Longleftrightarrow\quad
	\varphi(g)\sim\varphi(g')\,,
\end{equation}
where the symbol ``$\sim$'' denotes the adjacency relation. 

The game consists in the following. The referee provides the players Alice and Bob with respective questions $x_A,x_B\in V=V(\G)\cup V(\H)$, where $V(\G)$ and $V(\H)$ are assumed to be disjoint. In return, Alice and Bob use a predetermined strategy $\S$ in order to produce some vertices $y_A,y_B\in V$, and they send $y_A,y_B$ to the referee, who finally verifies if the players won the game. 
The first condition they need to satisfy is that $x_A$ and $y_A$ have to be in different vertex sets, and similarly for $x_B$ and $y_B$, meaning that
\begin{equation} \label{eq:rule-1-of-graph-isom-game}
x_A\in V(\G) \Leftrightarrow y_A\in V(\H)
\quad\quad\text{and}\quad\quad
x_B\in V(\G) \Leftrightarrow y_B\in V(\H)\,,
\end{equation}
otherwise they lose the game. Now, assuming that this condition holds, only one vertex among $x_A$ and $y_A$ is in $V(\G)$, let us call it $g_A\in V(\G)$, and the other $h_A\in V(\H)$, and similarly for $g_B\in V(\G)$ and $h_B\in V(\H)$.
The second condition for Alice and Bob to win the game is that $g_A$ has the same relation to $g_B$ as $h_A$ has to $h_B$, that is the three following equivalences are satisfied:
\begin{equation}
	g_A = g_B \Leftrightarrow h_A = h_B\,,
	\quad\quad
	g_A \sim g_B \Leftrightarrow h_A \sim h_B\,,
	\quad\quad 
	g_A \not\simeq g_B \Leftrightarrow h_A \not\simeq h_B\,,
\end{equation}
where the symbol ``$\not\simeq$'' means neither equal nor adjacent.\footnote{\label{footnote:the-inputs-are-not-necessarily-in-G}In this work, we use the original definition from~\cite{AMRSSV19}. There exists a simpler variant of the graph isomorphism game, in which $x_A$ and $x_B$ are always given in $V(\G)$. As explained in \cite[Remark~2.3]{Roberson-Schmidt-21}, in the classical and quantum settings, if the graphs $\G$ and $\H$ are assumed to have the same number of vertices, then this simpler version is equivalent to the original version, but they differ in the non-signalling setting. Notably, we will need throughout this manuscript a characterization of perfect non-signalling strategies in terms of common equitable partition and fractional isomorphism, which holds only in the original setting.} The game is described in terms of the types of strategies that the players are allowed to use. For instance, in the \emph{quantum graph isomorphism game}, Alice and Bob can use quantum strategies to produce $y_A,y_B$ out of $x_A,x_B$, and similarly for the other types of strategies described in \refprop[Figure]{fig:famous-types-of-strategies}. It is possible to prove that if there exists a perfect strategy for this game, then the graphs $\G$ and $\H$ have to have the same number of vertices, see~\cite[page~302]{AMRSSV19}.

If the graphs $\G$ and $\H$ are actually isomorphic, with bijection $\varphi$, Alice and Bob can perfectly win the game by simply answering $h_A=\varphi(g_A)$ and $h_B=\varphi(g_B)$. Conversely, assume that this game admits a perfect deterministic strategy, in the sense of the definition in \refprop[Figure]{fig:famous-types-of-strategies}, then the deterministic behavior of Alice to produce $h_A$ out of $g_A$ defines an isomorphism between the graphs $\G$ and $\H$. As a result, the isomorphism $\G\cong\H$ holds if and only if Alice and Bob can perfectly win the deterministic isomorphism game. More generally, we can extend this result by convexity to the set of classical strategies $\L$, and we have again that $\G\cong\H$ if and only if Alice and Bob can perfectly win the classical isomorphism game.

Now, even if $\G$ and $\H$ are not isomorphic, Alice and Bob can try to mimic it to the referee, using their nonlocal box to correlate the answer. We say that $\G$ and $\H$ are quantum isomorphic, denoted $\G\cong_q\H$, if Alice and Bob can perfectly win the game using quantum strategies; and similarly $\G$ and $\H$ are non-signalling isomorphic, denoted $\G\cong_{ns}\H$, using non-signalling strategies, in the sense as described in \refprop[Figure]{fig:characterizations-of-the-different-types-of-isomorphism}, see also~\cite{AMRSSV19} for details. These equivalence relations are a relaxed version of the usual isomorphism $\cong$ in the sense that:
$$
	\G\cong\H 
	\quad\Longrightarrow\quad
	\G\cong_q\H 
	\quad\Longrightarrow\quad
	\G\cong_{ns}\H \,.
$$
Man\v{c}inska and Roberson proved that, surprisingly, quantum isomorphism is characterized in terms of counting homomorphisms from planar graphs~\cite{MR20}, and with a larger team they showed that non-signalling is equivalent to fractional isomorphism~\cite{AMRSSV19}. These two results, in addition to many others, are summarized in \refprop[Figure]{fig:characterizations-of-the-different-types-of-isomorphism}. 
For the sake of completeness, we recall that the adjacency matrix $A_\G$ of a graph $\G$ with $n$ vertices $g_1,\dots,g_n$ is an $n\times n$ matrix defined using the set of edges of $\G$, where the coefficient $a_{ij}$ of the matrix is set to $1$ if $g_i\sim g_j$, and to $0$ otherwise. The notion of graph homomorphism will be recalled in \refprop[Section]{section:graph-homomorphism-game}.
Note also that Reference~\cite{AMRSSV19} gives examples of graphs $\G,\H$ such that $\G\cong_q\H$ but $\G\not\cong\H$, and others such that $\G\cong_{ns}\H$ but $\G\not\cong_q\H$, and they prove that the problem of determining whether $\G\cong_q\H$ is undecidable.
Other related results may be found in~\cite{CY23,FWM23}.

\begin{figure}[h]
	\footnotesize
	\centering
	$\begin{array}{| c | c | c |}
		\hline
		\rowcolor{gray1}
		\textbf{\small \vspace{0.5cm}Isom.\vspace{0.5cm}} & \textbf{\small Adjacency matrices} & \textbf{\small Homomorphism countings} \\
		\hline
		\hline
		\G \cong \H &  
		\begin{matrix} 
			\exists \text{ permutation matrix $u$}\\
			\text{s.t. $A_\G u = u A_\H$~\cite[Lem~3.1]{AMRSSV19}} \\ 
			(\text{equiv.: } \exists \text{ quantum permutation matrix $u$} \\ \text{with commuting entries}\\
			\text{s.t. $A_\G u = u A_\H$ \text{ \cite[Thm~II.1]{MR20}} }) 
		\end{matrix}
		& 
		\begin{matrix}
			\bullet~\forall \text{ graph } \K,\\ 
			\# \Hom(\K, \G) \!=\! \# \Hom(\K, \H)
			\text{ \cite[Eq~(5)]{Lovasz67}}\\
			\bullet~\forall \text{ graph }\K,\\ 
			\# \Hom(\G, \K) = \# \Hom(\H, \K)
			\text{ \cite{CV93}}
		\end{matrix}\\
		\hline 
		\G \cong_{q} \H & 
		\begin{matrix}
			\exists \text{ quantum permutation matrix $u$}\\ \text{s.t. $A_\G u = u A_\H$ \cite[Thm~4.4]{LMR20} } 
		\end{matrix}
		& 
		\begin{matrix}
			\forall \text{ planar graph }\K,
			\\ 
			\# \Hom(\K, \G) = \# \Hom(\K, \H)
			\\
			\text{ \cite[Main~Thm]{MR20}} 
		\end{matrix}
		\\
		\hline
		\begin{matrix}
			\G \cong^{D}_{ns} \H \\
			\text{[new, \refprop[Sec.]{section:vertex-distance-game}]}
		\end{matrix}	
		& 
		\begin{matrix} 
			\exists \text{ bistochastic matrix $u$}\\ 
			\text{s.t. $A^{(t)}_\G u = u A^{(t)}_\H$ $\forall t\leq D$~[new, \refprop[Thm.]{thm:Characterization-of-perfect-NS-strategies}]}\\ 
			\text{(i.e. $D$-fractionally isomorphic)} 
		\end{matrix} 
		& 
		\text{\large?}
		\\
		\hline
		\G \cong_{ns} \H 
		& 
		\begin{matrix} 
			\exists \text{ bistochastic matrix $u$}\\ 
			\text{s.t. $A_\G u = u A_\H$ \cite[Thm~4.5]{AMRSSV19}}\\ 
			\text{(i.e. fractionally isomorphic)} 
		\end{matrix} 
		& 
		\begin{matrix}
			\forall \text{ tree }\K,
			\\ 
			\# \Hom(\K, \G) \!=\! \# \Hom(\K, \H) \text{ \cite[Thm 1]{DGR18}}
		\end{matrix}
		\\
		\hline
	\end{array}$
	\caption{Characterization of different types of isomorphism.}
	\label{fig:characterizations-of-the-different-types-of-isomorphism}
\end{figure}

	\subsection{Key Ideas}
	\label{subsec:preliminary-result-for-graph-isom-game}
	
In this subsection, we work on a simple case to present the key ideas that will be generalized in the theorem of \refprop[Subsection]{subsec:link-with-communication-complexity}. The assumption here is that $\H$ admits exactly two connected components that are both complete.

\begin{definition}   \label{def:graph-and-diameter}
We denote by $\C_n$ the \emph{cycle} of size $n\geq1$, \ie the finite undirected graph with vertices $v_1,\dots,v_n$ and edges $v_i\sim v_{i+1}$ for $1\leq i\leq n-1$ and $v_n\sim v_1$.
We denote by $\K_n$ the \emph{complete graph} of size $n\geq1$, \ie the finite undirected graph with vertices $v_1,\dots,v_n$ and edges $v_i\sim v_{j}$ for any $i\neq j$.
We also define the \emph{path graph}, denoted $\L_n$, as the cycle $\C_n$ from which we remove one edge.
The distance $d$ between two vertices $v_1, v_2$ in a graph $\G$ is defined as the smallest number of edges of a path connecting $v_1$ to $v_2$ over all possible paths. By convention, it is taken to be $\infty$ when there is no path connecting the vertices. Here, we call \emph{diameter} of a graph $\G$, denoted $\diam(\G)$, the largest distance between two connected vertices $g_1, g_2$ of $\G$ (in this definition, the diameter of a finite graph is finite, even if it is not connected).
\end{definition}

\begin{theorem}[Collapse of CC]  \label{theo:toy-example-thm}
Let $\G$ and $\H$ be two graphs such that 
$\diam(\G)\geq2$, and that $\H$ admits exactly two connected components $\H_0$ and $\H_1$, which are both complete. Then any perfect strategy $\S$ for the graph isomorphism game $\G\cong_{ns} \H$ collapses communication complexity.
\end{theorem}

\begin{proof}
	The proof simply consists in simulating the $\PR$ box from the strategy $\S$, in the sense that from inputs $x,y\in\{0,1\}$, we want to produce outputs $a,b\in\{0,1\}$ with the same behavior as the $\PR$ would do, \ie to obtain $a\oplus b = xy$ in a non-signalling way. It is enough to simulate $\PR$ since this nonlocal box is known to collapse communication complexity~\cite{vD99}. 
	As the diameter is $\diam(\G)\geq2$, the graph $\G$ admits the path graph with three vertices $\G_0=\L_3$ as a subgraph. 
	We will use the protocol described in \refprop[Figure]{fig:simulating-PR-from-isom-game}. 
	In this protocol, bits $x$ and $y$ are given to Alice and Bob respectively. They apply the respective processes $A_1$ and $B_1$ as described in item~(b) and obtain vertices $g_A$ and $g_B$ in $V(\G_0)$. They use these two vertices in the graph isomorphism game of $(\G,\H)$ and receive outputs $h_A,h_B\in V(\H)$. Finally, they process these vertices with $A_2$ and $B_2$ as described in item~(c) to obtain their output $a,b\in\{0,1\}$.
	\begin{figure}[h]
		\centering
		\subfile{Figures/Simulation-of-PR/Simulation-of-PR}
		\begin{tikzpicture}[line cap=round,line join=round,>=triangle 45,x=1.0cm,y=1.0cm,every node/.style={scale=0.7}, scale=0.68,
my-arrow/.style={-{Classical TikZ Rightarrow[length=0.7mm]}, line width=1},
my-arrow-blue/.style={my-arrow, color=\boxcolor}
]

\definecolor{mycolor}{RGB}{30, 70, 120} 
\definecolor{mycolor2}{HTML}{53257F} 
\definecolor{mycolor3}{RGB}{191, 95, 0}
\definecolor{mycolor4}{RGB}{0, 96, 81}  
\definecolor{mycolor5}{RGB}{96, 0, 81}
\definecolor{mycolor6}{RGB}{150, 30, 30} 

\newcommand{\myheight}{4.5}
\newcommand{\mywidth}{-0.6}
\newcommand{\arrowwidth}{1.4}

\newcommand{\xshift}{5.3}
\newcommand{\ycenter}{-2.0*0.5 -1.2*0.5- 0.15*0.5 + \myheight*0.5}

\newcommand{\miniboxheight}{0.45}
\newcommand{\miniboxwidth}{0.5}

\newcommand{\yycenter}{-2.0*0.25 -1.2*0.25- 0.15*0.25 + \myheight*0.25 +\miniboxheight*0.5 - 1.2*0.5- 0.15*0.5 + \myheight*0.5}
\newcommand{\yyycenter}{-2.0*0.25 -1.2*0.25- 0.15*0.25 + \myheight*0.25 - \miniboxheight*0.5 - 2.0*0.5}

\draw (-0.2+\xshift-1.5, -1.45 + \myheight + 0.05) node {\Large(b)};


\fill[color=\boxcolor, fill=\boxcolor, fill opacity=0.40] (-0.2+\xshift-\miniboxwidth, \yycenter - \miniboxheight) -- (-0.2+\xshift+\miniboxwidth, \yycenter - \miniboxheight) -- (-0.2+\xshift+\miniboxwidth,\yycenter + \miniboxheight) -- (-0.2+\xshift-\miniboxwidth, \yycenter + \miniboxheight) -- cycle;
\draw[line width=1.3pt, color=\boxcolor] (-0.2+\xshift-\miniboxwidth, \yycenter - \miniboxheight) rectangle (-0.2+\xshift+\miniboxwidth, \yycenter + \miniboxheight) ;
\draw[color=\boxcolor] (-0.2+\xshift, \yycenter) node {$A_1$};

\fill[color=\boxcolor, fill=\boxcolor, fill opacity=0.40] (3.2 + \mywidth+\xshift -\miniboxwidth, \yycenter - \miniboxheight) -- (3.2 + \mywidth+\xshift +\miniboxwidth, \yycenter - \miniboxheight) -- (3.2 + \mywidth+\xshift +\miniboxwidth,\yycenter + \miniboxheight) -- (3.2 + \mywidth+\xshift -\miniboxwidth, \yycenter + \miniboxheight) -- cycle;
\draw[line width=1.3pt, color=\boxcolor] (3.2 + \mywidth+\xshift -\miniboxwidth, \yycenter - \miniboxheight) rectangle (3.2 + \mywidth+\xshift +\miniboxwidth, \yycenter + \miniboxheight) ;
\draw[color=\boxcolor] (3.2 + \mywidth+\xshift, \yycenter) node {$B_1$};


\newcommand{\mylinewidth}{1pt}
\newcommand{\mycirclesize}{0.12}
\newcommand{\mystep}{1}
\newcommand{\graphwidth}{-0.2}
\newcommand{\xbegin}{-0.2+\xshift-\graphwidth}
\newcommand{\xend}{3.2 + \mywidth+\xshift+\graphwidth}
\newcommand{\xmiddle}{-0.2*0.5+\xshift*0.5 + 3.2*0.5 + \mywidth*0.5+\xshift*0.5}
\newcommand{\ytop}{0}
\newcommand{\ybottom}{-1.8}

\draw[line width=\mylinewidth] (\xbegin, \ybottom) -- (\xmiddle, \ytop) -- (\xend, \ybottom) -- cycle;
\draw[fill=black] (\xbegin, \ybottom) circle (\mycirclesize);
\draw[fill=black] (\xmiddle, \ytop) circle (\mycirclesize);
\draw[fill=black] (\xend, \ybottom) circle (\mycirclesize);
        
\draw (\xmiddle, \ytop*0.5 + \ybottom*0.5-0.2) node{\small $\K_3$};
        

\newcommand{\yendarrow}{0.18}

\draw[my-arrow] (-0.2+\xshift,-1.45 + \myheight) -- node[above, yshift=0.2cm]{\footnotesize$x$} (-0.2+\xshift, \yycenter + \miniboxheight+0.07);

\draw[my-arrow] (3.2 + \mywidth+\xshift,-1.45 + \myheight) -- node[above, yshift=0.2cm]{\footnotesize$y$} (3.2 + \mywidth+\xshift, \yycenter + \miniboxheight+0.07);

\draw[my-arrow-blue] (-0.2+\xshift-0.1,\yycenter - \miniboxheight) .. controls +(down:2em) and +(left:2em)..  (\xbegin - \yendarrow, \ybottom);
\draw (\xbegin+0.18, \ytop-0.35) node[color=\boxcolor]{\scriptsize$g_{\!A}$ if $x\!=\!0$};

\draw[my-arrow-blue] (-0.2+\xshift +0.1,\yycenter - \miniboxheight) node[below right, xshift=0.1cm]{\scriptsize$g_{\!A}$ if $x\!=\!1$} .. controls +(down:2em) and +(up:2em)..  (\xmiddle-0.1, \ytop+\yendarrow);

\draw[my-arrow-blue] (3.2 + \mywidth+\xshift-0.1, \yycenter - \miniboxheight) .. controls +(down:2em) and +(up:2em).. (\xmiddle+0.1, \ytop+\yendarrow) node[above right]{\scriptsize$g_{\!B}$ if $y\!=\!1$};

\draw[my-arrow-blue] (3.2 + \mywidth+\xshift+0.1, \yycenter - \miniboxheight) .. controls +(down:2em) and +(right:2em).. (\xend + \yendarrow, \ybottom);
\draw (\xend+1.33, \ytop-0.35)  node[color=\boxcolor]{\scriptsize$g_{\!B}$ if $y\!=\!0$};

\end{tikzpicture}\!
		\begin{tikzpicture}[line cap=round,line join=round,>=triangle 45,x=1.0cm,y=1.0cm,every node/.style={scale=0.7}, scale=0.68,
my-arrow/.style={-{Classical TikZ Rightarrow[length=0.7mm]}, line width=1},
my-arrow-blue/.style={my-arrow, color=\boxcolor}
]

\definecolor{mycolor}{RGB}{30, 70, 120} 
\definecolor{mycolor2}{HTML}{53257F} 
\definecolor{mycolor3}{RGB}{191, 95, 0}
\definecolor{mycolor4}{RGB}{0, 96, 81}  
\definecolor{mycolor5}{RGB}{96, 0, 81}
\definecolor{mycolor6}{RGB}{150, 30, 30} 

\newcommand{\myheight}{4.5}
\newcommand{\mywidth}{-0.6}
\newcommand{\arrowwidth}{1.4}

\newcommand{\ycenter}{0}

\newcommand{\boxheight}{0.45}
\newcommand{\boxwidth}{0.5}
\newcommand{\boxspacing}{1.8}

\draw (-\boxspacing-0.9, 3.4) node {\Large(c)};


\fill[color=\boxcolor, fill=\boxcolor, fill opacity=0.40] (-\boxspacing-\boxwidth, \ycenter- \boxheight) -- (-\boxspacing+\boxwidth,  \ycenter- \boxheight) -- (-\boxspacing+\boxwidth, \ycenter+\boxheight) -- (-\boxspacing-\boxwidth, \ycenter+\boxheight) -- cycle;
\draw[line width=1.3pt, color=\boxcolor] (-\boxspacing-\boxwidth, \ycenter-\boxheight) rectangle (-\boxspacing+\boxwidth, \ycenter+\boxheight) ;
\draw[color=\boxcolor] (-\boxspacing, \ycenter) node {$A_2$};

\fill[color=\boxcolor, fill=\boxcolor, fill opacity=0.40] (\boxspacing -\boxwidth, \ycenter - \boxheight) -- (\boxspacing+\boxwidth, \ycenter - \boxheight) -- (\boxspacing+\boxwidth, \ycenter + \boxheight) -- (\boxspacing-\boxwidth, \ycenter + \boxheight) -- cycle;
\draw[line width=1.3pt, color=\boxcolor] (\boxspacing-\boxwidth, \ycenter - \boxheight) rectangle (\boxspacing+\boxwidth, \ycenter + \boxheight) ;
\draw[color=\boxcolor] (\boxspacing, \ycenter) node {$B_2$};


\newcommand{\mycirclesize}{0.12}
\newcommand{\mylinewidth}{1pt}

\newcommand{\graphsize}{3.1cm}
\newcommand{\graphxshift}{-0.1cm}
\newcommand{\graphyshift}{1.3cm}

\draw[line width=\mylinewidth] (\graphxshift, \graphyshift) -- (\graphxshift, \graphyshift+\graphsize);
\draw[fill=black] (\graphxshift, \graphyshift+\graphsize) circle (\mycirclesize) node[right, xshift=0.1cm]{$h_1$};
\draw[fill=black] (\graphxshift, \graphyshift) circle (\mycirclesize) node[right, xshift=0.1cm]{$h_2$};

\draw (0,5.1) node {$\K_2$};

\newcommand{\xbrace}{1}
\newcommand{\ybracetop}{4.5}
\newcommand{\ybracebottom}{1.2}
\draw[decorate,decoration={brace}, color=\boxcolor, line width=1] (\xbrace, \ybracetop) -- (\xbrace, \ybracebottom);
\draw[decorate,decoration={brace}, color=\boxcolor, line width=1] (-\xbrace, \ybracebottom) -- (-\xbrace, \ybracetop);

\draw[my-arrow-blue] (\xbrace+0.15, \ybracebottom*0.5 + \ybracetop*0.5) .. controls +(right:2em) and +(up:2em).. node[right]{$h_B$} (\boxspacing, \ycenter + \boxheight+0.1);

\draw[my-arrow-blue] (-\xbrace-0.15, \ybracebottom*0.5 + \ybracetop*0.5) .. controls +(left:2em) and +(up:2em).. node[left]{$h_A$} (-\boxspacing, \ycenter + \boxheight+0.1);

\newcommand{\arrowlength}{0.5}
\draw[my-arrow] (-\boxspacing, \ycenter - \boxheight-0.05) -- (-\boxspacing, \ycenter - \boxheight-0.07-\arrowlength) node[below]{\scriptsize$a\!=\!i\!-\!1$\, s.t.\, $h_A\!=\!h_i$};
\draw[my-arrow] (\boxspacing, \ycenter - \boxheight-0.05) -- (\boxspacing, \ycenter - \boxheight-0.07-\arrowlength) node[below, xshift=-0.1cm]{\scriptsize$b\!=\!j\!\!\!\mod\!2$\, s.t.\, $h_B\!=\!h_j$};

\end{tikzpicture}
		\caption{
			Illustration of the proof of \refprop[Theorem]{theo:toy-example-thm}.
			In item~(a), we simulate a $\PR$ box from a perfect $\NS$-strategy for the graph isomorphism game, called ``Isom. Game'' in the figure, together with the local processes $A_1,A_2,B_1,B_2$ that are described in items~(b) and~(c).
			In item~(b), the graph $\G_0$ is a subgraph of $\G$, in which Alice and Bob choose some input vertices $g_A$ and $g_B$. 
			In item~(c), the graphs $\H_1$ and $\H_2$ are the two connected components of $\H$, from which Alice and Bob receive some output vertices $h_A$ and $h_B$.
		}
		\label{fig:simulating-PR-from-isom-game}
	\end{figure}
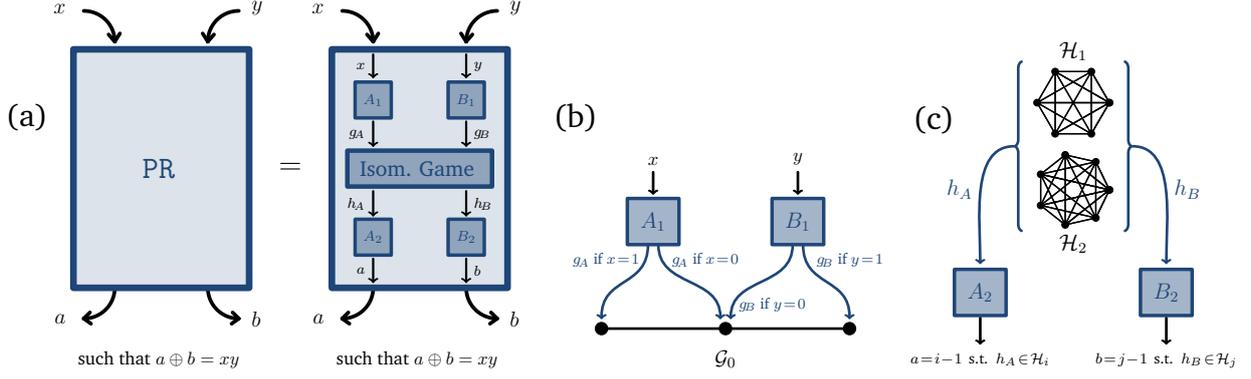
Let us prove that this protocol indeed simulates the $\PR$ box. 
On the one hand, if $xy=1$, then $x=1=y$, which gives $g_A\not\simeq g_B$ and therefore $h_A\not\simeq h_B$. It yields that the vertices $h_A$ and $h_B$ are in different components $\H_i$ and $\H_{i\oplus1}$, so $a\oplus b = i \oplus i \oplus 1 = 1 = xy$ as wanted. 
On the other hand, if $xy=0$, we have $x=0$ or $y=0$, so necessarily $g_A=g_B$ or $g_A\sim g_B$, and therefore $h_A=h_B$ or $h_A\sim h_B$. It follows that the vertices $h_A$ and $h_B$ are both in the same component $\H_i$, and $a\oplus b = i\oplus i=0 = xy$ as expected as well. In addition, note that this protocol does \emph{not} signal between Alice and Bob. Hence the $\PR$ is perfectly simulated, and there is a collapse of communication complexity.
\end{proof}

\newpage
\begin{example}
Any perfect strategy for the isomorphism game $\C_6 \cong_{ns} \K_3\sqcup\K_3$ allows to perfectly simulate the $\PR$ box and to collapse CC. 
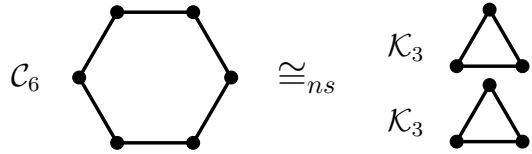
\footnote{\label{footnote:ns-isom-is-like-common-equitable-partition}
	As later detailed in \refprop[Subsection]{subsec:symmetric-strategies}, two finite undirected graphs $\G$ and $\H$ with the same number of vertices are $\NS$-isomorphic \iff they admit a common equitable partition.
	A sufficient condition for the latter condition is that the graphs 
	are regular with the same degree, that is each vertex has a fixed constant number of neighbors. In particular, it holds $\C_6 \cong_{ns} \K_3\sqcup\K_3$.} 
\begin{center}
	
\begin{tikzpicture}[line cap=round,line join=round,>=triangle 45,x=1.0cm,y=1.0cm,every node/.style={scale=1}, scale=1]

\newcommand{\mylinewidth}{1.3pt}
\newcommand{\mycirclesize}{0.085}

\draw (-1.7,0) node {$\C_6$};

\foreach \x in {0,60,...,300} {
        \draw[line width=\mylinewidth] (\x:1 cm) -- (\x + 60:1 cm);
        \draw[fill=black] (\x:1 cm) circle (\mycirclesize);
        }

\draw (2,0) node {\large$\cong_{ns}$};

\newcommand{\rotation}{-30}
\newcommand{\resize}{0.5}
\newcommand{\xshift}{4.4cm}
\newcommand{\yshift}{0.4cm}
\newcommand{\yyshift}{0.6cm}

\draw (3.3,\yshift) node {$\K_3$};
\draw (3.3,-\yyshift) node {$\K_3$};

\foreach \x in {0+\rotation,120+\rotation, 240+\rotation} {
        \draw[line width=\mylinewidth, xshift=\xshift, yshift=\yshift] (\x : \resize cm) -- (\x + 120 : \resize cm);
        \draw[fill=black, xshift=\xshift, yshift=\yshift] (\x: \resize cm) circle (\mycirclesize);
        }
        
\foreach \x in {0+\rotation,120+\rotation, 240+\rotation} {
        \draw[line width=\mylinewidth, xshift=\xshift, yshift=-\yyshift] (\x : \resize cm) -- (\x + 120 : \resize cm);
        \draw[fill=black, xshift=\xshift, yshift=-\yyshift] (\x: \resize cm) circle (\mycirclesize);
        }


\end{tikzpicture}

\end{center}
\end{example}

This result can be generalized to non-perfect strategies $\S$ as follows:

\begin{corollary}[Collapse With Non-Perfect Strategies] 
	\label{coro:toy-example-thm}
	With the same assumptions as in \hyperref[theo:toy-example-thm]{\normalfont Theorem~\ref{theo:toy-example-thm}}, any  strategy $\S$ that succeeds with probability $\pp>\frac{3+\sqrt{6}}{6}\approx0.91$ at the graph isomorphism game $\G\cong_{ns} \H$ collapses communication complexity.
\end{corollary}

\begin{proof}
	We use the same protocol as in \refprop[Theorem]{theo:toy-example-thm}. It simulates a $\PR$ box with probability $\pp$, which is known to collapse communication complexity from Brassard, Buhrman, Linden, Méthot, Tapp, and Unger~\cite{BBLMTU06}.
\end{proof}

These results can be generalized to more than two connected components in $\H$, based on the assumption that Alice and Bob are given access to a perfect $\NS$-strategy for the 2-coloring game of $\K_N$, which is always granted when $N=2$, see \refprop[Theorem]{thm:combining-isomorphism-game-and-coloring-game}.

	 \subsection{Symmetric Strategies}
	 \label{subsec:symmetric-strategies}

We define a new type of perfect strategy for the graph isomorphism game, that we call \emph{symmetric strategies}, for which we show a collapse of communication complexity in the next subsection.  

\begin{definition}
An $\NS$-strategy $\S$ for the graph isomorphism game $(\G, \H)$ is said \emph{symmetric} from $\G$ to the components of $\H$ if it is perfect (\ie the winning probability is $1$) and if there exist a disjoint decomposition $\H=\mathcal \H_1\sqcup \mathcal \H_2$ and some constants $\eta, \nu_{g_A,g_B}\in[0,1]$ such that:
\[
	\forall g_A,g_B\in \G,\quad
	\left\{
	\begin{array}{l}
		{\PP}_\S\big(h_A\in \H_1,\,h_B\in \H_2\,\big|\, g_A,\,g_B\big) 
		\,=\,
		{\PP}_\S\big(h_A\in \H_2,\, h_B\in \H_1\,\big|\, g_A,\,g_B\big) 
		=: \nu_{g_A,g_B}
		\\
		{\PP}_\S\big(h_A\in \H_1,\,h_B\in \H_1\,\big|\, g_A,\,g_B\big) 
		\,=\,
		\eta - \nu_{g_A,g_B}\,.
	\end{array}
	\right.
\]
\end{definition}

Examples of graphs admitting symmetric strategies can be found in \refprop[Example]{ex: decomposition of cycles}.
For the sake of the next proposition, we recall that two finite undirected graphs $\G$ and $\H$ are $\NS$-isomorphic \iff they are fractionally isomorphic~\cite{AMRSSV19}, \iff they admit a common equitable partition~\cite{RSU94}, whose definition is recalled below.

\paragraph*{Common Equitable Partition~\cite{RSU94}.} \label{para:defition-of-common-equitable-partition}
Given a graph $\G$, 
define a partition $\Cpartition=\{C_1, \dots, C_k\}$ of its vertices, that is subsets $C_i\subseteq V(\G)$ such that every vertex $g$ of $\G$ belongs to exactly one set $C_{i_g}$, which may be viewed as assigning a (unique) color $C_{i_g}$ to each vertex.
We say that this partition is \emph{equitable} if, for all $i,j\in[k]$, any vertex of color $C_i$ admits precisely a fixed number, denoted $c_{ij}$, of neighbors colored with $C_j$. 
Note that $c_{ij}$ and $c_{ji}$ may differ, but the equality $c_{ij} |C_i| = c_{ji} |C_j|$ always holds (see \refprop[Lemma]{lem:relation-betwenn-ct_ij-and-ct_ji} for a proof of a generalized result).
A trivial example of an equitable partition is the minimal partition $\Cpartition=\{\{g\}\,:\,g\in V(\G)\}$, where to each vertex a different color is assigned and where the matrix $[c_{ij}]_{i,j}$ is the adjacency matrix. 
Another example is the maximal partition $\Cpartition=\{V(\G)\}$, which is equitable \iff the graph $\G$ is regular of degree $c_{11}$.
Now, we say that two graphs $\G$ and $\H$ admit a \emph{common equitable partition} if they admit equitable partitions of same length $\Cpartition=\{C_1, \dots, C_k\}$ and $\Dpartition=\{D_1, \dots, D_k\}$ such that the cells have same size $|C_i|=|D_i|=:n_i$ and the partition parameters coincide $c_{ij}=d_{ij}$ for all $i,j\in[k]$. 
When such partitions exist, we may concisely describe them in terms of the parameters $\big(k, (n_i), (c_{ij})\big)$ and, when the context is clear, we may consider $\Cpartition$ as an equitable partition of both $\G$ and $\H$. 
As mentioned above, we will use the fact that $\G\cong_{ns}\H$ \iff the graphs admit a common equitable partition~\cite{RSU94}. For instance the graphs $\G=\C_6$ and $\H=\C_3\sqcup\C_3$ are both $2$-regular, so they admit admit a common equitable partition $\big(k=1, (n_1=6), (c_{11}=2) \big)$, which is why they are $\NS$-isomorphic. 

\begin{proposition}[Existence of Symmetric Strategies]   \label{prop: sufficient condition}
Let $\G\cong_{ns} \H$ such that $\H$ is not connected: $\H=\H_1\sqcup \H_2$. Denote the partitions $\Cpartition=\{C_1, \dots, C_k\}$ and $\Dpartition=\{D_1, \dots, D_k\}$ forming a common equitable partition for $\G$ and $\H$, and assume that the proportion of vertices of $\H_1$ assigned to $D_i$ is independent of $i$:
\begin{equation}   \label{eq: assumption sufficient condition proposition}   
			\tag*{(H)}
	\forall i,j\in[k],\quad\quad
	\frac{|D_i\cap \H_1|}{|D_i|}
	\,=\,
	\frac{|D_j\cap \H_1|}{|D_j|}\,.
\end{equation} 
Then the isomorphism game of $(\G,\H)$ admits a symmetric strategy.
\end{proposition}

\begin{proof}
Let $\big(k,\, [n_1, \dots, n_k],\, [c_{ij}]_{1\leq i,j\leq k} \big)$ be the parameters of a common equitable partition for $\G$ and $\H$, and consider $\overline{c_{ij}}:= n_j - c_{ij} - \delta_{ij}$ the number of non-neighbours a vertex of $C_i$ has in $C_j$ for all $i,j$, where $\delta_{ij}$ is the Kronecker delta function.  
We define the following strategy $\S$ as in~\cite[Lemma 4.4]{AMRSSV19} for which the authors prove it is perfect for the isomorphism game of $(\G,\H)$:
\begin{equation}  \label{eq:expression-of-P_S}
	\PP_\S(h_A,h_B\,|\, g_A,g_B)
	\,:=\,
	\left\{
	\begin{array}{c l}
		1/n_i & \text{if $g_A=g_B$ and $h_A=h_B$ and $(\star)$},\\
		1/n_i c_{ij} & \text{if $g_A\sim g_B$ and $h_A\sim h_B$ and $(\star)$},\\
		1/n_i \overline{c_{ij}} & \text{if $g_A\not\simeq g_B$ and $h_A\not\simeq h_B$ and $(\star)$},\\
		0 & \text{otherwise},
	\end{array}
	\right.
\end{equation}
where $(\star)$ denotes the condition ``$g_A\in C_i,\, g_B\in C_j,\, h_A\in D_i,\, h_B\in D_j$". 
Note that $\PP_\S$ is well defined because in each case the division by zero is prevented using the condition of occurrence. Let us show that $\S$ is symmetric in two steps.
\\
$\bullet$ First, we compute the constants $\nu_{g_A,g_B}:={\PP}_\S\big(h_A\in \H_1,\,h_B\in \H_2\,\big|\, g_A,\,g_B\big)$. Notice that $\nu_{g_A,g_B}=0$ when $g_A=g_B$ or $g_A\sim g_B$ 
by disconnectedness of $\H_1$ and $\H_2$. Now if $g_A\not\simeq g_B$ for some $g_A\in C_i$ and $g_B\in C_j$, then:
\begin{align*}
	\PP_\S\big( h_A\in\H_1, h_B\in\H_2 \,\big|\, g_A\not\simeq g_B\big)\hspace{0.2cm}
	&=
	\hspace{-0.2cm}\sum_{(h_A,h_B)\in\H_1\times\H_2} \PP_\S(h_A, h_B\,|\, g_A\not\simeq g_B)
	\\
	&=
	\sum_{\tiny\begin{array}{c} (h_A,h_B)\in\H_1\times\H_2 \\ \text{s.t. } h_A\in D_i, h_B\in D_j 
	\end{array}} 
	\frac{1}{n_i\, \overline{c_{ij}}} \\
	&=
	\quad
	\frac{|D_i\cap \H_1| \times |D_j\cap \H_2|}{n_i\, \overline{c_{ij}}}\,.
\end{align*}
But as $|D_i|=|D_i\cap \H_1| + |D_i\cap \H_2|$ for all $i$, we see that the assumption~\ref{eq: assumption sufficient condition proposition} is equivalent to saying $|D_i\cap \H_1| \times |D_j\cap \H_2| = |D_i\cap \H_2| \times |D_j\cap \H_1|$.
Therefore, the above quantity also equals $\PP_\S\big( h_A\in\H_2, h_B\in\H_1 \,\big|\, g_A, g_B)$, and we obtain:
\[
	\quad\quad
	\nu_{g_A,g_B}
	\,=\,
	\left\{
	\begin{array}{cl}
		\displaystyle{|D_i\cap \H_1| \times |D_j\cap \H_2|}/{n_i\, \overline{c_{ij}}} & \text{if $g_A\not\simeq g_B$ and $g_A\in C_i$ and $g_B\in C_j$},\\
		0 & \text{otherwise}.
	\end{array}
	\right.
\]
$\bullet$ Second, we compute $\eta := {\PP}_\S\big(h_A\in \H_1,\,h_B\in \H_1\,\big|\, g_A,\,g_B\big) + \nu_{g_A,g_B}$, which should be independent of $g_A,g_B$. Let $g_A\in C_i$ and $g_B\in C_j$. We split the study into three cases.\\
$(i)$~If $g_A=g_B$, then:
\begin{align*}
	\PP_\S\big( (h_A,h_B)\in{\H_1}^2 \,\big|\, g_A=g_B\big)\hspace{0.2cm}
	&=	
	\hspace{-0.2cm}\sum_{(h_A,h_B)\in{\H_1}^2} \PP_\S(h_A, h_B\,|\, g_A= g_B)
	\\
	&=
	\sum_{\tiny\begin{array}{c} h\in\H_1 \\ \text{s.t. } h\in D_i \end{array}} 
	\underbrace{\PP_\S(h, h\,|\, g_B=g_B)}_{=1/n_i}
	\\
	&=
	\quad
	\frac{|D_i\cap \H_1|}{n_i}
	\quad=\quad
	\eta - \nu_{g_A,g_B}\,,
\end{align*}
where we fixed $\eta:=\frac{|D_i\cap \H_1|}{n_i }$, and where we have $n_i\neq0$ because $g_A\in C_i$. 
Let us verify that this $\eta$ is appropriate in the other cases as well.\\
$(ii)$~If $g_A\sim g_B$, then:
\begin{align*}
	\PP_\S\big( (h_A,h_B)\in{\H_1}^2 \,\big|\, g_A\sim g_B\big)\hspace{0.2cm}
	&=	
	\hspace{-0.2cm}\sum_{(h_A,h_B)\in{\H_1}^2} \PP_\S(h_A, h_B\,|\, g_A\sim g_B)
	\\
	&=
	\hspace{-0.3cm}\sum_{\tiny\begin{array}{c} (h_A,h_B)\in{\H_1}^2 \\ \text{s.t. } h_A\in D_i, \\ h_B\in D_j \cap N(h_A) \end{array}} 
	\underbrace{\PP_\S(h_A, h_B\,|\, g_A\sim g_B)}_{=1/n_i c_{ij}}\,,
	\intertext{where $N(h_A)$ is the set of adjacent vertices to $h_A$ and where $c_{ij}\neq 0$ because $g_A\sim g_B$, so:}
	&= 
	\hspace{-0.1cm}\sum_{h_A\in D_i\cap \H_1}  \frac{\big|D_j\cap N(h_A)\cap \H_1\big|}{n_i\,c_{ij}} \,.
	\intertext{But by disconnectedness of $\H_1$ and $\H_2$ we have $|D_j\cap N(h_A)\cap \H_1| = |D_j\cap N(h_A)|$
	, which is equal to $c_{ij}$ by definition. Hence, it simplifies with the coefficient in the denominator, so we obtain:}
	&= 
	\hspace{-0.1cm}\sum_{h_A\in D_i\cap \H_1}\frac{1}{n_i}
	\quad= \quad
	 \frac{|D_i\cap \H_1|}{n_i}\
	 \quad=\quad
	\eta - \nu_{g_A,g_B}\,.
\end{align*}
$(iii)$~If $g_A\not\simeq g_B$:
\begin{align*}
	\PP_\S\big( (h_A,h_B)\in{\H_1}^2 \,\big|\, g_A\not\simeq g_B\big)\hspace{0.2cm}
	&=	
	\sum_{(h_A,h_B)\in{\H_1}^2} \PP_\S(h_A, h_B\,|\, g_A\not\simeq g_B)
	\\
	&=
	\hspace{-0.6cm}\sum_{\tiny\begin{array}{c} (h_A,h_B)\in{\H_1}^2 \\ \text{s.t. } h_A\in D_i, \\ h_B\in D_j \backslash(N(h_A)\cup\{h_A\}) \end{array}} 
	\hspace{-0.5cm}
	\underbrace{\PP_\S(h_A, h_B\,|\, g_A\not\simeq g_B)}_{=1/n_i \overline{c_{ij}}}\,,
	\intertext{where $\overline{c_{ij}}\neq 0$ because $g_A\not\simeq g_B$, so:}
	&=
	\sum_{h_A\in D_i\cap \H_1} \frac{\big|D_j\cap \H_1\backslash(N(h_A)\cup\{h_A\})\big|}{n_i \,\overline{c_{ij}}}\,.
	\intertext{But by definition $n_j := |D_j| = |D_j\cap \H_2| + |D_j\cap \H_1\backslash(N(h_A)\cup\{h_A\})| + |D_j\cap \H_1\cap(N(h_A)\cup\{h_A\})|$, where the last term is $c_{ij} + \delta_{ij}$. After reordering the terms, it yields that the second term is $|D_j\cap \H_1\backslash(N(h_A)\cup\{h_A\})| = \overline{c_{ij}} - |D_j\cap \H_2|$, therefore:
	}
	&=
	\sum_{h_A\in D_i\cap \H_1} \frac{\overline{c_{ij}} - |D_j\cap \H_2|}{n_i \,\overline{c_{ij}}}
	\\
	&=
	\frac{|D_i\cap \H_1|}{n_i} - \frac{|D_i\cap \H_1| \times |D_j\cap \H_2|}{n_i \,\overline{c_{ij}}}
	\\
	&=
	\eta - \nu_{g_A,g_B}\,.
\end{align*}
Hence, the coefficient $\eta$ is the same in the three cases and independent of $g_A,g_B$, so we proved that $\S$ is indeed symmetric.	
\end{proof}

\begin{corollary}
Let $\G, \H$ be two graphs of degree $d$ such that 
$\H$ is not connected.
Then $\G\cong_{ns} \H$ and this isomorphism game admits a symmetric strategy.
\end{corollary}

\begin{proof}
Consider the common equitable partition given by the parameters $\big(k=1,\, n_1=|V(\G)|,\, c_{11} = d\big)$. Deduce that $\G\cong_{ns} \H$ by \cite{AMRSSV19}, and then conclude using \refprop[Proposition]{prop: sufficient condition}, because the above hypothesis~\ref{eq: assumption sufficient condition proposition} is obviously satisfied when $k=1$.
\end{proof}

\begin{example}  \label{ex: decomposition of cycles}
For any integer decomposition $M = m_1+ \dots + m_K$ with $m_i\geq3$ and $K\geq 2$, the previous corollary tells us that the following isomorphism of cycles holds: 
\[
	\C_M 
	\quad\cong_{ns}\quad
	\C_{m_1} \sqcup \dots \sqcup \C_{m_K}\,,
\]
and that the associated isomorphism game admits a symmetric strategy.
\end{example}

	\subsection{Existence of Perfect Non-Signalling Strategies that Collapse CC}
	\label{subsec:link-with-communication-complexity}

In this subsection, under some conditions, we prove that the graph isomorphism game $(\G,\H)$ admits a perfect non-signalling strategy that collapses communication complexity. In the next subsection, we will add conditions on $\H$ to obtain that all perfect non-signalling strategies are collapsing.

\begin{theorem}[Collapse of CC]  \label{theo: collapse of CC}
Let $\G\cong_{ns} \H$
such that $\diam(\G)\geq2$
and such that $\H$ is not connected: $\H=\H_1\sqcup \H_2$, where each of $\H_1$ and $\H_2$ may possibly be decomposed in several connected components. 
Denote the partitions $\Cpartition=\{C_1, \dots, C_k\}$ and $\Dpartition=\{D_1, \dots, D_k\}$ forming a common equitable partition for $\G$ and $\H$, and assume the hypothesis~\ref{eq: assumption sufficient condition proposition} as in \refprop[Proposition]{prop: sufficient condition}\,.  
Then the isomorphism game of $(\G,\H)$ admits a perfect strategy that collapses communication complexity.
\end{theorem}

To prove the theorem, we define a noisy version of the $\PR$ box, that we denote $\PR_{\alpha, \beta} := \alpha\,\PR + \beta\,\P_0 + (1-\alpha-\beta)\,\P_1$, which is the convex combination with coefficients $\alpha,\beta\geq0$ of the boxes $\PR$, $\P_0$ and $\P_1$ defined in \refprop[Figure]{fig:famous-types-of-strategies}\,. This noisy box is known to collapse communication complexity as long as $\alpha>0$~\cite{BMRC19, BBCNP23}, so the idea is to first prove that $\PR_{\alpha,\beta}$ can be generated from a perfect strategy $\S$ for the isomorphism game:

\begin{lemma}  \label{theo: simulate P alpha beta}
Let $\G,\H$ two graphs such that $\diam(\G)\geq2$ and such that $\H$ is not connected: $\H=\H_1\sqcup \H_2$. Assume $\G\cong_{ns}\H$ for some strategy $\S$ that is symmetric from $\G_0$ to the components of $\H$, and suppose that:
\begin{equation} \label{eq: assumption Thm}
\nu_{g_1, g_3} >0\,.
\end{equation}
Then the box $\PR_{\alpha,\beta}$  is perfectly simulated with  $\alpha=2\,\nu_{g_1, g_3}>0$ and some $\beta\geq0$.
\end{lemma}

\begin{proof}
The protocol in this proof is inspired by the one of \refprop[Theorem]{theo:toy-example-thm}. 
As the diameter is $\diam(\G)\geq2$, the graph $\G$ admits the path graph with three vertices $\G_0=\L_3$ as a subgraph, whose vertices are called $g_1,g_2,g_3$ each one being connected to the next one. 
We proceed with the same protocol as described in \refprop[Figure]{fig:simulating-PR-from-isom-game}, the only difference being that $\H_1$ and $\H_2$ are not necessarily complete and not even connected.
Denote $\P(a,b\,|\,x,y)$ the nonlocal box induced by this protocol. Let us prove that $\P=\PR_{\alpha,\beta}$ for some $\alpha,\beta\in[0,1]$. To this end, we compare the correlation tables of $\P$ and $\PR_{\alpha,\beta}$, defined as follows:
\begin{align*}
	M_\P
	:= &
	\footnotesize
	\begin{bmatrix}
		\P(0,0\,|\,0,0) & \P(0,1\,|\,0,0) & \P(1,0\,|\,0,0) & \P(1,1\,|\,0,0) \\
		\P(0,0\,|\,0,1) & \P(0,1\,|\,0,1) & \P(1,0\,|\,0,1) & \P(1,1\,|\,0,1) \\
		\P(0,0\,|\,1,0) & \P(0,1\,|\,1,0) & \P(1,0\,|\,1,0) & \P(1,1\,|\,1,0) \\
		\P(0,0\,|\,1,1) & \P(0,1\,|\,1,1) & \P(1,0\,|\,1,1) & \P(1,1\,|\,1,1)
	\end{bmatrix}\,,
	\\
\intertext{and similarly defined for $\PR_{\alpha,\beta}$. On the one hand, by symmetry of the strategy $\S$, the correlation table of $\P$ can be computed explicitely:}
	M_\P
	= &
	\begin{bmatrix}
		\eta - \nu_{g_2,g_2} & \nu_{g_2,g_2} & \nu_{g_2,g_2} & 1 - \eta - \nu_{g_2,g_2} \\
		\eta - \nu_{g_2,g_3} & \nu_{g_2,g_3} & \nu_{g_2,g_3} & 1 - \eta - \nu_{g_2,g_3} \\
		\eta - \nu_{g_1,g_2} & \nu_{g_1,g_2} & \nu_{g_1,g_2} & 1 - \eta - \nu_{g_1,g_2} \\
		\eta - \nu_{g_1,g_3} & \nu_{g_1,g_3} & \nu_{g_1,g_3} & 1 - \eta - \nu_{g_1,g_3} \\
	\end{bmatrix} \,.
\intertext{
	But, we see that if $x=0$ or $y=0$, then the inputs $g_A,g_B\in V(\G)$ of Alice and Bob in $\S$ are either equal or adjacent. It turns out that the outputs $h_A, h_B\in V(\H)$ of $\S$ are in the same connected component, therefore $a=b$ almost surely and $\P(a\neq b \,|\, x=0 \text{ or } y=0)=0$. 
	Hence, in the first three lines of the matrix, we have $\nu_{g_2,g_2} = \nu_{g_2,g_3} = \nu_{g_1,g_2} = 0$, which gives:}	
	M_\P= &
	\begin{bmatrix}
		\eta  & 0 & 0 & 1 - \eta \\
		\eta  & 0 & 0 & 1 - \eta \\
		\eta  & 0 & 0 & 1 - \eta \\
		\eta - \nu_{g_1,g_3} & \nu_{g_1,g_3} & \nu_{g_1,g_3} & 1 - \eta - \nu_{g_1,g_3} \\
	\end{bmatrix} \,.
\end{align*}
On the other hand, by linearity, the correlation table of the box $\PR_{\alpha, \beta}$ is the convex combination of the correlation tables of $\PR$, $\P_0$, $\P_1$ with coefficients $\alpha,\beta\in[0,1]$ fixed above, so:
\[
	M_{\PR_{\alpha,\beta}}
	\,=\,
	\begin{bmatrix}
		\frac{\alpha}{2} + \beta & 0 & 0 & 1 - \frac{\alpha}{2} - \beta\\
		\frac{\alpha}{2} + \beta & 0 & 0 & 1 - \frac{\alpha}{2} - \beta\\
		\frac{\alpha}{2} + \beta & 0 & 0 & 1 - \frac{\alpha}{2} - \beta\\
		\beta & \frac{\alpha}{2} & \frac{\alpha}{2} & 1 - \alpha - \beta
	\end{bmatrix} \,.
\]
Now, taking $\alpha := 2\,\nu_{g_1,g_3}$ and $\beta:= \eta - \nu_{g_1,g_3}$, we obtain $\P=\PR_{\alpha,\beta}$ as wanted. 
\end{proof}

\begin{proof}[{ Proof (Theorem~\ref{theo: collapse of CC})}] 
	To invoke the former lemma, we need to prove the existence of a symmetric strategy, which was the purpose of \refprop[Proposition]{prop: sufficient condition}.
	We can apply this proposition because all its assumptions are found in the theorem as well. It yields a symmetric strategy from $\G$ to the components of $\H$, and in particular, its restriction to $\G_0$ is also symmetric. 
	Moreover, assumption~\eqref{eq: assumption Thm} in \refprop[Lemma]{theo: simulate P alpha beta} is also satisfied because,
	using the computations in the proof of \refprop[Proposition]{prop: sufficient condition}\,, we have:
\[
	\nu_{g_1,g_3}
	\,=\,
	\frac{|D_i\cap \H_1| \times |D_j\cap \H_2|}{n_i\, \overline{c_{ij}}}
	\,>\,
	0\,,
\]
where $i$ and $j$ are such that $g_1\in C_i$ and $g_3\in C_j$.
Thus we can apply \refprop[Lemma]{theo: simulate P alpha beta} and we can simulate $\PR_{\alpha, \beta}$ for some $\alpha>0$. But the box $\PR_{\alpha,\beta}$ with $\alpha>0$ is known to be collapsing~\cite{BMRC19, BBCNP23}. Hence we deduce the existence of a protocol that collapses CC.
\end{proof}

\begin{corollary}[Collapse of CC]
Each of the $\NS$-isomorphisms given in \refprop[Example]{ex: decomposition of cycles} admits a perfect strategy $\S$ that allows to perfectly produce a box $\PR_{\alpha,\beta}$ with $\alpha>0$ and therefore to collapse communication complexity.\qed
\end{corollary}

	\subsection{All Perfect Non-Signalling Strategies Collapse CC}
	\label{para:All-Perfect-Strategies-Collapse-CC}

In \refprop[Theorem]{theo: collapse of CC} above, the statement was that \emph{some} perfect strategies collapse communication complexity. Now, if we add regularity and transitivity conditions on $\H$, we obtain that \emph{every} perfect strategy collapses communication complexity. First, we recall the definition of the automorphism group of a graph.

\subsubsection{Automorphism Group} The \emph{automorphism group} of $\H$, denoted $\Aut(\H)$, is the set of all bijective maps $\varphi:\H\to\H$ that preserve the adjacency relation, meaning that $h_1\sim h_2$ \iff $\varphi(h_1)\sim\varphi(h_2)$. As a consequence, any automorphism $\varphi\in\Aut(\H)$ also preserves the relation ``$\not\simeq$'', and therefore a graph $\H$ and its complement $\H^c$ have the same automorphism group: $\Aut(\H)=\Aut(\H^c)$. 
Moreover, the composition of two automorphisms is again an automorphism, and $\varphi^{-1}\in\Aut(\H)$, which endows the set $\Aut(\H)$ with a group structure.
For instance, the automorphism group of the complete graph $\K_N$ is the symmetric group $\Aut(\K_N)=\Sym_N$ of order $N!$, and the one of the cycle $\C_N$ is the dihedral group $\Aut(\C_N)=D_N$ of order $2N$.
We refer to~\cite{Godsil-Royle-01} for more details on automorphism groups.

\subsubsection{Graph Transitivity} The graph $\H$ is said to be \emph{vertex-transitive} if for all vertices $h,h'\in V(\H)$, there exists a graph automorphism $\varphi\in\Aut(\H)$ such that $\varphi(h)=h'$. 
Similarly, we say that $\H$ is \emph{edge-transitive} if for all edges $h_1\sim h_2,\,h_1'\sim h_2'\in E(\H)$, there exists a graph automorphism $\varphi\in\Aut(\H)$ such that $\varphi(h_1)=h_1'$ and $\varphi(h_2)=h_2'$. 
We refer to~\cite{Godsil-Royle-01} for more details on graph transitivity, and to~\cite{Gauyacq-97, SVW19} for related notions.
In the definition below, we strengthen the vertex- and edge- transitivity of a graph $\H$ and its complement $\H^c$ in what we call the \emph{strong transitivity}:

\begin{definition}[Strongly Transitive]   \label{def:strongly-transitive}
We say that a graph $\H$ is \emph{strongly transitive} if there exists a subset of the automorphism group $S\subseteq\Aut(\H)$ such that the three following conditions hold:\vspace{-5pt}
\begin{enumerate}[label=(\roman*),itemsep=-3pt]
	\item \label{item:def-strong-transitive-1} There is a constant $d_1\geq1$ such that for all vertices $h,h'\in V(\H)$, there exist exactly $d_1$ automorphisms $\varphi\in S$ such that $\varphi(h)=h'$.
	\item \label{item:def-strong-transitive-2} There is a constant $d_2\geq1$ such that for all edges $h_1\sim h_2,\,h_1'\sim h_2'\in E(\H)$, there exist exactly $d_2$ automorphisms $\varphi\in S$ such that $\varphi(h_1)=h_1'$ and $\varphi(h_2)=h_2'$.
	\item \label{item:def-strong-transitive-3} There is a constant $d_3\geq1$ such that for all edges in the complement graph $h_1\sim h_2,\,h_1'\sim h_2'\in E(\H^c)$, there exist exactly $d_3$ automorphisms $\varphi\in S$ such that $\varphi(h_1)=h_1'$ and $\varphi(h_2)=h_2'$.
\end{enumerate}
\end{definition}

Notice that strong transitivity implies vertex-transitivity and edge-transitivity of both $\H$ and its complement $\H^c$, since it is possible to pick one among the respective $d_1$, $d_2$, $d_3$ automorphisms $\varphi\in S$ satisfying the wanted condition for each vertices and edges.
Note that if $\H$ is strongly transitive, then its complement $\H^c$ is also strongly transitive.
Note also that $S$ cannot be the empty set $\emptyset$ because of condition~\ref{item:def-strong-transitive-1}, unless the graph $\H$ is itself empty.

Let us prove the following characterization of strong transitivity, and then provide some examples of strongly transitive graphs. First recall that a graph $\H$ is called \emph{distance-transitive} if for any $d\in\NN$ and any two pairs $(h_1,h_2)$ and $(h_1',h_2')$ of vertices $h_1,h_2,h_1',h_2'\in V(\H)$ with distance $d(h_1,h_2)=d(h_1',h_2')=d$, there is an automorphism $\varphi$ of $\H$ such that $\varphi(h_1)=h_1'$ and $\varphi(h_2)=h_2'$.

\begin{lemma}[Characterization of Strong Transitivity]  \label{lem:characterization-of-strong-transitivity}
	A graph $\H$ is strongly transitive \iff it is distance-transitive and its diameter satisfies $\diam(\H)\leq 2$. Moreover, we may always choose $S=\Aut(\H)$ in \refprop[Definition]{def:strongly-transitive}\,.
\end{lemma}
\begin{proof}
Assume that $\H$ is strongly transitive in the sense of \refprop[Definition]{def:strongly-transitive}\,.
First, we prove distance-transitivity for three instances of $d\in\NN$: 
vertex-transitivity ($d=0$) is a particular case of item~\ref{item:def-strong-transitive-1} of the definition;
edge-transitivity ($d=1$) is a particular case of item~\ref{item:def-strong-transitive-2}; 
and in any other case ($d\geq2$), vertices at distance $d$ in $\H$ are adjacent in the complement graph $\H^c$, so the existence of automorphism $\varphi$ follows from item~\ref{item:def-strong-transitive-3}; hence the distance-transitivity.
 We then prove that all vertices in $\H$ have distance at most $2$. Assume by contradiction that there are two vertices $h_1,h_2\in V(\H)$ with $d(h_1,h_2)>2$. Hence, there is a path from $h_1$ to $h_2$ of length greater than two, which needs to pass through a vertex $h_3\in V(\H)$ with $d(h_1,h_3)=2$. Now, as $h_1\sim h_2$ and $h_1\sim h_3$ are edges of the complement graph $\H^c$, item~\ref{item:def-strong-transitive-3} of \hyperref[{def:strongly-transitive}]{Definition~\ref{def:strongly-transitive}} tells us that we can find an automorphism $\varphi\in S$ with $\varphi(h_1)=h_1$ and $\varphi(h_2)=h_3$, and as automorphisms preserve distances, we have $d(h_1,h_2)=d(\varphi(h_1),\varphi(h_2))=d(h_1,h_3)$. But this contradicts $d(h_1,h_2)> d(h_1,h_3)$, so we have $\diam(\H)\leq 2$ as claimed. 

Conversely, assume that $\H$ is distance-transitive with $\diam(\H)\leq 2$, and choose $S=\Aut(\H)$. We prove the three items of \refprop[Definition]{def:strongly-transitive} in the canonical order. Given vertices $h,h'\in V(\H)$ denote by $a_{h,h'}$ the number of automorphisms $\varphi\in \Aut(\H)$ with $\varphi(h)=h'$. It is nonzero since $\H$ is distance-transitive so in particular vertex-transitive. Now, given two further vertices  $k,k'\in V(\H)$, there are automorphisms $\varphi_1,\varphi_2\in \Aut(\H)$ with $\varphi_1(h)=k$ and $\varphi_2(h')=k'$. For any automorphism $\psi$ with $\psi(k)=k'$, the map $\varphi_2^{-1}\circ\psi\circ\varphi_1$ yields an automorphism mapping $h$ to $h'$. This shows $a_{k,k'}\leq a_{h,h'}$. Now, by the symmetry of the argument, this is actually equality, and we may set $d_1:=a_{h,h'}\geq1$, thus proving item~\ref{item:def-strong-transitive-1}.
Then, for item~\ref{item:def-strong-transitive-2}, the proof is very similar, using the edge-transitivity of $\H$. 
Finally, for item~\ref{item:def-strong-transitive-3}, note that edges in the complement graph $\H^c$ correspond to pairs of vertices in $\H$ at distance $2$, so once again distance-transitivity allows us to conclude with the same argument as for item~\ref{item:def-strong-transitive-1}.
\end{proof}

\begin{example}
From this characterization, we deduce that the following graphs are examples of strongly transitive graphs, among others:
the complete graphs $\K_N$ and their complement $\K_N^c$ for any $N\geq0$ (note that the empty graph $\K_N^c$ has diameter $0$ with our convention), 
the path graphs $\L_N$ for $N\leq3$, 
the cycle graphs $\C_N$ for $N\leq 5$, 
the complete bipartite graph $\K_{3,3}$, 
and the famous Petersen graph. 
Moreover, several finite groups may be written as the automorphism group of a distance-transitive graph with diameter $2$, see details in~\cite{Godsil-Royle-01}. 
\end{example}

We prove that when $\H$ is strongly transitive, there is a strong connection between cardinalities, which will be useful in the proof of the collapse of CC in \refprop[Theorem]{theo:collapse-for-ALL-strategies}.

\begin{lemma}   \label{lem:strong-transitivity-implies-conditions-on-cardinals}
Let $\H$ be a strongly transitive graph different from the complete graph $\K_N$ and its complement $\K_N^c$, together with its associated subset $S\subseteq \Aut(\H)$ and parameters $(d_1,d_2,d_3)$. Then the size of the set $S$ is necessarily
$$
	|S|
	\,=\,
	d_1\,|V(\H)|
	\,=\,
	2\,d_2\,|E(\H)|
	\,=\,
	2\,d_3\,|E(\H^c)|\,.
$$
\end{lemma}

\begin{proof}
We prove the first equality by showing the two inequalities. Let us index $h_1, \dots, h_n$ the vertices of $\H$, where $n=|V(\H)|$.
On the one hand, using condition~\ref{item:def-strong-transitive-1}, we know that there are exactly $d_1$ automorphisms $\varphi\in S$ sending $h_1$ to itself, and again $d_1$ other automorphisms sending $h_1$ to $h_2$, so on and so forth until $h_n$. Note also that a given automorphism $\varphi\in\Aut(\H)$ cannot send $h_1$ to two different vertices. It yields that the set $S$ contains at least $d_1n$ elements.
On the other hand, each element $\varphi$ of $S$ necessarily sends $h_1$ to a vertex $h_i$ of $\H$, so $S$ contains at most $d_1 n$ elements, which gives the first equality.
For the other two equalities, proceed similarly using conditions~\ref{item:def-strong-transitive-2} and~\ref{item:def-strong-transitive-3}, where the factor ``$2$'' comes from the fact that graphs are undirected, so the two relations $h_1\sim h_2$ and $h_2\sim h_1$ are counted as only one edge.
Note also that the condition $\H\neq \K_N$ and $\K\neq\K_N^c$ prevents the sets $E(\H)$ and $E(\H^c)$ to be empty.
Hence the wanted chain of equalities.
\end{proof}

\subsubsection{Collapse of Communication Complexity} In the theorem below, we combine this notion of strong transitivity with $d$-regularity of $\H$ to obtain a collapse of CC. The key idea will be to compute an expectation $\mathbb{E}$ over $\varphi$ uniformly sampled in a subset of the automorphism group $S\subseteq\Aut(\H)$ and to use the strong transitivity to obtain a symmetric strategy that collapses CC. Recall that $\H$ is said to be $d$-regular if every vertex is connected to exactly $d$ other vertices.

\begin{theorem}[All Perfect Strategies Collapse CC]  \label{theo:collapse-for-ALL-strategies}
Let $\G$ and $\H$ be two graphs such that the conditions of \refprop[Theorem]{theo: collapse of CC} hold. Assume moreover that $\H$ is strongly transitive and $d$-regular, and that the players share randomness. Then \emph{every} perfect non-signalling strategy for the isomorphism game associated with $\G\cong_{ns}\H$ collapses communication complexity.
\end{theorem}

\begin{proof}
Denote $\P$ an arbitrary perfect strategy for $\G\cong_{ns}\H$, and denote $S\subseteq\Aut(\H)$ and $(d_1,d_2,d_3)$ as given in the definition of strong transitivity. 
We will post-process $\P$ in order to generate a symmetric strategy, which will allow us to generate a collapsing nonlocal box.
The two players Alice and Bob can use their shared randomness to pick uniformly at random an automorphism $\varphi\in S$ that is known by the two of them.
They apply the following protocol: once they receive the outputs $h_A,h_B$ from $\P$, they compute the images $h_A'=\varphi(h_A)$ and $h_B'=\varphi(h_B)$ and they call $\Ptilde(h_A',h_B'\,|\,g_A,g_B)$ the new strategy, in other words this is the following expectation:
\[
	\Ptilde(h_A',h_B'\,|\,g_A,g_B)
	\,=\,
	\underset{\varphi\in S}{\EE}\, \P\big( \varphi^{-1}(h_A'),\varphi^{-1}(h_B')\,\big|\, g_A,g_B \big)\,.
\]
Observe that $\Ptilde$ is non-signalling by the composition of a non-signalling strategy with a non-signalling post-process.
Let us compute the expression of $\Ptilde$.
First, if $g_A=g_B$, then:
\begin{align*}
	\Ptilde(h_A',h_B'\,|\,g_A,g_A)
	&\,=\,
	\frac{1}{d_1\,|V(\H)|} \sum_{\varphi\in S}\, \P\big( \varphi^{-1}(h_A'),\varphi^{-1}(h_B')\,\big|\, g_A,g_A \big)
	\\
	&\,=\,
	\frac{1}{d_1\,|V(\H)|} \sum_{\varphi\in S}\, \P\big( \varphi^{-1}(h_A'),\varphi^{-1}(h_A')\,\big|\, g_A,g_A \big) \, \delta_{\varphi^{-1}(h_A')=\varphi^{-1}(h_B')}
	\\
	&\,=\,
	\frac{1}{d_1\,|V(\H)|} \,\, d_1\underbrace{\sum_{h\in V(\H)}\, \P\big( h, h\,\big|\, g_A,g_A \big)}_{=1} \, \delta_{h_A'=h_B'}
	\,=\,
	\frac{\delta_{h_A'=h_B'}}{|V(\H)|}\,,
\end{align*}
where the first equality follows from the definition of $\Ptilde$ combined with \refprop[Lemma]{lem:strong-transitivity-implies-conditions-on-cardinals}, the second one from the rules of the isomorphism game, and the third one from condition~\ref{item:def-strong-transitive-1} in the definition of strong transitivity of $\H$; moreover, the Kronecker delta condition is changed using the bijectivity property of an automorphism $\varphi$, and the underbrace equality ``$=1$'' comes from the rules of the isomorphism game.
Second, if $g_A\sim g_B$, then similarly:
\begin{align*}
	\Ptilde(h_A',h_B'\,|\,g_A,g_B)
	&\,=\,
	\frac{1}{2\,d_2\,|E(\H)|} \sum_{\varphi\in S}\, \P\big( \varphi^{-1}(h_A'),\varphi^{-1}(h_B')\,\big|\, g_A,g_B \big)
	\\
	&\,=\,
	\frac{1}{2\,d_2\,|E(\H)|}\,\, d_2\! \underbrace{\sum_{h_1\sim h_2\in V(\H)}\, \P\big( h_1, h_2\,\big|\, g_A,g_B \big)}_{=1}\,\delta_{h_A'\sim h_B'}
	\,=\,
	\frac{\delta_{h_A'\sim h_B'}}{2\,|E(\H)|}\,.
\end{align*}
Third, if $g_A\not\simeq g_B$, then we proceed similarly, simply replacing ``$d_2$'' by ``$d_3$'', ``$\H$'' by ``$\H^c$'', and ``$\sim$'' by ``$\not\simeq$'', and we obtain $\Ptilde=\delta_{h_A'\not\simeq h_B'}/2|E(\H^c)|$.
To sum up, we have:
\[
	\Ptilde(h_A',h_B'\,|\,g_A,g_B)
	\,=\,
	\left\{
	\begin{array}{ll}
		1/|V(\H)| & \text{if $g_A=g_B$ and $h_A'=h_B'$}\,,\\
		1/2|E(\H)| & \text{if $g_A\sim g_B$ and $h_A'\sim h_B'$}\,,\\
		1/2|E(\H^c)| & \text{if $g_A\not\simeq g_B$ and $h_A'\not\simeq h_B'$}\,,\\
		0 & \text{otherwise}\,.
	\end{array}
	\right.
\]
Now, when comparing this expression of $\Ptilde$ with the expression of $\PP_\S$ in equation~\eqref{eq:expression-of-P_S} of the proof of \refprop[Proposition]{prop: sufficient condition}, we see that they coincide in the case of the maximal partition $\Dpartition'=\{V(\H)\}$ with parameters $(k=1, n_1=|V(\H)|, c_{11}=d)$, where $d$ is the parameter of regularity of $\H$ by assumption.
Then, following the same proof, it turns out that the strategy $\Ptilde$ is perfect for the isomorphism game associated with $\G\cong_{ns}\H$, and that it is symmetric with parameter: 
$$
	\nu_{g_1,g_3}
	\,=\,
	\frac{|\H_1|\times|\H_2|}{|V(\H)|\times |E(\H^c)|}>0\,,
$$
where the denominator is not zero because $\H$ contains several connected components, so it is not complete and $|E(\H^c)|>0$.
Therefore, we can apply \refprop[Lemma]{theo: simulate P alpha beta} and we can simulate $\PR_{\alpha, \beta}$ for some $\alpha>0$. Hence, as in the proof of \refprop[Theorem]{theo: collapse of CC} we conclude the existence of a non-signalling protocol that collapses communication complexity.
\end{proof}

Finally, as quantum correlations cannot collapse CC~\cite{CvDNT99}, it yields:

\begin{corollary}[These Strategies are not Quantum]  \label{coro:cannot-be-quantum}
A perfect $\NS$-strategy $\S$ for the graph isomorphism game satisfying the conditions of \refprop[Theorem]{theo:collapse-for-ALL-strategies} cannot be quantum. \qed
\end{corollary}

\bib

\section{Graph Coloring Game}
	\label{section:graph-homomorphism-game}
	
		\subsection{Definition of the Graph Homomorphism and the Graph Coloring Games}
		
Here, we introduce two non-local games, namely the \emph{graph homomorphism game} and the \emph{graph coloring game}, the latter being a valuable particular case of the former.

\paragraph{Graph Homomorphism Game~\cite{MR16}.}
Similarly to the isomorphism game, in this game associated with some graphs $(\G,\H)$, the players try to convince a referee that they know a homomorphism $\varphi:\G\to\H$. Recall that a map $\varphi:V(\G)\to V(\H)$ is a graph homomorphism if it preserves the adjacency, meaning that an edge $g\sim g'$ in $\G$ is sent to an edge $\varphi(g)\sim\varphi(g')$ in $\H$. Note that a graph isomorphism additionally requires that $\varphi$ is bijective and that the converse also holds, as stated in equation~\eqref{eq:def-of-a-graph-isomorphism}.
Note also that the composition of two graph homomorphisms is again a graph homomorphism itself -- this defines a category.
The nonlocal game associated with this notion is naturally called \emph{(graph) homomorphism game}. This game consists in giving respective vertices $g_A,g_B\in V(\G)$ to Alice and Bob, they answer some vertices $h_A,h_B\in V(\H)$, and the referee declares that they win the game \iff they satisfy the following conditions:
\[
	g_A=g_B\Longrightarrow h_A=h_B\,,
	\quad\quad\quad
	g_A\sim g_B\Longrightarrow h_A\sim h_B\,.
\]
As opposed to the isomorphism game, in this game it may happen that $g_A\not\sim g_B$ but still $h_A=h_B$ or $h_A\sim h_B$.
We denote $\G\rightarrow\H$ if Alice and Bob can perfectly win the game using classical resources, which is equivalent to saying that there exists a graph homomorphism from $\G$ to $\H$.
We similarly denote $\G\rightarrow_q\H$ and $\G\rightarrow_{ns}\H$ when the game can be perfectly won using quantum and non-signalling resources respectively.

\paragraph{Graph Coloring Game~\cite{Cameron-Newman-Montanaro-etal-06}.}
Interestingly, when $\H=\K_N$ is complete, the homomorphism game corresponds to proving to a referee that the graph $\G$ is $N$-colorable, so this particular case is called the \emph{(graph) coloring game}. Indeed, we say that a graph $\G$ is $N$-colorable if, from a set of $N$ different colors, we can assign one color to each vertex of $\G$ so that no adjacent vertex shares the same color. This is the exact same constraint for a graph homomorphism $\G\to\K_N$, since two adjacent vertices in $\G$ are sent to adjacent vertices in $\K_N$, \ie to two different ``colors''.
The complete graph $\K_M$ is $N$-colorable \iff $M\leq N$, but interestingly when we consider the coloring game with non-signalling strategies, Alice and Bob are able to pretend to the referee that they know an $N$-coloring for $\K_M$:
		
\begin{example}
	\label{ex:perfect-strategies-for-homomorphism-game-K_N-to-K_2-exist}
	We have $\K_M\to_{ns}\K_N$ for any $M,N\geq2$, that is $\K_M$ admits a perfect non-signalling strategy for the $N$-coloring game:
	$$
	\PP_\S\big(h_A,\,h_B\,\big|\,g_A,\,g_B\big)
	\,=\,
	\left\{
	\begin{array}{cl}
		1/N & \text{if $h_A=h_B$ and $g_A=g_B$,}\\
		1/N(N-1) & \text{if $h_A\sim h_B$ and $g_A\sim g_B$,}\\
		0 & \text{otherwise.}\\
	\end{array}
	\right.
	$$
	This strategy $\S$ indeed satisfies the rules of the homomorphism game by construction. Moreover, it is a well-defined probability distribution: it is non-negative by construction; and for all fixed $g_A,g_B\in V(\K_M)$, it sums to $1$ over $h_A,h_B\in V(\K_N)$ because (i)~for $g_A=g_B$, see that $\K_N$ admits exactly $N$ vertices, (ii)~for $g_A\sim g_B$, there are exactly $N^2-N=N(N-1)$ pairs $(h_A,h_B)\in V(\K_N)^2$ such that $h_A\neq h_B$, and (iii)~the case $g_A\not\simeq g_B$ never happens in $\K_M$. Finally, we prove that this strategy is indeed non-signalling because Alice's marginal is independent of Bob's input $g_B$:
	\begin{align*}
		\sum_{h_A}\, \PP_\S(h_A,\,h_B\,|\,g_A,\,g_B)
		&\,=\,
		\frac1N \sum_{h_A}\delta_{h_A=h_B} \delta_{g_A=g_B}
		+ \frac1{N(N-1)} \sum_{h_A}\delta_{h_A\sim h_B} \delta_{g_A\sim g_B}
		\\
		&\,=\,
		\frac1N\, \delta_{g_A=g_B}
		+ \frac{N-1}{N(N-1)}\, \delta_{g_A\sim g_B}
		\,=\,
		\frac1N\,,
	\end{align*}
	where the last line holds because $\K_N$ is complete so exactly one of the Kronecker deltas is~$1$ and the other is zero,
	and similarly for Bob's marginal.
\end{example}

		\subsection{Link with Communication Complexity}
	
We begin with the following simple result about a protocol generating a perfect $\PR$ box. Note that this protocol is slightly different than the one we used for the isomorphism game, but it has the same taste.
	
\begin{lemma}[Simulation of a $\PR$ Box]  \label{lemma: K3 to K2 implies PR simulated}
	Any perfect non-signalling strategy for the $2$-coloring game of $\K_3$ allows to perfectly simulate the $\PR$ box. 
\end{lemma}

\begin{proof}
	Consider the protocol described in \refprop[Figure]{fig:simulating-PR-from-coloring-game}.
	We check that it indeed produces a $\PR$ box. 
	On the one hand, if $x=1=y$, then Alice and Bob input the same vertex. Therefore they obtain the same vertex in $\K_2$ and it yields $a\oplus b=1=xy$ as expected.
	On the other hand, if $x=0$ or $y=0$, then Alice and Bob input adjacent vertices. So they receive two different vertices in $\K_2$ and $a\oplus b = 0 = xy$ as wanted. 	
\end{proof}

	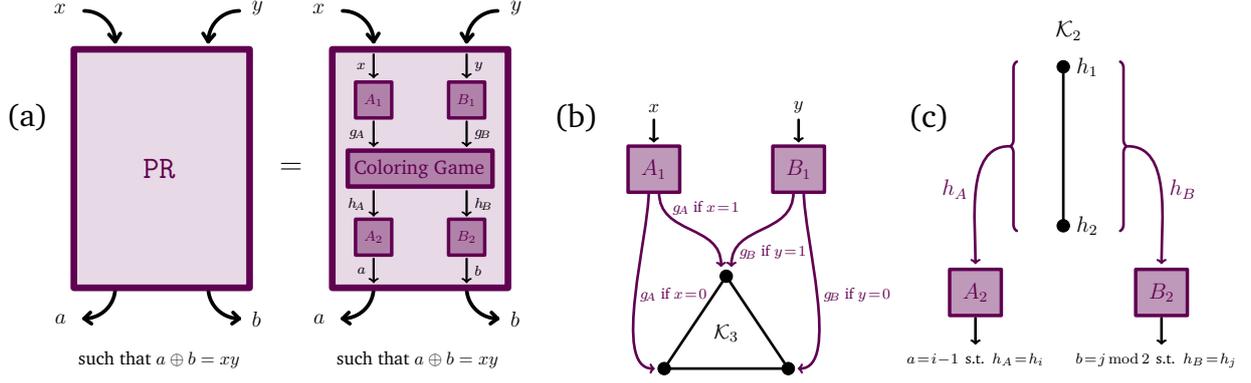
\begin{figure}
		\newcommand{\boxcolor}{mycolor5}
		\centering
		\subfile{Figures/Simulation-of-PR/Simulation-of-PR}
		\begin{tikzpicture}[line cap=round,line join=round,>=triangle 45,x=1.0cm,y=1.0cm,every node/.style={scale=0.7}, scale=0.68,
my-arrow/.style={-{Classical TikZ Rightarrow[length=0.7mm]}, line width=1},
my-arrow-blue/.style={my-arrow, color=\boxcolor}
]

\definecolor{mycolor}{RGB}{30, 70, 120} 
\definecolor{mycolor2}{HTML}{53257F} 
\definecolor{mycolor3}{RGB}{191, 95, 0}
\definecolor{mycolor4}{RGB}{0, 96, 81}  
\definecolor{mycolor5}{RGB}{96, 0, 81}
\definecolor{mycolor6}{RGB}{150, 30, 30} 

\newcommand{\myheight}{4.5}
\newcommand{\mywidth}{-0.6}
\newcommand{\arrowwidth}{1.4}

\newcommand{\xshift}{5.3}
\newcommand{\ycenter}{-2.0*0.5 -1.2*0.5- 0.15*0.5 + \myheight*0.5}

\newcommand{\miniboxheight}{0.45}
\newcommand{\miniboxwidth}{0.5}

\newcommand{\yycenter}{-2.0*0.25 -1.2*0.25- 0.15*0.25 + \myheight*0.25 +\miniboxheight*0.5 - 1.2*0.5- 0.15*0.5 + \myheight*0.5}
\newcommand{\yyycenter}{-2.0*0.25 -1.2*0.25- 0.15*0.25 + \myheight*0.25 - \miniboxheight*0.5 - 2.0*0.5}

\draw (-0.2+\xshift-1.5, -1.45 + \myheight + 0.05) node {\Large(b)};


\fill[color=\boxcolor, fill=\boxcolor, fill opacity=0.40] (-0.2+\xshift-\miniboxwidth, \yycenter - \miniboxheight) -- (-0.2+\xshift+\miniboxwidth, \yycenter - \miniboxheight) -- (-0.2+\xshift+\miniboxwidth,\yycenter + \miniboxheight) -- (-0.2+\xshift-\miniboxwidth, \yycenter + \miniboxheight) -- cycle;
\draw[line width=1.3pt, color=\boxcolor] (-0.2+\xshift-\miniboxwidth, \yycenter - \miniboxheight) rectangle (-0.2+\xshift+\miniboxwidth, \yycenter + \miniboxheight) ;
\draw[color=\boxcolor] (-0.2+\xshift, \yycenter) node {$A_1$};

\fill[color=\boxcolor, fill=\boxcolor, fill opacity=0.40] (3.2 + \mywidth+\xshift -\miniboxwidth, \yycenter - \miniboxheight) -- (3.2 + \mywidth+\xshift +\miniboxwidth, \yycenter - \miniboxheight) -- (3.2 + \mywidth+\xshift +\miniboxwidth,\yycenter + \miniboxheight) -- (3.2 + \mywidth+\xshift -\miniboxwidth, \yycenter + \miniboxheight) -- cycle;
\draw[line width=1.3pt, color=\boxcolor] (3.2 + \mywidth+\xshift -\miniboxwidth, \yycenter - \miniboxheight) rectangle (3.2 + \mywidth+\xshift +\miniboxwidth, \yycenter + \miniboxheight) ;
\draw[color=\boxcolor] (3.2 + \mywidth+\xshift, \yycenter) node {$B_1$};


\newcommand{\mylinewidth}{1pt}
\newcommand{\mycirclesize}{0.12}
\newcommand{\mystep}{1}
\newcommand{\graphwidth}{-0.2}
\newcommand{\xbegin}{-0.2+\xshift-\graphwidth}
\newcommand{\xend}{3.2 + \mywidth+\xshift+\graphwidth}
\newcommand{\xmiddle}{-0.2*0.5+\xshift*0.5 + 3.2*0.5 + \mywidth*0.5+\xshift*0.5}
\newcommand{\ytop}{0}
\newcommand{\ybottom}{-1.8}

\draw[line width=\mylinewidth] (\xbegin, \ybottom) -- (\xmiddle, \ytop) -- (\xend, \ybottom) -- cycle;
\draw[fill=black] (\xbegin, \ybottom) circle (\mycirclesize);
\draw[fill=black] (\xmiddle, \ytop) circle (\mycirclesize);
\draw[fill=black] (\xend, \ybottom) circle (\mycirclesize);
        
\draw (\xmiddle, \ytop*0.5 + \ybottom*0.5-0.2) node{\small $\K_3$};
        

\newcommand{\yendarrow}{0.18}

\draw[my-arrow] (-0.2+\xshift,-1.45 + \myheight) -- node[above, yshift=0.2cm]{\footnotesize$x$} (-0.2+\xshift, \yycenter + \miniboxheight+0.07);

\draw[my-arrow] (3.2 + \mywidth+\xshift,-1.45 + \myheight) -- node[above, yshift=0.2cm]{\footnotesize$y$} (3.2 + \mywidth+\xshift, \yycenter + \miniboxheight+0.07);

\draw[my-arrow-blue] (-0.2+\xshift-0.1,\yycenter - \miniboxheight) .. controls +(down:2em) and +(left:2em)..  (\xbegin - \yendarrow, \ybottom);
\draw (\xbegin+0.18, \ytop-0.35) node[color=\boxcolor]{\scriptsize$g_{\!A}$ if $x\!=\!0$};

\draw[my-arrow-blue] (-0.2+\xshift +0.1,\yycenter - \miniboxheight) node[below right, xshift=0.1cm]{\scriptsize$g_{\!A}$ if $x\!=\!1$} .. controls +(down:2em) and +(up:2em)..  (\xmiddle-0.1, \ytop+\yendarrow);

\draw[my-arrow-blue] (3.2 + \mywidth+\xshift-0.1, \yycenter - \miniboxheight) .. controls +(down:2em) and +(up:2em).. (\xmiddle+0.1, \ytop+\yendarrow) node[above right]{\scriptsize$g_{\!B}$ if $y\!=\!1$};

\draw[my-arrow-blue] (3.2 + \mywidth+\xshift+0.1, \yycenter - \miniboxheight) .. controls +(down:2em) and +(right:2em).. (\xend + \yendarrow, \ybottom);
\draw (\xend+1.33, \ytop-0.35)  node[color=\boxcolor]{\scriptsize$g_{\!B}$ if $y\!=\!0$};

\end{tikzpicture}\!\!\!
		\begin{tikzpicture}[line cap=round,line join=round,>=triangle 45,x=1.0cm,y=1.0cm,every node/.style={scale=0.7}, scale=0.68,
my-arrow/.style={-{Classical TikZ Rightarrow[length=0.7mm]}, line width=1},
my-arrow-blue/.style={my-arrow, color=\boxcolor}
]

\definecolor{mycolor}{RGB}{30, 70, 120} 
\definecolor{mycolor2}{HTML}{53257F} 
\definecolor{mycolor3}{RGB}{191, 95, 0}
\definecolor{mycolor4}{RGB}{0, 96, 81}  
\definecolor{mycolor5}{RGB}{96, 0, 81}
\definecolor{mycolor6}{RGB}{150, 30, 30} 

\newcommand{\myheight}{4.5}
\newcommand{\mywidth}{-0.6}
\newcommand{\arrowwidth}{1.4}

\newcommand{\ycenter}{0}

\newcommand{\boxheight}{0.45}
\newcommand{\boxwidth}{0.5}
\newcommand{\boxspacing}{1.8}

\draw (-\boxspacing-0.9, 3.4) node {\Large(c)};


\fill[color=\boxcolor, fill=\boxcolor, fill opacity=0.40] (-\boxspacing-\boxwidth, \ycenter- \boxheight) -- (-\boxspacing+\boxwidth,  \ycenter- \boxheight) -- (-\boxspacing+\boxwidth, \ycenter+\boxheight) -- (-\boxspacing-\boxwidth, \ycenter+\boxheight) -- cycle;
\draw[line width=1.3pt, color=\boxcolor] (-\boxspacing-\boxwidth, \ycenter-\boxheight) rectangle (-\boxspacing+\boxwidth, \ycenter+\boxheight) ;
\draw[color=\boxcolor] (-\boxspacing, \ycenter) node {$A_2$};

\fill[color=\boxcolor, fill=\boxcolor, fill opacity=0.40] (\boxspacing -\boxwidth, \ycenter - \boxheight) -- (\boxspacing+\boxwidth, \ycenter - \boxheight) -- (\boxspacing+\boxwidth, \ycenter + \boxheight) -- (\boxspacing-\boxwidth, \ycenter + \boxheight) -- cycle;
\draw[line width=1.3pt, color=\boxcolor] (\boxspacing-\boxwidth, \ycenter - \boxheight) rectangle (\boxspacing+\boxwidth, \ycenter + \boxheight) ;
\draw[color=\boxcolor] (\boxspacing, \ycenter) node {$B_2$};


\newcommand{\mycirclesize}{0.12}
\newcommand{\mylinewidth}{1pt}

\newcommand{\graphsize}{3.1cm}
\newcommand{\graphxshift}{-0.1cm}
\newcommand{\graphyshift}{1.3cm}

\draw[line width=\mylinewidth] (\graphxshift, \graphyshift) -- (\graphxshift, \graphyshift+\graphsize);
\draw[fill=black] (\graphxshift, \graphyshift+\graphsize) circle (\mycirclesize) node[right, xshift=0.1cm]{$h_1$};
\draw[fill=black] (\graphxshift, \graphyshift) circle (\mycirclesize) node[right, xshift=0.1cm]{$h_2$};

\draw (0,5.1) node {$\K_2$};

\newcommand{\xbrace}{1}
\newcommand{\ybracetop}{4.5}
\newcommand{\ybracebottom}{1.2}
\draw[decorate,decoration={brace}, color=\boxcolor, line width=1] (\xbrace, \ybracetop) -- (\xbrace, \ybracebottom);
\draw[decorate,decoration={brace}, color=\boxcolor, line width=1] (-\xbrace, \ybracebottom) -- (-\xbrace, \ybracetop);

\draw[my-arrow-blue] (\xbrace+0.15, \ybracebottom*0.5 + \ybracetop*0.5) .. controls +(right:2em) and +(up:2em).. node[right]{$h_B$} (\boxspacing, \ycenter + \boxheight+0.1);

\draw[my-arrow-blue] (-\xbrace-0.15, \ybracebottom*0.5 + \ybracetop*0.5) .. controls +(left:2em) and +(up:2em).. node[left]{$h_A$} (-\boxspacing, \ycenter + \boxheight+0.1);

\newcommand{\arrowlength}{0.5}
\draw[my-arrow] (-\boxspacing, \ycenter - \boxheight-0.05) -- (-\boxspacing, \ycenter - \boxheight-0.07-\arrowlength) node[below]{\scriptsize$a\!=\!i\!-\!1$\, s.t.\, $h_A\!=\!h_i$};
\draw[my-arrow] (\boxspacing, \ycenter - \boxheight-0.05) -- (\boxspacing, \ycenter - \boxheight-0.07-\arrowlength) node[below, xshift=-0.1cm]{\scriptsize$b\!=\!j\!\!\!\mod\!2$\, s.t.\, $h_B\!=\!h_j$};

\end{tikzpicture}
		\caption{
			Illustration of the proof of \refprop[Lemma]{lemma: K3 to K2 implies PR simulated}.
			In item~(a), we simulate a $\PR$ box from a perfect $\NS$-strategy for the graph coloring game, called ``Coloring Game'' in the figure, together with the local processes $A_1,A_2,B_1,B_2$ that are described in items~(b) and~(c).
			In item~(b), given $x$ and $y$, Alice and Bob choose some input vertices $g_A$ and $g_B$ in $\K_3$. 
			In item~(c), Alice and Bob receive some output vertices $h_A$ and $h_B$ from $\K_2$, and they choose $a$ and $b$ accordingly.
		}
		\label{fig:simulating-PR-from-coloring-game}
	\end{figure}

More generally, note that if we factorize the homomorphism $\K_3\to_{ns}\K_2$ by a graph $\G$, \ie if we have $\K_3\to_{ns} \G \to_{ns} \K_2$, we obtain from the lemma:

\begin{proposition}[Simulation of a $\PR$ Box]
	Given a perfect strategy for each of the homomorphism games $\K_3\to_{ns} \G$ and $\G\to_{ns} \K_2$, the $\PR$ box is perfectly simulable. \qed
\end{proposition}

Recall from~\cite{BBLMTU06} that whenever the $\PR$ box is simulated with probability $>\frac{3+\sqrt{6}}{6}$, there is a collapse of communication complexity. 
Hence, combining this fact with the former proposition, we obtain:

\begin{theorem}[Collapse of CC]  \label{thm:collapse-of-CC-for-the-homomorphism-game}
	Let $\G$ be a finite undirected graph, and
	let $\pp,\qq\leq1$ such that $\pp\qq>\frac{3+\sqrt{6}}{6}\approx0.91$.
	Then, any strategy winning the homomorphism game $\K_3\to_{ns} \G$ with probability~$\pp$, combined with a non-signalling strategy winning the $2$-coloring game of $\G$ with probability~$\qq$,
	induce a collapse of communication complexity. \qed
\end{theorem}

\begin{example} Let $\pp>\frac{3+\sqrt{6}}{6}$. The following examples are consequences of the theorem: \vspace{-0.1cm}
\begin{itemize}\setlength\itemsep{0em}
	\item As cycles of even length $\C_{2N}$ and path graphs $\L_N$ are $2$-colorable for any $N\geq1$, we have that any strategy winning the homomorphism game $\K_3\to_{ns} \C_{2N}$ or $\K_3\to_{ns} \L_{N}$ with probability at least $\pp$ allows to collapse CC.

	\item As $\K_3$ is $N$-colorable for any $N\geq 3$, we have that any non-signalling strategy winning the $2$-coloring game of $\K_N$ with probability $\pp$ allows to collapse CC.
\end{itemize}
\end{example}

\begin{proposition}[Graph Sequence]
	If we have the following decomposition:
	\[
		\G=:\G_1 \twoheadrightarrow \dots \twoheadrightarrow \G_n \twoheadrightarrow \K_3\to_{ns} \H_1 \to_{ns} \dots \to_{ns} \H_m \to_{ns} \K_2\,,
	\]
	where ``$\G\twoheadrightarrow \H$" denotes surjectivity,
	then the $\PR$ box is perfectly simulable and therefore there is a collapse of CC.
\end{proposition}

\begin{proof}
	Denote $g_1,g_2,g_3$ the three vertices of $\K_3$. By surjectivity of the first $n$ maps, there exist some vertices $a_1,a_2,a_3$ in $\G_1$ that are (deterministically) mapped to $g_1,g_2,g_3$  respectively in $\K_3$.
	We do a similar protocol as in \refprop[Lemma]{lemma: K3 to K2 implies PR simulated}: upon receiving $x$, Alice chooses $a_1$ if $x=0$ or $a_2$ otherwise, and upon receiving $y$ Bob chooses $a_3$ if $y=0$ or $a_2$ otherwise. 
	It produces in $\K_3$ the same scenario as in the protocol of \refprop[Lemma]{lemma: K3 to K2 implies PR simulated}.
	Then, the composition of the last $(m+1)$ morphisms simulates a morphism $\K_3\to_{ns} \K_2$, so $\PR$ is perfectly simulated.	
\end{proof}

	\subsection{Combining Graph Isomomorphism and Graph Coloring Strategies}

We present a result that generalizes \refprop[Corollary]{coro:toy-example-thm}~~to more than two connected components in $\H$, based on the assumption that Alice and Bob are given access to a perfect $\NS$-strategy for the 2-coloring game of $\K_N$, which is possible thanks to \refprop[Example]{ex:perfect-strategies-for-homomorphism-game-K_N-to-K_2-exist}.

\begin{theorem}[Collapse of CC]  \label{thm:combining-isomorphism-game-and-coloring-game}
	Let $\G$ and $\H$ be such that $\diam(\G)\geq2$, and that $\H$ admits exactly $N$ connected components $\H_1, \dots, \H_N$, all being complete. Then, given any  strategy $\S$ winning the graph isomorphism game $\G\cong_{ns} \H$ with probability $\pp$, combined with an $\NS$-strategy winning the $2$-coloring game of $\K_N$ with probability $\qq$ such that $\pp\qq>\frac{3+\sqrt{6}}{6}\approx0.91$, there is a collapse of communication complexity.
\end{theorem}

\begin{proof}
	We proceed similarly as in the proof of \refprop[Theorem]{theo:toy-example-thm}, but here the choice of $a$ and $b$ is given by the coloring of the components $\H_i$, $\H_j$ containing respectively $h_A$, $h_B$. 
	By assumption, Alice and Bob are given access to an $\NS$-strategy at the 2-coloring game of $\K_N$, so they can use it to simulate a coloring of the components of $\H$: they can assign different colors to two different components and to assign simultaneously the same color if they are given the same component. 
	Based on this ability, if the component $\H_i$ containing $h_A$ is of the first color, Alice assigns $a=0$, otherwise she assigns $a=1$, and similarly for Bob. 
	It yields $a\neq b$
	\iff Alice and Bob have different colors,
	\iff $h_A$ and $h_B$ are in different connected components of $\H$ with probability $\qq$,
	\iff $g_A\not\simeq g_B$ with probability $\pp$ because of the completeness of the components of $\H$,
	\iff $x=y=1$ in the protocol of \refprop[Figure]{fig:simulating-PR-from-isom-game}. Hence, the relation $a\oplus b=xy$ 
	is satisfied with probability $\pp\qq$, the $\PR$ box is simulated with the same probability and thanks to~\cite{BBLMTU06}, we conclude that there is a collapse of communication complexity.
	\end{proof}

\bib

\section{Vertex Distance Game~[new]}
\label{section:vertex-distance-game}

In this section, we introduce and study a generalization of the graph isomorphism game (\refprop[Section]{section:graph-isom-game}). We name it the \emph{vertex distance game}.

\subsection{Definition of the Game}

We introduce a new nonlocal game that we call \emph{vertex distance game} with parameter $D\in\N$, or simply \emph{$D$-distance game}. Given two graphs $\G$ and $\H$ with disjoint vertex sets, two question vertices are chosen by a referee $x_A, x_B\in V=V(\G)\cup V(\H)$ and distributed to space-like separated players Alice and Bob who are not allowed to communicate. 
See footnote~\ref{footnote:the-inputs-are-not-necessarily-in-G} at page~\pageref{footnote:the-inputs-are-not-necessarily-in-G} to understand why we choose the inputs $x_A,x_B$ in $V$ and not simply in $V(\G)$.
In order to win the game, Alice and Bob try to output vertices $y_A, y_B\in V$ satisfying two conditions. The first one is the same first rule as~\eqref{eq:rule-1-of-graph-isom-game} for the graph isomorphism game:
\begin{equation}   \label{eq:first-condition-of-the-D-distance-game}
	x_A\in V(\G) \Leftrightarrow y_A\in V(\H)
	\quad\quad\text{and}\quad\quad
	x_B\in V(\G) \Leftrightarrow y_B\in V(\H)\,.
\end{equation}
Assuming this relation holds, we relabel the vertices into $g_A,g_B,h_A,h_B$ as in the isomorphism game: only one vertex among $x_A$ and $y_A$ is in $V(\G)$, let us call it $g_A\in V(\G)$, and the other $h_A\in V(\H)$, and similarly for $g_B\in V(\G)$ and $h_B\in V(\H)$.
Then, the second condition is that distances are preserved until $D$:
\begin{equation}  \label{eq:second-condition-of-the-D-distance-game}
d(h_A, h_B) = \left\{
\begin{array}{cl}
d(g_A, g_B)
& 
\text{if $d(g_A, g_B) \leq D$}\,,
\\
> D
& 
\text{otherwise}\,.
\end{array}
\right.
\end{equation}
We write $\G\!\cong^D\!\!\H$, and we say that the graphs $\G$ and $\H$ are $D$-isomorphic, if there exists a perfect classical strategy for the $D$-distance game, and similarly $\cong^D_q$ and $\cong_{ns}^D$ with perfect quantum and non-signalling strategies. 
These notations will make sense with regard to \refprop[Example]{ex:basic-cases-for-the-vertex-distance-game} because this game generalizes the isomorphism game.
Note that:
\[
	\dots
	\quad\Longrightarrow\quad
	\G\cong^{D=2}_s\H
	\quad\Longrightarrow\quad
	\G\cong^{D=1}_s\H
	\quad\Longrightarrow\quad
	\G\cong^{D=0}_s\H\,,
\]
for any strategy type $s$ such as classical, quantum, non-signalling, or any other type. 
Note that we do not need to assume that $|V(\G)|=|V(\H)|$ since it is a consequence of the setting, see equation~\eqref{eq:G-and-H-have-the-same-cardinality} below.
Three cases are noticeable:

\begin{example} \label{ex:basic-cases-for-the-vertex-distance-game}
\begin{enumerate}
	\item The case $D=0$ corresponds to the Graph Bisynchronous Game~\cite{PR21}, where we require that same vertices $g_A=g_B$ are mapped to same vertices $h_A=h_B$, and that different vertices are mapped to different vertices.
	In particular, if we consider the graphs $\G=\K_M$ and $\H=\K_N$, the case $D=0$ exactly corresponds to the $N$-coloring game of $\K_M$.
	\item The case $D=1$ corresponds precisely to the Graph Isomorphism Game introduced in \refprop[Section]{section:graph-isom-game}, where the relations $\{=, \sim, \not\simeq\}$ are preserved. Hence the vertex distance game is a generalization of the graph isomorphism game.
	\item The case $D=\diam(\H)$ corresponds to:
	\[
	d(h_A, h_B) = \left\{
	\begin{array}{cl}
	d(g_A, g_B)
	& 
	\text{if $d(g_A, g_B) \leq \diam(\H)$,}
	\\
	\infty
	& 
	\text{otherwise.}
	\end{array}
	\right.
	\]
	Note that it requires that $\H$ admits at least two connected components if the diameter of $\G$ is larger than the one of $\H$. 
	We will particularly be interested in this third case in what follows.
\end{enumerate}
\end{example}

\begin{remark}   
	\label{rmk:construction-of-G_t}
	There is another way to define the vertex distance game.
	From a given graph $\G$, construct $\G_t$ the graph with the same vertex set as $\G$, 
	by putting an edge between two vertices in $\G_t$ if the distance in $\G$ is exactly $t$.
	Then, observe that winning the $D$-distance game is equivalent to winning the graph isomorphism games of $(\G_t, \H_t)$ for all $t\leq D$.
\end{remark}

We will need the following lemma that generalizes~\cite[Lemma~4.1]{AMRSSV19}:

\begin{lemma}  \label{lem:first-properties-of-the-D-distance-game}
	If $\P\in\NS$ is a perfect non-signalling strategy for the $D$-distance game of $(\G,\H)$, then for all $g\in V(\G)$ and  $h\in V(\H)$:
	\vspace{-5pt}
	\begin{enumerate}[label=(\roman*),itemsep=-3pt]
		\item \label{item:first-property-of-the-D-distance-game-1} $\sum_{h\in V(\H)} \P(h,h\,|\,g,g)
			=1
			= \sum_{g\in V(\G)} \P(h,h\,|\,g,g)$\,,
		\item \label{item:first-property-of-the-D-distance-game-2} $\P(h,h\,|\,g,g)
			=\P(g,h\,|\,h,g)
			=\P(h,g\,|\,g,h)
			=\P(g,g\,|\,h,h)$\,.
	\end{enumerate}
\end{lemma}

\begin{proof}
	The first equality of item~\ref{item:first-property-of-the-D-distance-game-1} follows from the first condition of the game in equation~\eqref{eq:first-condition-of-the-D-distance-game}, which states that both outputs have to be in $V(\H)$ if both inputs are in $V(\G)$, combined with the condition in equation~\eqref{eq:second-condition-of-the-D-distance-game} with distance $0$ stating that equality is preserved. The second equality is similarly shown.
	As for item~\ref{item:first-property-of-the-D-distance-game-2}, if we denote $V=V(\G)\cup V(\H)$, the winning conditions of the game together with the non-signalling condition give rise to:
	\[
		\P(h,h\,|\,g,g)
		\,=\,
		\sum_{y\in V} \P(y,h\,|\,g,g)
		\,=\,
		\sum_{y\in V}\P(y,h\,|\,h,g)
		\,=\,
		\P(g,h\,|\,h,g)\,.
	\]
	We obtain the other equalities with a similar method.
\end{proof}

In particular, by combining items~\ref{item:first-property-of-the-D-distance-game-1} and~\ref{item:first-property-of-the-D-distance-game-2}, we obtain that the vertex sets have the same cardinality:
\begin{align} \label{eq:G-and-H-have-the-same-cardinality}
\begin{split}
	|V(\G)|
	\,=\,
	\sum_{g\in V(\G)} 1
	&\,=\,
	\sum_{g\in V(\G)} \sum_{h\in V(\H)} \P(h,h\,|\,g,g)
	\\
	&\,=\,
	\sum_{h\in V(\H)} \sum_{g\in V(\G)} \P(g,g\,|\,h,h)
	\,=\,
	\sum_{h\in V(\H)} 1
	\,=\,
	|V(\H)|\,,
\end{split}
\end{align}
which was not obvious at first glance.

		\subsection{Characterization of Perfect Classical and Quantum Strategies}
		\label{subsec:characterizing-perfect-strategies-at-the-vertex-distance-game}

Surprisingly, as shown in the following proposition, the case $D=1$ may be extended to any $D\geq1$ when strategies are deterministic, classical, or quantum. On the contrary, we will see in \refprop[Subsection]{subsec:example-of-D-isom-but-not-(D+1)-isom} that it is not the case for non-signalling strategies.

\begin{proposition}[Characterization]
	\label{prop:same-perfect-classical-and-quantum-strategies}
For any $D\geq1$, perfect deterministic/classical/quantum strategies coincide for the graph isomorphism game and the $D$-distance game:
\begin{align*}
	\G\cong\H
	\quad\quad&\Longleftrightarrow\quad\quad
	\forall D\geq1, \quad
	\G\cong^D\H
	&\Longleftrightarrow\quad\quad
	\exists D\geq1, \quad
	\G\cong^D\H\,,
	\\
	\G\cong_q\H
	\quad\quad&\Longleftrightarrow\quad\quad
	\forall D\geq1, \quad
	\G\cong^D_q\H
	&\Longleftrightarrow\quad\quad
	\exists D\geq1, \quad
	\G\cong^D_q\H\,.
\end{align*}
\end{proposition}

\begin{proof}
Let $D\geq1$.
On the one hand, as the isomorphism game consists in preserving distances $\{0,1,>\!1\}$, we see that any perfect strategy for the $D$-distance game is also perfect for the isomorphism game, whether it is deterministic, classical, quantum or non-signalling.
On the other hand:
\begin{enumerate}[label=(\alph*)]
\item \label{bullet:perfect-deterministic-strategies} Perfect deterministic strategies for the isomorphism game preserve the distances. Indeed, if $D'$ is the distance separating some vertices $g_A$ and $g_B$ in $\G$, then there is a path $(g_0, g_1, \dots, g_{D'})$ in $\G$, where $g_0=g_A$, $g_d=g_B$ and $g_i\sim g_{i+1}$ for all $i$. It follows that:
	\begin{align*}
		d(g_A, g_B) 
		&\,=\, 
		d(g_0, g_1) + \dots + d(g_{D-1}, g_{D'})
		\\
		&\,=\,
		d(h_0, h_1) + \dots + d(h_{{D'}-1}, h_{D'})
		\, \geq\,
		d(h_0, h_{D'})\,,
	\end{align*}
	where $h_i$ is the image of $g_i$ under the bijection $\varphi:\G\to\H$ induced by the deterministic strategy, and where the last inequality holds by the triangular inequality. We note that $h_0$, $h_D$ are the respective images of $g_A$, $g_B$.
	Assume by contradiction that the last inequality is strict. Then there is a path in $\H$ connecting $h_0$ to $h_{D'}$ of length $D''<{D'}$. Applying $\varphi^{-1}$, we obtain a path in $\G$ connecting $g_A$ to $g_B$ of length $D''<{D'}$, which contradicts the fact that ${D'}$ is the distance between $g_A$ and $g_B$. Thus $d(g_A, g_B) =d(h_0, h_{D'})$ and we have the desired result.
	\item A classical strategy is a convex combination of deterministic strategies:
		\[
			\P(h_A, h_B \,|\, g_A, g_B)
			\,=\,
			\sum_i p_i \,\P^{\text{det}}_i(h_A, h_B \,|\, g_A, g_B)\,,
		\]
		with $\sum_i p_i = 1$ and the sum index is finite. It means that we apply the strategy $\P^{\text{det}}_i$ with probability $p_i$. If $\P$ is a perfect classical strategy, then it wins with probability one, and therefore each $\P^{\text{det}}_i$ also has to be perfect. So by item~\ref{bullet:perfect-deterministic-strategies}, each $\P^{\text{det}}_i$ preserves the distance, thus $\P$ also preserves the distance.

	\item Let $\P$ be a perfect quantum strategy for the isomorphism game. This proof is an easy adaptation of~\cite[Theorem~1.1]{Schmidt2020}. 
	We want to show that distance is preserved, \ie that the event ``$d(h_A,h_B)\neq d(g_A,g_B)$'' has zero probability. 
	Using~\cite[Theorem~5.14]{AMRSSV19}, we know that there exists a $C^*$-algebra $\A$ with a tracial state $\tau$ and projections $E_{gh}\in\A$ for $g\in V(\G)$ and $h\in V(\H)$ such that $u$ is a quantum permutation matrix and:
	\begin{equation}   \label{eq:product-of-u-is-null-when-not-relations}
	u_{g_Ah_A}u_{g_Bh_B}\,=\,0
	\quad\quad
	\text{if \quad $\rel(g_A,g_B)\neq\rel(h_A,h_B)$}\,,
	\end{equation}
	where the function $\rel(g_A,g_B)$ takes value $0$ if $g_A=g_B$, value $1$ if $g_A\sim g_B$, and value $2$ if $g_A\not\simeq g_B$.
	Using these notations, the correlation $\P$ is then of the form 
	\begin{equation}  \label{eq:P-in-terms-of-tau-and-u}
	\P(h_A,h_B\,|\,g_A,g_B)=\tau(u_{g_Ah_A}u_{g_Bh_B})\,.
	\end{equation}
	Now, as the isomorphism game is equivalent to the $1$-distance game, we already know that distances $0$ and $1$ are preserved. It remains to show the results for distances $\geq2$. 
	Consider vertices $g_A,g_B,h_A,h_B$ such that $2\leq d(g_A,g_B)=t<d(h_A,h_B)$. We will show that then $\P(h_A,h_B\,|\,g_A,g_B)=0$.
	By definition, there exist a path in $\G$ with adjacent vertices $g_1,\dots,g_t\in V(\G)$ going from $g_1=g_A$ to $g_t=g_B$,
	but no such path exists in $\H$ from $h_A$ to $h_B$. So for all $h_1,\dots,h_t\in V(\H)$ such that $h_1=h_A$ and $h_t=h_B$, we infer there exists at least one index $s\in\{1,\dots,t-1\}$ such that $h_s\not\sim h_{s+1}$ but $g_s\sim g_{s+1}$.
	By equation~\eqref{eq:product-of-u-is-null-when-not-relations}, we deduce that $u_{g_sh_s}u_{g_{s+1}h_{s+1}}=0$, and therefore:
	\begin{align*}
		u_{g_Ah_A}u_{g_Bh_B}
		&\,=\,
		u_{g_Ah_A} \cdot \IdentityMatrix \cdot ... \cdot \IdentityMatrix \cdot u_{g_Bh_B}\\
		&\,=\,
		\sum_{h_2,...,h_{t-1}\in V(\H)} u_{g_Ah_A} u_{g_2h_2} \dots u_{g_{t-1}h_{t-1}} u_{g_Bh_B}
		\,=\,
		0\,.
	\end{align*}
	Thus using equation~\eqref{eq:P-in-terms-of-tau-and-u}, we obtain that the event ``$d(h_A,h_B)\neq d(g_A,g_B)$'' has probability probability zero, and similarly for the event with the opposite inequality. \qedhere
\end{enumerate}
\end{proof}

A direct consequence of \refprop[Proposition]{prop:same-perfect-classical-and-quantum-strategies} and the results~\cite{AMRSSV19, Lovasz67, CV93, LMR20, MR20} displayed in \refprop[Figure]{fig:characterizations-of-the-different-types-of-isomorphism} is the following lists of characterizations:

\begin{corollary}[Classical Strategies]
	Let $D\geq1$. The following are equivalent:\vspace{-5pt}
	\begin{enumerate}[label=(\roman*),itemsep=-3pt]
		\item $\G\cong^D\H$.
		\item There exists a permutation matrix $P$ such that $A_\G P=PA_\H$.
		\item For any graph $\K$, we have $\#\Hom(\K, \G)=\#\Hom(\K,\H)$.
		\item For any graph $\K$, we have $\#\Hom(\G,\K)=\#\Hom(\H,\K)$.
	\end{enumerate}
\end{corollary}

\begin{corollary}[Quantum Strategies]
	Let $D\geq1$. The following are equivalent:\vspace{-5pt}
	\begin{enumerate}[label=(\roman*),itemsep=-3pt]
		\item $\G\cong^D_q\H$.
		\item There exists a quantum permutation matrix $P$ such that $A_\G P=PA_\H$.
		\item For all planar graph $\K$, we have $\#\Hom(\K, \G)=\#\Hom(\K,\H)$.
	\end{enumerate}
\end{corollary}

		\subsection{Characterization of Perfect Non-Signalling Strategies}
		\label{subsec:perfect-NS-strategies}

In this subsection, we generalize the results that $\G\cong_{ns}\H$ 
\iff $\G$ and $\H$ are fractionally isomorphic~\cite{AMRSSV19},
\iff they admit a common equitable partition~\cite{RSU94}. 
Along this subsection, we relax the definitions of fractional isomorphism and of common equitable partition with a parameter $D$, and the combination of all lemmata leads to the following theorem:

\begin{theorem}[Characterization of  Perfect $\NS$-Strategies]
	\label{thm:Characterization-of-perfect-NS-strategies}
	Let $\G$ and $\H$ be two graphs and $D\geq0$ be an integer. The following are equivalent:\vspace{-5pt}
	\begin{enumerate}[label=(\roman*),itemsep=-3pt]
		\item \label{item:Characterization-of-perfect-NS-strategies-1} $\G\cong_{ns}^D\H$\,,
		\item \label{item:Characterization-of-perfect-NS-strategies-2} $\G\cong_{frac}^D\H$\,,
		\item \label{item:Characterization-of-perfect-NS-strategies-3} There exists a $D$-common equitable partition of $(\G,\H)$.
	\end{enumerate}
\end{theorem}

\begin{remark}[Characterization for the Game with Inputs in $\G$]
	If we consider the similar -- yet different -- $D$-distance game where the inputs $x_A,x_B$ are always in $V(\G)$ instead of $V$, then a similar proof shows the following statement.
	Let $\G,\H$ two graphs with the same number of vertices, and let $D\geq0$ be an integer. Then, the following are equivalent:\vspace{-0.1cm}
	\begin{enumerate}[label=(\roman*),itemsep=-3pt]
		\item[(i')] This version of the game admits a perfect strategy $\P\in\NS$ such that the flipped correlation $\P'(g_A,g_B\,|\,h_A,h_B):=\P(h_A,h_B\,|\,g_A,g_B)$ is also in $\NS$.
		\item $\G\cong_{frac}^D\H$.
		\item There exists a $D$-common equitable partition of $(\G,\H)$.
	\end{enumerate}
\end{remark}

\subsubsection{Generalized Fractional Isomorphism}
\label{para:Generalized-Fractionnal-Isomorphism}
Let $D\geq0$. Two graphs $\G$ and $\H$ are said to be \emph{$D$-fractionally isomorphic}, denoted $\G\cong_{frac}^D\H$, if there exists a bistochastic matrix $u\in\mathcal M_n(\R)$ (that is a matrix whose entries are non-negative and whose rows and columns sum to one) such that for all distances $t\leq D$ we have:
\begin{equation}  \label{eq:definition-of-D-fractionally-isomorphism}
\forall\,g\in V(\G),\,\, \forall\,h\in V(\H),\quad\quad
\sum_{h'\in C(h,t)} u_{gh'}
\,=\,
\sum_{g'\in C(g,t)} u_{g'h}\,,
\end{equation}
where $C(g,t)$ is the circle of radius $t$ in $\G$ centered at $g$, \ie the set of neighbors of $g$ in $\G$ at distance exactly $t$.
Note that in the case $D=1$, we retrieve the usual notion of fractional isomorphism, because the condition in the equation amounts to $\sum_{h'\sim h} u_{gh'}
\,=\,
\sum_{g'\sim g} u_{g'h}$, which is equivalent to saying that the adjacency matrices satisfy $u\, A_\G = A_\H\, u$. We can rephrase equation~\eqref{eq:definition-of-D-fractionally-isomorphism} in terms of a generalization of the adjacency matrix. We call the matrix $A_\G^{(D)}$ the \emph{$D$-adjacency matrix} of a graph $\G$, whose coefficients $a_{ij}$ are $1$ if the distance between $g_i$ and $g_j$ satisfies $d(g_i,g_j)=D$, and $0$ otherwise. We have the equivalence:
\begin{equation} \label{eq:characterization-of-D-fractionnal-isom-in-terms-of-the-D-adjacency-matrix}
	\text{Equation~\eqref{eq:definition-of-D-fractionally-isomorphism}}
	\quad\quad\Longleftrightarrow\quad\quad
	u\, A_\G^{(t)} = A_\H^{(t)}\, u\,.
\end{equation}
Note that the $D$-adjacency matrix may be expressed in terms of the (usual) powers $A_\G^t$ of the adjacency matrix. The coefficients of the latter matrix may be interpreted as taking value $1$ \iff there exists a path of length $t$ in $\G$ joining the corresponding vertices. We see that a coefficient is $1$ in the $D$-adjacency matrix $A_\G^{(D)}$ \iff there exists a path of length $D$ in $\G$ joining the corresponding vertices, but no path of length $t<D$. 
In other words $A_\G^{(t)}$ is the adjacency matrix of the graph $\G_t$ as described in \refprop[Remark]{rmk:construction-of-G_t}.
We have the following relation:
\[
	A_\G^{(D)}
	\,=\,
	\Big(A_\G^D\div A_\G^D\Big) \circledast 
	\Big(\1-A_\G^{D-1}\div A_\G^{D-1}\Big) 
	\circledast 
	...
	\circledast 
	\Big(\1-A_\G^{0}\div A_\G^{0}\Big)\,,
\]
where $\div$ and $\circledast$ are the element-wise division and multiplication of matrices (\aka the Hadamard division and product, or Schur product), and where $\1$ is the matrix with all entries $1$ of the same size as $A_\G$.
Observe that $A_\G^{(0)}=\IdentityMatrix$ the identity matrix, and $A_\G^{(1)}=A_\G$ the adjacency matrix, and $A_\G^{(D)}=\0$ the zero matrix for all $D> \diam(\G)$ because the graph $\G$ admits no vertices with such a distance $D$. 
Note that the $D$-adjacency matrix may be equivalently recursively defined:
\[
	A_\G^{(D)}
	\,=\,
	\Big(A_\G^D\div A_\G^D\Big) \circledast 
	\left(\1-\sum_{t=0}^{D-1} A_\G^{(t)}\right)\,,
\]
because two vertices of $\G$ have distance $D$ \iff there is a path of length $D$ joining them and they do not have distance $t\leq D-1$. 
Note that we will provide in \hyperref[{subsec:example-of-D-isom-but-not-(D+1)-isom}]{Subsection~\ref{subsec:example-of-D-isom-but-not-(D+1)-isom}} an example of sequence of graphs $(\G_D,\H_D)$ such that $\G_D\cong_{frac}^D\H_D$ but $\G_D\not\cong_{frac}^{(D+1)}\H_D$.
Here is a sufficient condition in order to have a $D$-fractional isomorphism:

\begin{lemma}   \label{lem:NS-D-isom-implies-D-fractional}
	If $\G\cong_{ns}^D\H$ for some integer $D\geq0$, 
	then $\G$ and $\H$ are $D$-fractionally isomorphic:
	$$
	\G\cong_{ns}^D\H
	\quad\Longrightarrow\quad
	\G\cong_{frac}^D\H\,.
	$$
\end{lemma}

\begin{proof}
	This proof generalizes~\cite[Lemma~4.2]{AMRSSV19}.
	We want to construct a bistochastic matrix $u$ such that equation~\eqref{eq:definition-of-D-fractionally-isomorphism} holds for all $t\leq D$. We will index the elements of the matrix $u$ by the vertices of $\G$ and $\H$, for instance $u_{gh}$.
	As the strategy $\P$ is a valid probability distribution, we can define $u_{gh}=\P(h,h\,|\,g,g)\geq0$, and using \refprop[Lemma]{lem:first-properties-of-the-D-distance-game}~\ref{item:first-property-of-the-D-distance-game-1}, we have that rows and columns sum to one, $\sum_h u_{gh}=1$ and $\sum_g u_{gh}=1$, so the matrix $u$ is bistochastic.
	Let us verify the equality of equation~\eqref{eq:definition-of-D-fractionally-isomorphism} for an arbitrary integer $t\leq D$ and vertices $g\in V(\G)$ and $h\in V(\H)$. We have:
	\begin{align*}
		\sum_{h'\in C(h,t)} u_{gh'}
		&\,=\,
		\sum_{h'\in C(h,t)} \P(h',h'\,|\,g,g)
		\,=\,
		\sum_{h'\in C(h,t)} \sum_{g'\in V(\G)} \P(h',h'\,|\,g,g')\,,
	\intertext{which holds because $\P$ is perfect so it satisfies the rule that the outputs need to be equal \iff the inputs $g$ and $g'$ are equal. 
	Now, using the non-signalling condition of $\P'\in\NS$ on Bob's marginal, we obtain:}
		&\,=\,
		\sum_{h'\in C(h,t)} \sum_{g'\in V(\G)} \P(h',h\,|\,g,g')
		\,=\,
		\sum_{h'\in C(h,t)} \sum_{g'\in C(g,t)} \P(h',h\,|\,g,g')\,,
	\intertext{where the last equality holds because $\P$ is perfect so it satisfies the rule that the the distance $t$ is the same for the outputs $h',h$ and the inputs $g,g'$. Then, we swap the two sums and we use that $\P\in\NS$ and similar arguments as before to derive what follows:}
		&\,=\,
		\sum_{g'\in C(g,t)} \sum_{h'\in C(h,t)} \P(h',h\,|\,g,g')
		\,=\,
		\sum_{g'\in C(g,t)} \sum_{h'\in V(\H)} \P(h',h\,|\,g,g')
		\\
		&\,=\,
		\sum_{g'\in C(g,t)} \sum_{h'\in V(\H)} \P(h',h\,|\,g',g')
		\,=\,
		\sum_{g'\in C(g,t)} \P(h,h\,|\,g',g')
		\,=\,
		\sum_{g'\in C(g,t)} u_{g'h}\,.
	\end{align*}
	Hence, equation~\eqref{eq:definition-of-D-fractionally-isomorphism} holds, and the graphs $\G$ and $\H$ are $D$-fractionally isomorphic.
\end{proof}

For \refprop[Theorem]{thm:Characterization-of-perfect-NS-strategies}, we also generalize the notion of common equitable partition. Recall that the usual notion of common equitable partition was defined on page~\pageref{para:defition-of-common-equitable-partition}.

\subsubsection{Generalized Common Equitable Partition}
Let $D\geq0$.
We say that two graphs $\G$ and $\H$ admit a \emph{$D$-common equitable partition} if they admit respective partitions $\Cpartition=(C_1, \dots, C_k)$ and $\Dpartition=(D_1, \dots, D_\ell)$ with the following common parameters:
\begin{align*}
	& k = \ell\,, \\
	\forall i\in\{1,\dots,k\}\quad& |C_i|=|D_i|=:n_i\,, \\
	\forall t\leq D, \forall i,j\in\{1,\dots,k\}, \forall g\in C_i, \forall h\in D_i, \quad& |C_j\cap C(g,t)|=|D_j\cap C(h,t)|=:c_{ij}^{(t)}\,.
\end{align*}
Note that the case $D=1$ corresponds exactly to the usual notion of common equitable partition. 
Note that $c_{ij}^{(0)}=\delta_{ij}$ is the Kronecker delta, and that $c_{ij}^{(t)}=0$ when $t>\min\{\diam(\G), \diam(\H)\}$.
We do not necessarily have $c_{ij}^{(t)}=c_{ji}^{(t)}$, but we always have the following relation:

\begin{lemma} \label{lem:relation-betwenn-ct_ij-and-ct_ji}
	If the graph $\G$ admits a $D$-equitable partition with parameters as above, 
	then:
	$$n_i\,c_{ij}^{(t)}=n_j\,c_{ji}^{(t)}\,.$$ 
\end{lemma}

\begin{proof}
	This proof is a generalization of~\cite[Section~2.1]{RSU94}.
	Up to reordering the rows and columns of the $t$-adjacency matrix $A_\G^{(t)}$, this matrix can be decomposed in blocks $A_{ij}^{(t)}$ of size $n_i\times n_j$ such that the rows sum to $c_{ij}^{(t)}$. 
	By symmetry of the $t$-adjacency matrix, the blocks satisfy $A_{ij}^{(t)\top}=A_{ji}^{(t)}$, so the columns of $A_{ij}^{(t)}$ sum to $c_{ji}^{(t)}$.
	Now, as the sum of all the elements of $A_{ij}^{(t)}$ equals both the sum of its rows and the sum of its columns, we obtain
	$
	n_i\,c_{ij}^{(t)}=n_j\,c_{ji}^{(t)}
	$,
	hence the wanted result.
\end{proof}

We prove that $D$-fractional isomorphic is a sufficient condition for the graphs to admit a $D$-common equitable partition:

\begin{lemma}   \label{lem:if-D-fractional-isom-then-D-common-equitable-partition}
	If $\G\cong_{frac}^D\H$ for some integer $D\geq0$\,,
	then there exists a $D$-common equitable partition of $(\G,\H)$.
\end{lemma}

\begin{proof}
	This proof generalizes~\cite[Theorem~2.2]{RSU94}.
	We use an equivalent characterization of $\cong_{frac}^D$ as the one given in equation~\eqref{eq:characterization-of-D-fractionnal-isom-in-terms-of-the-D-adjacency-matrix}, \ie there exists a bistochastic matrix $u$ such that for all $t\leq D$:
	\begin{equation} \label{eq:in-the-proof-of-lemma-3-6}
	A_\G^{(t)}\, u = u\,A_\H^{(t)}\,.
	\end{equation}
	From this matrix $u$, we define a partition $\Cpartition$ on $V(\G)$ based on the following equivalence relation:
	$$
	g\leftrightarrow g'
	\quad\Longleftrightarrow\quad
	\begin{matrix}
	\text{there exists a ``link'' from $g$ to $g'$, \ie $\exists n, g_1, ..., g_n, h_1, ..., h_n$ such that:}
	\\
	\text{$g_1=g$, \quad\quad $g_n=g'$,\quad\quad $u_{g_1h_1}\cdot u_{g_2h_1}\cdot u_{g_2h_2}\cdot ... \cdot u_{g_nh_n}>0$\,,}
	\end{matrix}
	$$
	and we similarly define a partiton $\Dpartition$ on $V(\H)$.
	By construction, up to reordering the rows and columns of $u$, these partitions $\Cpartition$ and $\Dpartition$ are in correspondence with a block decomposition $u=U_1\oplus\dots\oplus U_k$, where each $U_i$ is an indecomposable $n_i\times m_i$ bistochastic matrix for some $n_i,m_i$. In particular, we have that both partitions $\Cpartition$ and $\Dpartition$ have $k$ cells and that each cell $C_i$ and $D_i$ has respective size $n_i$ and $m_i$. 
	Using the fact that $u$ is bistochastic, we have:
	\[
		m_i
		\,=\,
		\sum_{h\in D_i} 1
		\,=\,
		\sum_{h\in D_i} \sum_{g\in C_i} u_{gh}
		\,=\,
		\sum_{g\in C_i}\sum_{h\in D_i} u_{gh}
		\,=\,
		\sum_{g\in C_i}1
		\,=\,
		n_i\,,
	\]
	hence $|C_i|=|D_i|=n_i$ as wanted. It remains to prove that $\Cpartition$ and $\Dpartition$ admit common parameters $c_{ij}^{(t)}$. Let $t\leq D$ and denote $A^{(t)}:=A_\G^{(t)}$ and $B:=A_\H^{(t)}$ the $t$-adjacency matrices of the graphs $\G$ and $\H$. Write $A^{(t)}$ with blocks $A_{ij}^{(t)}$ of size $n_i\times n_j$ induced by the partition $\Cpartition$, and similarly for $B^{(t)}$ with blocks $B_{ij}^{(t)}$ of the same size. From equation~\eqref{eq:in-the-proof-of-lemma-3-6}, we deduce the following relations:
	\begin{equation}  \label{eq:lemma-3-6-*}
		\forall i,j\in[k],\quad\quad
		A_{ij}^{(t)}\,U_j
		\,=\,
		U_i\,B_{ij}^{(t)}\,.
	\end{equation}
	As well, after swapping $i$ and $j$, we have $A_{ji}^{(t)}\,U_i=U_j\,B_{ji}^{(t)}$, and taking the transpose we obtain:
	\begin{equation} \label{eq:lemma-3-6-**}
		\forall i,j\in[k],\quad\quad
		U_i^\top\,A_{ij}^{(t)}
		\,=\,
		B_{ij}^{(t)}\,U_j^\top\,.
	\end{equation}
	Let $v_{ij}^{(t)}:=A_{ij}^{(t)}\,\1$ be the vector such that each coordinate corresponds to a $g\in C_i$ and represents the number of $g'\in C_j$ at distance exactly $t$ of $g$, where $\1$ denotes the vector of appropriate size with all entries $1$. In order to have a ``$D$-equitable'' partition, we want that all the coordinates of the vector $v_{ij}^{(t)}$ have the same value $c_{ij}^{(t)}\in\R$, \ie that $v_{ij}^{(t)}=c_{ij}^{(t)}\,\1$. 
	Similarly, we define the vector $w_{ij}^{(t)}:=B_{ij}^{(t)}\,\1$, and in order to have a $D$-``common'' equitable partition, we want to prove that: 
	\begin{equation} \label{eq:lemma-3-6-***}
	v_{ij}^{(t)}=w_{ij}^{(t)}=c_{ij}^{(t)}\,\1\,.
	\end{equation}
	Using the fact that $U_j$ is bistochastic and then equation~\eqref{eq:lemma-3-6-*}, we have:
	\[
		v_{ij}^{(t)}
		\,:=\,
		A_{ij}^{(t)}\,\1
		\,=\,
		A_{ij}^{(t)}\,U_j\,\1
		\,=\,
		U_i\,B_{ij}^{(t)}\,\1
		\,=\,
		U_i\,w_{ij}^{(t)}\,,
	\]
	and similarly, using the fact that $U_j^\top$ is bistochastic and then equation~\eqref{eq:lemma-3-6-**}, we get:
	\[
		w_{ij}^{(t)}
		\,:=\,
		B_{ij}^{(t)}\,\1
		\,=\,
		B_{ij}^{(t)}\,U_j^\top\,\1
		\,=\,
		U_i^\top\,A_{ij}^{(t)}\,\1
		\,=\,
		U_i^\top\,v_{ij}^{(t)}\,.
	\]
	Now, from those two relations, we can apply~\cite[Lem~2.3]{RSU94} and conclude that there exists a constant $c_{ij}^{(t)}\in\R$ such that equation~\eqref{eq:lemma-3-6-***} is satisfied. Moreover, by construction of $v_{ij}^{(t)}$, we know that $c_{ij}^{(t)} = |C_j\cap C(g,t)|\in\N$ for any $g\in C_i$. This yields the desired $D$-common equitable partition.
\end{proof}

\subsubsection{$D$-Common Equitable Partition Implies $D$-$\NS$-Isomorphism}
Lastly, we prove that the generalized notion of common equitable partition is sufficient in order to have a perfect non-signalling strategy for the $D$-distance game:

\begin{lemma}   \label{lem:D-common-equitable-partition-implies-D-NS-isom}
	Let $D\geq0$.
	If $(\G,\H)$ admits a $D$-common equitable partition,
	then $\G\cong_{ns}^D\H$.
\end{lemma}

\begin{proof}
	This proof generalizes~\cite[Lemma~4.4]{AMRSSV19}.
	Denote $\big( k, [n_1, \dots, n_k], [c_{ij}^{(t)}]\big)$, for $t\in\{0,\dots, D\}$, the parameters of the given $D$-common equitable partition of $(\G,\H)$, and consider $\overline{c_{ij}}:=n_j - \sum_{t=0}^Dc_{ij}^{(t)}$ the number of elements in $C_j$ that are at distance $>D$ of a fixed element in $C_i$\,.
	We consider the following correlation:
	\[
		\P(h_A, h_B\,|\, g_A, g_B)
		\,=\,
		\left\{
		\begin{array}{cl}
			1/n_i c_{ij}^{(t)} & \text{if $d(h_A,h_B)=t=d(g_A,g_B)$ and $t\leq D$ and $(\star)$\,,}\\
			1/n_i \overline{c_{ij}} & \text{if $d(h_A,h_B)>D$ and $d(g_A,g_B)>D$ and $(\star)$\,,}\\
			0 & \text{otherwise\,,}
		\end{array}
		\right.
	\]
	where the condition $(\star)$ stands for ``$g_A\in C_i, g_B\in C_j, h_A\in D_i, h_B\in D_j$''.
	Moreover, define:
	\[
		\P(h,h'\,|\,g,g')
			=\P(g,h'\,|\,h,g')
			=\P(h,g'\,|\,g,h')
			=\P(g,g'\,|\,h,h')\,,
	\]
	for all $g,g'\in V(\G)$ and all $h,h'\in V(\H)$, and set $\P=0$ in all the cases not yet accounted for.
	By construction, this is a perfect strategy for the $D$-distance game because the probability of losing is zero, so it remains to show that $\P\in\NS$.
	First of all, the non-negativity condition is satisfied by the construction of $\P$. Let $V=V(\G)\cup V(\H)$. 
	Let us check the marginal condition in the case where both inputs $x_A,x_B$ are in $V(\G)$: fix $y_A=h_A\in D_i$ and $x_A=g_A\in C_i$ and $x_B=g_B\in C_j$\,. 
	On the one hand, if $d(g_A,g_B)=t\leq D$, then:
	\[
		\sum_{y\in V} \P(h_A,y\,|\,g_A,g_B)
		\,=\,
		\sum_{h_B\in D_j \cap C(h_A,t)} \frac{1}{n_i\,c_{ij}^{(t)}}
		\,=\,
		\frac{|D_j \cap C(h_A,t)|}{n_i\,c_{ij}^{(t)}}
		\,=\,
		\frac{1}{n_i}\,,
	\]
	because $c_{ij}^{(t)}=|D_j \cap C(h_A,t)|$. 
	On the other hand, if $d(g_A,g_B)>D$, then:
	\[
		\sum_{y\in V} \P(h_A,y\,|\,g_A,g_B)
		\,=\,
		\sum_{h_B\in D_j \cap B(h_A,D)^c} \frac{1}{n_i\,\overline{c_{ij}}}
		\,=\,
		\frac{|D_j \cap B(h_A,D)^c|}{n_i\,\overline{c_{ij}}}
		\,=\,
		\frac{1}{n_i}\,,
	\]
	because $\overline{c_{ij}}=|D_j \cap B(h_A,D)^c|$, 
	where $B(h_A,D)^c$ is the complement of the ball centered at $h_A$ of radius $D$, \ie the element of $V(\H)$ that are at distance $>D$ of $h_A$. In both equations, the result does not depend on $g_B$, hence Alice's marginal $\P(h_A\,|\,g_A)$ is well-defined. Similarly, Bob's marginal $\P(h_B\,|\,g_B)=1/n_j$ is also well-defined using the relation 
	$n_i\,c_{ij}^{(t)}=n_j\,c_{ji}^{(t)}$ from \refprop[Lemma]{lem:relation-betwenn-ct_ij-and-ct_ji}.
	A similar proof works in all the other choices of $x_A,x_B\in V$, using the fact that the parameters $n_i$, and $c_{ij}^{(t)}$, and $\overline{c_{ij}}$ are ``common'' for $\G$ and $\H$, and we have $\P(g_A\,|\,h_A)=1/n_i$ and $\P(g_B\,|\,h_B)=1/n_j$.
	Finally, for any $x_A\in C_i\subseteq V(\G)$ and $x_B\in V$, the normalization condition is verified by summing the marginals:
	\[
		\sum_{y_A,y_B\in V} \P(y_A,y_B\,|\,x_A,x_B)
		\,=\,
		\sum_{y_A\in D_i} \P(y_A\,|\,x_A)
		\,=\,
		\sum_{y_A\in D_i} \frac{1}{n_i}
		\,=\,
		\frac{|D_i|}{n_i}
		\,=\,
		1\,,
	\]
	and similarly in the case $x_A\in D_i\subseteq V(\H)$.
	We therefore obtain the wanted result. 
\end{proof}

The above \refprop[Lemmata]{lem:NS-D-isom-implies-D-fractional}, \ref{lem:if-D-fractional-isom-then-D-common-equitable-partition}, and \ref{lem:D-common-equitable-partition-implies-D-NS-isom} prove the respective implications \ref{item:Characterization-of-perfect-NS-strategies-1}$\Rightarrow$\ref{item:Characterization-of-perfect-NS-strategies-2}, \ref{item:Characterization-of-perfect-NS-strategies-2}$\Rightarrow$\ref{item:Characterization-of-perfect-NS-strategies-3}, and \ref{item:Characterization-of-perfect-NS-strategies-3}$\Rightarrow$\ref{item:Characterization-of-perfect-NS-strategies-1} of \refprop[Theorem]{thm:Characterization-of-perfect-NS-strategies}, hence we obtain the wanted characterization of perfect non-signalling strategies for the $D$-distance game in terms of $D$-fractional isomorphism and of $D$-common equitable partition.

	\subsection{Example of $D$-Isomorphic but not $(D+1)$-Isomorphic Graphs}
	\label{subsec:example-of-D-isom-but-not-(D+1)-isom}

In this subsection, we construct a sequence of graphs $(\G_D,\H_D)$ that are $D$-isomorphic but not $(D+1)$-isomorphic in the sense of the generalized fractional isomorphism defined on page~\pageref{para:Generalized-Fractionnal-Isomorphism}.

We label the vertices of the cycle $\C_n$ from $0$ to $n-1$ clockwise.
The adjacency matrix of this graph is the matrix $A_n:=(a_{ij})$ such that $a_{ij}=1$ if $j=i\pm1\,[n]$, and $a_{ij}=0$ otherwise, where $[n]$ denotes the congruence modulo $n$. 
More generally, for any $t<\frac{n}{2}$, its $t$-adjacency matrix is 
\[A_n^{(t)}:=(a_{ij}^{(t)})_{i,j=0,\ldots,n-1}\,, \quad\text{where}\quad a_{ij}^{(t)}:=\begin{cases} 1&j=i+t\,[n]\,,\\1&j=i-t\,[n]\,,\\0&\textnormal{otherwise}\,.\end{cases}\]
Note that for $t=1$ we recover $A_n=A_n^{(1)}$.
Denote with $\C_{2n}':=\C_n\sqcup \C_n$ the disjoint union of two cycles $\C_n$ on $n$ points. Using block matrix notation, its adjacency and $t$-adjacency matrices are
\[B_{2n}:=\begin{pmatrix} A_n&\0\\\0&A_n\end{pmatrix}, \quad B_{2n}^{(t)}:=\begin{pmatrix} A_n^{(t)}&\0\\\0&A_n^{(t)}\end{pmatrix},\]
for any $t<\frac{n}{2}$.
Finally, let $u_{2n}$ be the following bistochastic matrix:
\[u_{2n}:=\frac{1}{2}\begin{pmatrix} \IdentityMatrix_n&\IdentityMatrix_n\\\IdentityMatrix_n&\IdentityMatrix_n\end{pmatrix},\]
where $\IdentityMatrix_n$ is the $n\times n$ identity matrix.

\begin{lemma}\label{Lem1}
Denoting by $e_0,\ldots, e_{2n-1}\in\CC^{2n}$ the canonical basis of $\CC^{2n}$, the matrices $A_{2n}^{(t)}$, $B_{2n}^{(t)}$, and $u_{2n}$, act as follows, for any $t<\frac{n}{2}$ and $k=0,\ldots,2n-1\,.$
\begin{enumerate}[label=(\alph*)]
\item \label{item:Lem1-a} $A_{2n}^{(t)}e_k=e_{k+t\,[2n]}+ e_{k-t\,[2n]}$.
\item \label{item:Lem1-b} $B_{2n}^{(t)}e_k=\begin{cases}e_{k+t\,[n]}+ e_{k-t\,[n]} &k<n\,,\\e_{(k+t\,[n])+n}+ e_{(k-t\,[n])+n} &k\geq n\,.\end{cases}$
\item \label{item:Lem1-c} $u_{2n} e_k=\frac{1}{2}(e_k + e_{k+n\,[2n]})$\,. 
\item \label{item:Lem1-d} In particular $u_{2n} e_k=u_{2n} e_{k+n}=\frac{1}{2}(e_k + e_{k+n})$, if $k<n$\,.
\end{enumerate} 
\end{lemma}
\begin{proof}
Items~\ref{item:Lem1-a}, \ref{item:Lem1-b} and~\ref{item:Lem1-c} follow directly from the definitions of the matrices. For~\ref{item:Lem1-d}, the result follows from item~\ref{item:Lem1-c} and using the fact that $k+n+n\,[2n]\equiv k\,[2n]$.
\end{proof}

We need the following simple facts from number theory.
\begin{lemma}\label{Lem2}
For $n, t\in\mathbb N$ with $t<\frac{n}{2}$ and $k\in\{0,\ldots, 2n-1\}$, we have\vspace{-2pt}
\begin{enumerate}[label=(\alph*),itemsep=-2pt]
\item \label{item:Lem2-a} $\{k+t\,[n], (k+t\,[n])+n\}=\{k+t\,[2n], k+t+n\,[2n]\}$\,,
\item \label{item:Lem2-b}  $\{k-t\,[n], (k-t\,[n])+n\}=\{k-t\,[2n], k-t+n\,[2n]\}$\,.
\end{enumerate}
\end{lemma}
\begin{proof}
For~\ref{item:Lem2-a}, we check by case distinction:
\begin{center}
\begin{tabular}{l|l|l|l}
&$1\leq k+t <n$&$n\leq k+t<2n$&$2n\leq k+t<3n$\\\hline
$k+t\,[n]$&$k+t$&$k+t-n$&$k+t-2n$\\
$(k+t\,[n])+n$&$k+t+n$&$k+t$&$k+t-n$\\\hline
$k+t\,[2n]$&$k+t$&$k+t$&$k+t-2n$\\
$ k+t+n\,[2n]$&$k+t+n$&$k+t-n$&$k+t-n$
\end{tabular}
\end{center}
Similarly for~\ref{item:Lem2-b}.
\end{proof}

The next proposition is a consequence of the preceding two lemmata.

\begin{proposition}\label{PropTIsomorphic}
For all $n, t\in\mathbb N$ with $t<\frac{n}{2}$, we have $u_{2n}A_{2n}^{(t)}=B_{2n}^{(t)}u_{2n}$\,, which means:
$$\C_{2n}\cong_{frac}^D\C_n\sqcup C_n\,,$$
for all $D<\frac{n}{2}$.
\end{proposition}
\begin{proof}
Let $k\in\{0,\ldots,2n-1\}$ with $k<n$. Then, by \refprop[Lemma]{Lem1}:
\begin{align*}
u_{2n}A_{2n}^{(t)}e_k
&\,=\,u_{2n}(e_{k+t\,[2n]}+ e_{k-t\,[2n]})
\,=\,\textstyle\frac{1}{2}(e_{k+t\,[2n]}+ e_{k+t+n\,[2n]}+e_{k-t\,[2n]}+e_{k-t+n\,[2n]})\,,
\end{align*}
and:
\begin{align*}
B_{2n}^{(t)}u_{2n}e_k
&=\textstyle\frac{1}{2}B_{2n}^{(t)}(e_k + e_{k+n})\\
&=\textstyle\frac{1}{2}(e_{k+t\,[n]}+e_{k-t\,[n]}+ e_{(k+n+t\,[n])+n}+e_{(k+n-t\,[n])+n})\\
&=\textstyle\frac{1}{2}(e_{k+t\,[n]}+ e_{(k+t\,[n])+n}+e_{k-t\,[n]}+e_{(k-t\,[n])+n})\,,
\end{align*}
which by \refprop[Lemma]{Lem2} shows the relation $u_{2n}A_{2n}^{(t)}e_k=B_{2n}^{(t)}u_{2n}e_k$. In the case $k\geq n$, we apply \hyperref[Lem1]{Lemma~\ref{Lem1}}~\ref{item:Lem1-d} and obtain the same result.
\end{proof}

We now give a criterion, when two graphs fail to be fractionally $D$-isomorphic.

\begin{lemma}\label{LemNoDistanceT}
Let $\G$ and $\H$ be two finite graphs with the same number of vertices. If $\diam(\G)\geq D>\diam(\H)$, then $\G$ and $\H$ are not $D$-fractionally isomorphic.
\end{lemma}
\begin{proof}
	Let $(u_{gh})$ a bistochastic matrix indexed by the vertices $g\in V(\G)$ and $h\in V(\H)$. 
	On the one hand, as $D>\diam(\H)$, the $D$-adjacency matrix of $\H$ is zero $A_\H^{(D)}=\0$, therefore:
	$$ 
	u\,A_\H^{(D)}
	\,=\,
	\0\,.
	$$
	On the other hand, we can find in $\G$ two vertices $g_1$ and $g_2$ at a distance exactly $D$, and as $\sum_{h} u_{g_2h}=1$ by bistochasticity, we know there exists at least one vertex $h_2\in V(\H)$ such that $u_{g_2h_2}>0$. It yields that the matrix $A_\G^{(D)}u$ admits a non-zero element:
	$$
		\big[ A_\G^{(D)}\,u \big]_{g_1,h_2}
		\,=\,
		\sum_{g'\in C(g_1,D)}u_{g'h_2}
		\,\geq\,
		u_{g_2h_2}
		\,>\,
		0\,.
	$$
	Hence $u\,A_\H^{(D)}\neq A_\G^{(D)}u$, and the graphs fail to be $D$-fractionally isomorphic.
\end{proof}

Now, combining \refprop[Proposition]{PropTIsomorphic} and \refprop[Lemma]{LemNoDistanceT}, we obtain:

\begin{proposition}\label{prop:ExampleDButNotDPlusOne}
Given $D\in\mathbb N$, the graphs $\C_{2(2D+1)}$ and $\C'_{2(2D+1)}=\C_{2D+1}\sqcup \C_{2D+1}$ are $D$-fractionally isomorphic but not $(D+1)$-fractionally isomorphic:
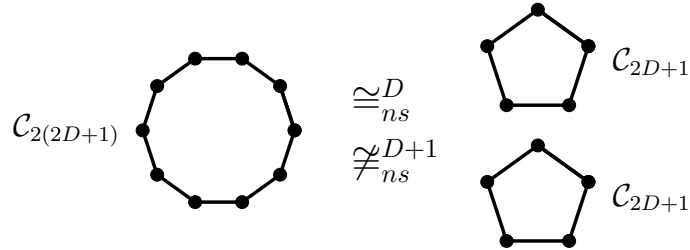
\begin{center}
	
\begin{tikzpicture}[line cap=round,line join=round,>=triangle 45,x=1.0cm,y=1.0cm,every node/.style={scale=1}, scale=1]

\newcommand{\mylinewidth}{1.3pt}
\newcommand{\mycirclesize}{0.085}

\newcommand{\cycle}[1]{
\def \myangle {360/#1}
\foreach \x in {0,1,...,#1} {
	\draw[line width=\mylinewidth, xshift=\graphxshift, yshift=\graphyshift] (\x*\myangle + \rotate : \graphsize) -- (\x*\myangle + \rotate + \myangle : \graphsize);
        \draw[fill=black, xshift=\graphxshift, yshift=\graphyshift] (\x*\myangle+\rotate : \graphsize) circle (\mycirclesize);
        }
}

\draw (-2,0) node {$\C_{2(2D+1)}$};

\newcommand{\rotate}{0}
\newcommand{\graphsize}{1cm}
\newcommand{\graphxshift}{0cm}
\newcommand{\graphyshift}{0cm}
\cycle{10} 

\draw (1.6, 0.4) node[right]{\large$\cong_{ns}^D$};
\draw (1.6, -0.4) node[right]{\large$\not\cong_{ns}^{D+1}$};

\renewcommand{\rotate}{18}
\renewcommand{\graphsize}{0.7cm}
\renewcommand{\graphxshift}{4.2cm}
\renewcommand{\graphyshift}{0.9cm}
\cycle{5} 
\draw (\graphxshift+1.5cm, \graphyshift) node {$\C_{2D+1}$};

\renewcommand{\graphyshift}{-0.9cm}
\cycle{5} 
\draw (\graphxshift+1.5cm, \graphyshift) node {$\C_{2D+1}$};

\end{tikzpicture}

\end{center}
\end{proposition}

\begin{remark}
	More precisely, the graphs $\G=\C_{2(2D+1)}$ and $\H=\C_{2D+1}\sqcup \C_{2D+1}$ are $t$-fractionally isomorphic for $t\leq D=\diam(\H)$ and $t> 2D+1=\diam(\G)$, and are not $t$-fractionally isomorphic for $D<t\leq 2D+1$.
\end{remark}

We obtain the chain of \emph{strict} implications drawn in \refprop[Figure]{fig:chain-of-strict-implications}.
As indicated with the question mark symbols, for the moment we have no example of two graphs that are $D$-fractionally isomorphic for all $D\in\NN$ but not quantum isomorphic. Note however, that there are graphs, which are $D$-fractionally isomorphic for all $D\in\NN$ but not isomorphic, thanks to the example of quantum isomorphic graphs that are non-isomorphic~\cite{AMRSSV19}.
We remark that if $\G\cong_{frac}^{D_0}\H$ for $D_0:=\max\big( \diam(\G), \diam(\H) \big)$, then $\G\cong_{frac}^{D}\H$ for all $D\in\NN$.

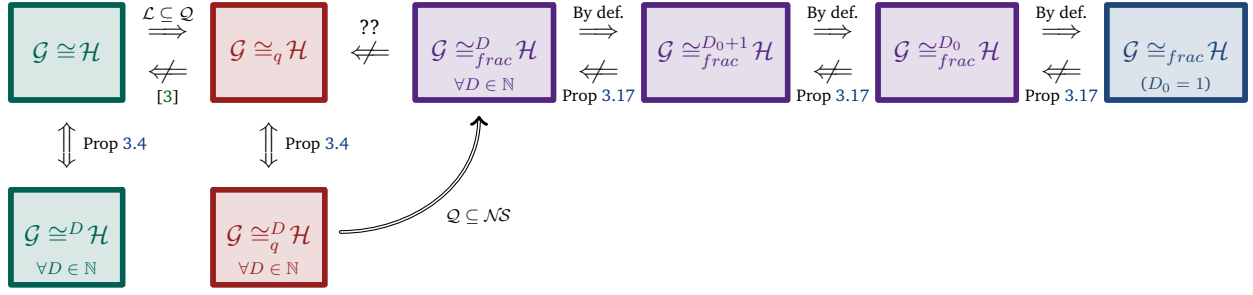
\begin{figure}[h]
	\centering
	\begin{tikzpicture}[line cap=round,line join=round,>=triangle 45,x=1.0cm,y=1.0cm,every node/.style={scale=0.8}, scale=0.7]

\newcommand{\myheight}{1.8}
\newcommand{\mywidthA}{2.1}
\newcommand{\mywidthB}{2.1}
\newcommand{\mywidthC}{2.6}
\newcommand{\mywidthD}{2.7}
\newcommand{\mywidthE}{2.6}
\newcommand{\mywidthF}{2.6}
\newcommand{\mysep}{1.7}
\newcommand{\myyshift}{3.4em}
\newcommand{\myxshift}{3.4em}

\definecolor{mycolor}{RGB}{30, 70, 120} 
\definecolor{mycolor2}{HTML}{53257F} 
\definecolor{mycolor3}{RGB}{191, 95, 0}
\definecolor{mycolor4}{RGB}{0, 96, 81}  
\definecolor{mycolor5}{RGB}{96, 0, 81}
\definecolor{mycolor6}{RGB}{150, 30, 30} 

\draw[line width=2.pt, color=mycolor4, fill=mycolor4, fill opacity=0.15] (0,0) rectangle ++(\mywidthA,\myheight) ;
\draw[line width=2.pt, color=mycolor4] (0,0) rectangle node {\color{mycolor4}$\G\cong\!\H$} ++(\mywidthA,\myheight) ;
\node (A) at (\mywidthA+0.5*\mysep, 0.7*\myheight) {$\Longrightarrow$};
\node[above=of A, yshift=-\myyshift] {\scriptsize$\L\subseteq\Q$};
\node (A') at (\mywidthA+0.5*\mysep, 0.3*\myheight) {$\centernot\Longleftarrow$};
\node[below=of A', yshift=\myyshift] {\scriptsize\cite{AMRSSV19}};

\draw[line width=2.pt, color=mycolor4, fill=mycolor4, fill opacity=0.15] (0,-\myheight-\mysep) rectangle ++(\mywidthA,\myheight) ;
\draw[line width=2.pt, color=mycolor4] (0,-\myheight-\mysep) rectangle node {\color{mycolor4}$\begin{matrix}~\\\G\cong^{D}\!\H\\\text{\scriptsize$\forall D\in\NN$}\end{matrix}$} ++(\mywidthA,\myheight) ;
\node (A'') at (0.5*\mywidthA, -0.5*\mysep) {$\Big\Updownarrow$};
\node[right=of A'', xshift=-\myxshift] {\scriptsize Prop~\ref{prop:same-perfect-classical-and-quantum-strategies}};

\draw[line width=2.pt, color=mycolor6, fill=mycolor6, fill opacity=0.15] (\mywidthA+\mysep,0) rectangle ++(\mywidthB,\myheight) ;
\draw[line width=2.pt, color=mycolor6] (\mywidthA+\mysep,0) rectangle node {\color{mycolor6}$\G\cong_q\!\H$} ++(\mywidthB,\myheight) ;
\node (A') at (\mywidthA+\mywidthB+1.5*\mysep, 0.5*\myheight) {$\centernot\Longleftarrow$};
\node[above=of A', yshift=-\myyshift] {\footnotesize??};

\draw[line width=2.pt, color=mycolor6, fill=mycolor6, fill opacity=0.15] (\mywidthA+\mysep,-\myheight-\mysep) rectangle ++(\mywidthA,\myheight) ;
\draw[line width=2.pt, color=mycolor6] (\mywidthA+\mysep,-\myheight-\mysep) rectangle node {\color{mycolor6}$\begin{matrix}~\\\G\cong_q^{D}\!\H\\\text{\scriptsize$\forall D\in\NN$}\end{matrix}$} ++(\mywidthA,\myheight) ;
\node (A'') at (\mywidthA+\mysep+0.5*\mywidthB, -0.5*\mysep) {$\Big\Updownarrow$};
\node[right=of A'', xshift=-\myxshift] {\scriptsize Prop~\ref{prop:same-perfect-classical-and-quantum-strategies}};

\draw[>=triangle 45, -{to[scale=2]}, double] (\mywidthA+\mywidthB+\mysep+0.3, -0.5*\myheight-\mysep+0.1) to [out=0,in=270] node[below right] {\scriptsize $\Q\subseteq\NS$} (\mywidthA+\mywidthB+0.5*\mywidthC+2*\mysep-0.1, -0.3);

\draw[line width=2.pt, color=mycolor2, fill=mycolor2, fill opacity=0.15] (\mywidthA+\mywidthB+2*\mysep,0) rectangle ++(\mywidthC,\myheight) ;
\draw[line width=2.pt, color=mycolor2] (\mywidthA+\mywidthB+2*\mysep,0) rectangle node {\color{mycolor2}$\begin{matrix}~\\\G\cong_{frac}^{D}\!\H\\\text{\scriptsize$\forall D\in\NN$}\end{matrix}$} ++(\mywidthC,\myheight) ;
\node (A) at (\mywidthA+\mywidthB+\mywidthC+2.5*\mysep, 0.7*\myheight) {$\Longrightarrow$};
\node[above=of A, yshift=-\myyshift] {\scriptsize By def.};
\node (A') at (\mywidthA+\mywidthB+\mywidthC+2.5*\mysep, 0.3*\myheight) {$\centernot\Longleftarrow$};
\node[below=of A', yshift=\myyshift] {\scriptsize Prop~\ref{prop:ExampleDButNotDPlusOne}};

\draw[line width=2.pt, color=mycolor2, fill=mycolor2, fill opacity=0.15] (\mywidthA+\mywidthB+\mywidthC+3*\mysep,0) rectangle ++(\mywidthD,\myheight) ;
\draw[line width=2.pt, color=mycolor2] (\mywidthA+\mywidthB+\mywidthC+3*\mysep,0) rectangle node {\color{mycolor2}$\G\cong_{frac}^{D_0+1}\!\H$} ++(\mywidthD,\myheight) ;
\node (A) at (\mywidthA+\mywidthB+\mywidthC+\mywidthD+3.5*\mysep, 0.7*\myheight) {$\Longrightarrow$};
\node[above=of A, yshift=-\myyshift] {\scriptsize By def.};
\node (A') at (\mywidthA+\mywidthB+\mywidthC+\mywidthD+3.5*\mysep, 0.3*\myheight) {$\centernot\Longleftarrow$};
\node[below=of A', yshift=\myyshift] {\scriptsize Prop~\ref{prop:ExampleDButNotDPlusOne}};

\draw[line width=2.pt, color=mycolor2, fill=mycolor2, fill opacity=0.15] (\mywidthA+\mywidthB+\mywidthC+\mywidthD+4*\mysep,0) rectangle ++(\mywidthE,\myheight) ;
\draw[line width=2.pt, color=mycolor2] (\mywidthA+\mywidthB+\mywidthC+\mywidthD+4*\mysep,0) rectangle node {\color{mycolor2}$\G\cong_{frac}^{D_0}\!\H$} ++(\mywidthE,\myheight) ;
\node (A) at (\mywidthA+\mywidthB+\mywidthC+\mywidthD+\mywidthE+4.5*\mysep, 0.7*\myheight) {$\Longrightarrow$};
\node[above=of A, yshift=-\myyshift] {\scriptsize By def.};
\node (A') at (\mywidthA+\mywidthB+\mywidthC+\mywidthD+\mywidthE+4.5*\mysep, 0.3*\myheight) {$\centernot\Longleftarrow$};
\node[below=of A', yshift=\myyshift] {\scriptsize Prop~\ref{prop:ExampleDButNotDPlusOne}};

\draw[line width=2.pt, color=mycolor, fill=mycolor, fill opacity=0.15] (\mywidthA+\mywidthB+\mywidthC+\mywidthD+\mywidthE+5*\mysep,0) rectangle ++(\mywidthF,\myheight) ;
\draw[line width=2.pt, color=mycolor] (\mywidthA+\mywidthB+\mywidthC+\mywidthD+\mywidthE+5*\mysep,0) rectangle node {\color{mycolor}$\begin{matrix}~\\\G\cong_{frac}\!\H\\\text{\scriptsize($D_0=1$)}\end{matrix}$} ++(\mywidthF,\myheight) ;

\end{tikzpicture}
	\caption{Chain of strict implications, with $D_0\geq2$ fixed. 
	} 
	\label{fig:chain-of-strict-implications}
\end{figure}

We conclude the subsection with the following theorem, which is a consequence of \refprop[Theorem]{thm:Characterization-of-perfect-NS-strategies} and \refprop[Figure]{fig:chain-of-strict-implications}, which tells us that we can distinguish perfect non-signalling strategies between the $D$-distance game and the graph isomorphism game, as opposed to the classical and quantum cases:

\begin{theorem}[$\NS$ is Finer Than $\L$ and $\Q$] \label{theo:strict-ns-D-isomorphisms}
	As opposed to the classical and quantum cases (\refprop[Subsection]{subsec:characterizing-perfect-strategies-at-the-vertex-distance-game}), the set of perfect non-signalling strategies for the $D$-distance game is strictly included in the set of perfect non-signalling strategies for the isomorphism game. \qed
\end{theorem}

\subsection{Links with Communication Complexity}

In this subsection, we give statements showing the collapse of communication complexity in various cases. These statements are mainly generalizations of results from the first two sections about the isomorphism and coloring games.

\subsubsection{Existence of Collapsing Non-Signalling Strategies} 
To show the existence of a perfect collapsing strategy, we want to adapt \refprop[Theorem]{theo: collapse of CC} to the $D$-distance game.
To this end, we generalize two results from the isomorphism game to the $D$-distance game. 
	First, see that \refprop[Proposition]{prop: sufficient condition} can be easily generalized using the characterization of perfect strategies in terms of $D$-common equitable partition (\hyperref[thm:Characterization-of-perfect-NS-strategies]{Theorem~\ref{thm:Characterization-of-perfect-NS-strategies}}), and it gives:
	
\begin{lemma}
	Let $\G\cong^D_{ns} \H$ such that $\H$ is not connected: $\H=\H_1\sqcup \H_2$. Denote the partitions $\Cpartition=\{C_1, \dots, C_k\}$ and $\Dpartition=\{D_1, \dots, D_k\}$ forming a $D$-common equitable partition for $\G$ and $\H$, and assume that the proportion of vertices of $\H_1$ assigned to $D_i$ is independent of $i$:
\begin{equation}   \label{eq:H-prime}
			\tag*{(H$'$)}
	\forall i,j\in[k],\quad\quad
	\frac{|D_i\cap \H_1|}{|D_i|}
	\,=\,
	\frac{|D_j\cap \H_1|}{|D_j|}\,.
\end{equation}   
Then the $D$-distance game of $(\G,\H)$ admits a symmetric perfect strategy of the following form:
	\[
		\PP_\S(h_A,h_B\,|\, g_A,g_B)
	\,:=\,
	\left\{
	\begin{array}{c l}
		1/n_i c^{(t)}_{ij} & \text{if $d(g_A, g_B)=d(h_A, h_B)=t\in\{0,\dots,D\}$ and $(\star)$},\\
		1/n_i \overline{c_{ij}} & \text{if $d(g_A, g_B)>D$ and $d(h_A, h_B)>D$ and $(\star)$},\\
		0 & \text{otherwise},
	\end{array}
	\right.
	\]
	where $(\star)$ denotes the condition ``$g_A\in C_i,\, g_B\in C_j,\, h_A\in D_i,\, h_B\in D_j$". \qed
\end{lemma}
	
	Then, using similar arguments as in the proof of \refprop[Proposition]{prop:easy-case-D-distance-game}, we observe that \hyperref[{theo: simulate P alpha beta}]{Lemma~\ref{theo: simulate P alpha beta}} can be generalized straightforwardly as follows:
	
\begin{lemma}  
Let $\G,\H$ two graphs such that $1\leq D<\diam(\G)$ and such that $\H$ is not connected: $\H=\H_1\sqcup \H_2$. There exists a path $\Path\subseteq\G$ of length $D+1$, for which we call $g_1$ and $g_3$ the extremal vertices. Assume $\G\cong^D_{ns}\H$ for some strategy $\S$ that is symmetric from $\Path$ to the components of $\H$, and suppose that:
\begin{equation*}
\nu_{g_1, g_3} >0\,.
\end{equation*}
Then the box $\PR_{\alpha,\beta}$  is perfectly simulated with  $\alpha=2\,\nu_{g_1, g_3}>0$ and some $\beta\geq0$.\qed
\end{lemma}

	Now, using these two generalized lemmata, the exact same proof as the one of \refprop[Theorem]{theo: collapse of CC} also gives the result for the $D$-distance game:

\begin{theorem}[Existence of Collapsing Strategies]   \label{theo:existence-of-collapsing-strategy-for-the-D-distance-game}
Let $\G\cong_{ns}^D \H$ for some $1\leq D<\diam(\G)$
and such that $\H$ is not connected: $\H=\H_1\sqcup \H_2$, where each of $\H_1$ and $\H_2$ may possibly be decomposed in several connected components. 
Denote the partitions $\Cpartition=\{C_1, \dots, C_k\}$ and $\Dpartition=\{D_1, \dots, D_k\}$ forming a $D$-common equitable partition for $\G$ and $\H$, and assume that condition~\ref{eq:H-prime} holds.
Then the $D$-distance game of $(\G,\H)$ admits a perfect strategy that collapses communication complexity. \qed
\end{theorem}

\subsubsection{All Perfect Non-Signalling Strategies Collapse CC} 

Now, we want to prove sufficient conditions so that all perfect strategies for the $D$-distance game collapse communication complexity.
We begin the study with the simple case where the graph $\H$ has a smaller diameter than $\G$. First, we assume that $\H$ admits exactly $2$ connected components, and then more generally $N$ connected components. 

\begin{proposition}[Collapse of CC]
\label{prop:easy-case-D-distance-game}
If $\diam(\G)>\diam(\H)\geq D\geq1$ and if $\H$ admits exactly two connected components, then any perfect $\NS$-strategy for the $D$-distance game collapses communication complexity.
\end{proposition}

\begin{proof}
By assumption, there exist vertices $g_1, g_3$ in $\G$ whose distance is exactly $\diam(\H)+1$. In a minimal path joining $g_1$ to $g_3$ in $\G$, consider $g_2$ at distance $D$ of $g_1$ and distance $\diam(\H)+1-D$ of $g_3$.
Assume that there exists a perfect strategy $\S$ for the $D$-distance game.
Similarly to the proof of \refprop[Theorem]{theo:toy-example-thm}, Alice and Bob will use this perfect strategy $\S$ as a black box to generate a $\PR$ box, which is known to collapse communication complexity~\cite{vD99}. 
Suppose Alice and Bob are given respective bits $x,y\in\{0,1\}$. They want to produce $a,b\in\{0,1\}$ without signalling such that $a\oplus b = xy$. If $x=0$, Alice chooses $g_A=g_2$, and if $x=1$, she chooses $g_A=g_1$. As for Bob, given respectively $y=0,1$, he chooses $g_B=g_2, g_3$. Alice and Bob input their choice $(g_A, g_B)$ in the strategy $\S$, which outputs some vertices $(h_A, h_B)$ of $\H$ satisfying the conditions of the $D$-distance game. Notice that $h_A$ and $h_B$ are in different connected components of $\H=\H_1\sqcup\H_2$ if and only if $x=y=1$. Upon receiving $h_A\in\H_i$, Alice produces the bit $a=i$, and similarly for Bob with $h_B\in\H_j$ and $b=j$. It follows that the relation $a\oplus b=xy$ is always satisfied, thus the $\PR$ box is perfectly simulated, and there is a collapse of communication complexity.
\end{proof}

\begin{remark}
	Actually, it is enough to have a noisy $\NS$-strategy winning the $D$-distance game with probability $\pp> \frac{3+\sqrt{6}}{6}$, since the same proof would generate a $\PR$ box with probability $\pp$ and therefor collapse CC by~\cite{BBLMTU06}.
\end{remark}

\begin{proposition}[Collapse of CC]
\label{prop:easy-case-D-distance-game-2}
If $\diam(\G)>\diam(\H)\geq D\geq1$ and if $\H$ admits exactly $N$ connected components, then any perfect $\NS$-strategy for the $D$-distance game, combined with a perfect $\NS$-strategy for the $2$-coloring game of $\K_N$, collapses communication complexity.
\end{proposition}	

\begin{proof}
Proceed as in the proof of \refprop[Theorem]{thm:combining-isomorphism-game-and-coloring-game} combined with \refprop[Proposition]{prop:easy-case-D-distance-game}.
\end{proof}

\begin{remark}
	Again, we can generalize this result to a noisy version: it is enough that we have $\pp$ and $\qq$ such that the product satisfies $\pp\qq>\frac{3+\sqrt{6}}{6}$, that the $\NS$-strategy for the $D$-distance game wins with probability $\pp$, and that the $\NS$-strategy for the $2$-coloring game of $\K_N$ wins with probability $\qq$.
\end{remark}

Finally, the following statement is a particular case of \refprop[Theorem]{theo:collapse-for-ALL-strategies}. 
Indeed, any perfect strategy for the $D$-distance game is perfect for the isomorphism game. In the theorem, we gave sufficient conditions on graphs so that all perfect strategies for the isomorphism game collapse CC. Hence, with the same conditions, we have that the result also holds for the $D$-distance game for any $D\geq 1$. Recall that the proof of this theorem was based on tools from graph automorphism theory and graph transitivity notions.

\begin{theorem}[Collapse of CC]  \label{theo:collapse-for-ALL-strategies-D-distance-game} 
Let $D\geq1$.
Let $\G\cong_{ns}^D \H$ such that $2\leq\diam(\G)$
and such that $\H$ is not connected: $\H=\H_1\sqcup \H_2$, where each of $\H_1$ and $\H_2$ may possibly be decomposed in several connected components. 
Let $\Cpartition=\{C_1, \dots, C_k\}$ and $\Dpartition=\{D_1, \dots, D_k\}$ form a $1$-common equitable partition for $\G$ and $\H$ such that condition~\ref{eq:H-prime} holds.
Assume moreover that $\H$ is strongly transitive and $d$-regular, and that the players share randomness. 
Then \emph{every} perfect non-signalling strategy for the $D$-distance game of $(\G,\H)$ collapses communication complexity. \qed
\end{theorem}

\bib

	\section*{Conclusion}
	\addcontentsline{toc}{section}{Conclusion}
	\label{sec:conclusion}

Our investigations contribute to the line of research that ``the quantum world is the best possible one'': indeed, we just proved for many instances that, going beyond the quantum case, some non-signalling correlations collapse CC -- and are hence physically unfeasible according to the present intuition of nature~\cite{vD99, BBLMTU06, BS09, BG15}. 
The importance of our results is emphasized by considering the many known impossibility results~\cite{BG15, Mori16, SWH20}.
However, notice that there is still a gap to be filled in between: there remain some non-signalling correlations for which we need to know if they are feasible according to the principle of CC.
See a recent overview of what remains to be shown in~\cite[Figure~2]{BBP23}.

In addition to the study of CC, our work also adds value to the study of non-signalling correlations themselves -- not only for promoting the quantum theory, but also as a field of research as such. Indeed,  we remark that although there is no difference between our new vertex $D$-distance game and the graph isomorphism game in terms of perfect classical or quantum strategies, the two games do not admit the same perfect non-signalling strategies. More precisely, we proved that it is even possible to build a sequence of graphs $(\mathcal G_D,\mathcal H_D)$ that are $D$-isomorphic but not $(D + 1)$-isomorphic in the sense of the vertex distance game. Therefore, non-signalling strategies yield a finer distinction of this non-local game than quantum and classical strategies.

We list a few open questions for future research: \vspace{-0.4cm}

\paragraph{Open Question~1.} While results~\ref{theo:toy-example-thm}, \ref{coro:toy-example-thm}, \ref{thm:collapse-of-CC-for-the-homomorphism-game}, \ref{thm:combining-isomorphism-game-and-coloring-game}, \ref{prop:easy-case-D-distance-game}, \ref{prop:easy-case-D-distance-game-2}, and~\ref{theo:collapse-for-ALL-strategies-D-distance-game} are in their optimal form in the sense that ``\emph{any} perfect non-signalling strategy collapses CC'', we see that results~\ref{theo: collapse of CC} and~\ref{theo:existence-of-collapsing-strategy-for-the-D-distance-game} await an improvement since they only state that ``\emph{there exists} a perfect non-signalling strategy which collapses CC''. The latter statement is much weaker and it would be of great interest to infer that \emph{any} perfect strategy collapses CC. Would it be possible to turn the quantifiers from ``there exists'' to ``for all'' in these results?\vspace{-0.4cm}

\paragraph{Open Question~2.} In \refprop[Figure]{fig:characterizations-of-the-different-types-of-isomorphism}, Lovasz-type characterizations of certain variants of graph isomorphisms are given in terms of homomorphism counts: classical isomorphism is characterized by homomorphism counts from all graphs, quantum isomorphism amounts to counts on planar graphs, and non-signalling/fractional isomorphism is given by counts on trees. 
What would be the correct type of homomorphism counts characterizing the perfect strategies for the new $D$-fractional isomorphism?
Intuitively, this should be a class of graphs in between planar graphs and trees, with a dependency on the parameter $D$.\vspace{-0.4cm}

\paragraph{Open Question~3.} In \refprop[Figure]{fig:chain-of-strict-implications}, we list the relevant notions of graph isomorphism presenting a chain of implications from classical isomorphism to quantum isomorphism and then to several variants of the $D$-fractional isomorphisms: first for all $D\in\mathbb N$, then for a fixed $D\in\mathbb N$, and lastly for $D=1$ which amounts to the previously known fractional isomorphism. We know in all but one cases that the implications are strict. The only missing implication is: are there graphs $\mathcal G$ and $\mathcal H$ which are $D$-fractional isomorphic for all $D\in\mathbb N$, but which are not quantum isomorphic?\vspace{-0.4cm}

\paragraph{Open Question~4.} In this work, we chose to study the principle of communication complexity. Although this principle alone cannot rule out the quantum set~\cite{NGHA15}, a clever idea would be to combine it with other principles, such as \emph{nonlocal computation}~\cite{LPSW07}, \emph{information causality}~\cite{PPKSWZ09, JGM23}, \emph{macroscopic locality}~\cite{NW09}, \emph{local orthogonality}~\cite{FSABCLA13}, \emph{nonlocality swapping}~\cite{SBP09}, \emph{many-box locality}~\cite{ZCBGS17}, in working towards a comprehensive information-based description of Quantum Mechanics. Therefore, an interesting work could be to consider the study of these graph games for other principles than communication complexity: for a different principle, do we have major improvements on the set of non-feasible correlations?
We leave these questions open for future work.

\bib

	\section*{Acknowledgements} 

This work began during the Focus Semester on Quantum Information held at Saarland University from September to December 2022. The authors thank Simon Schmidt for many insightful discussions and for pointing out the relation of strong transitivity with distance-transitivity, see \refprop[Lemma]{lem:characterization-of-strong-transitivity}.
They also thank Anne Broadbent, Ion Nechita, and Clément Pellegrini for many constructive comments.
P.B. acknowledges the support of the Institute for Quantum Technologies in Occitanie. M.W. has been supported by the SFB-TRR 195, the Heisenberg program of the DFG, and a joint OPUS-LAP grant with Adam Skalski. 
This article is part of the PhD thesis of P.B.

\addcontentsline{toc}{section}{References}

\printbibliography


\end{document}